\documentclass[a4paper,12pt]{article}
\pdfoutput=1
\usepackage{jheppub}

\usepackage[T1]{fontenc} 

\usepackage{amssymb,amsmath,bm,bbm,natbib}
\usepackage{color}
\usepackage{slashed}
\usepackage{graphics}
\usepackage{graphicx}
\usepackage{inputenc}
\usepackage[caption=false]{subfig}
\usepackage{url}
\usepackage{dsfont}
\usepackage{float} 
\usepackage{cancel}
\usepackage{xcolor,colortbl}
\usepackage{rotating}
\usepackage[export]{adjustbox}
\usepackage{subfig}
\usepackage{textcomp}

\newcommand{\eq}[1]{Eq.~\eqref{#1}}

\newcommand{\modified}[1]{{\color{magenta}{#1}}}
\definecolor{Gray}{gray}{0.7}

\preprint{
	AJB-21-3,    
CERN-2021-56, PSI-PR-21-06,   \begin{flushright} ZU-TH  15/20 \end{flushright}}

	\title{\boldmath Global Analysis of Leptophilic $Z^\prime$ Bosons}
	
\author[a]{Andrzej~J.~Buras,}

\author[b,c,d]{Andreas Crivellin,}

\author[c,d]{Fiona Kirk,}
		
\author[c,d]{Claudio Andrea Manzari,}

\author[c,d]{Marc Montull}

\affiliation[a]{TUM Institute for Advanced Study, Lichtenbergstr. 2a, D--85747 Garching, Germany}
\affiliation[b]{CERN Theory Division, CH--1211 Geneva 23, Switzerland}
\affiliation[c]{Physik-Institut, Universit\"at Z\"urich, Winterthurerstrasse 190, CH--8057 Z\"urich, Switzerland}
\affiliation[d]{Paul Scherrer Institut, CH--5232 Villigen PSI, Switzerland}
	
\emailAdd{andrzej.buras@tum.de}
\emailAdd{andreas.crivellin@cern.ch}
\emailAdd{fiona.kirk@psi.ch}
\emailAdd{claudioandrea.manzari@physik.uzh.ch}
\emailAdd{marc.montull@psi.ch}
\abstract
{New neutral heavy gauge bosons ($Z^\prime$) are predicted within many extensions of the Standard Model. While in case they couple to quarks the LHC bounds are very stringent, leptophilic $Z^\prime$ bosons (even with sizable couplings) can be much lighter and therefore lead to interesting quantum effects in precision observables {(like $(g-2)_\mu$)} and generate flavour violating decays of charged leptons. In particular, $\ell\to\ell^\prime\nu\bar\nu$ decays, anomalous magnetic moments of charged leptons, $\ell\to\ell^\prime\gamma$ and $\ell\to3\ell^\prime$ decays place stringent limits on leptophilic $Z^\prime$ bosons. Furthermore, in case of mixing $Z^\prime$ with the SM $Z$, $Z$ pole observables are affected. In light of these many observables we perform a global fit to leptophilic $Z^\prime$ models with the main goal of finding the bounds for the $Z^\prime$ couplings to leptons. To this end we consider a number of scenarios for these couplings.  While in generic scenarios correlations are weak, this changes once additional constraints on the couplings are imposed. In particular, if one considers an $L_\mu-L_\tau$ symmetry broken only by left-handed rotations, or considers the case of $\tau-\mu$ couplings only. In the latter setup, on can explain  the $(g-2)_\mu$ anomaly and the hint for lepton flavour universality violation in $\tau\to\mu\nu\bar\nu/\tau\to e\nu\bar\nu$ without violating bounds from electroweak precision observables. }

\begin{document} 
\maketitle
\flushbottom

\newpage

\section{Introduction}
\label{Introduction}

In 2012, the LHC confirmed the predictions of the Standard Model (SM) of particle physics by discovering the (Brout-Englert) Higgs boson~\cite{Aad:2012tfa,Chatrchyan:2012xdj}. However, so far no particles beyond the ones of the SM have been observed in high energy searches. In particular, the bounds from di-jet~\cite{Aaboud:2017yvp,Sirunyan:2018xlo} and di-lepton~\cite{Aad:2019fac,CMS:2019tbu,Sirunyan:2021khd} searches on particles that can be produced resonantly in the s-channel are very stringent. This also puts tight bounds on heavy neutral gauge bosons ($Z^\prime$s), which are predicted by many new physics models (see \cite{Langacker:2000ju,Langacker:2008yv,Buras:2012jb,Altmannshofer:2015mqa,Crivellin:2015era,Allanach:2019mfl,Buras:2020xsm}), in case they have sizable couplings to quarks.
\smallskip

However, in case the resonances are neutral and couple (to a good approximation) only to leptons, mostly LEP searches apply and the bounds are much weaker~\cite{Schael:2013ita}, i.e. significantly below the TeV scale. Therefore, such leptophilic $Z^\prime$ bosons can have sizable couplings while at the same time being quite light. They can thus lead to relevant quantum corrections to leptonic precision observables and generate lepton flavour violating decays of leptons that are extremely suppressed in the SM since they vanish in the limit of massless neutrinos. The $(g-2)_\mu$ discrepancy~\cite{Bennett:2006fi,Mohr:2015ccw}, recently reinforced by the $g-2$ experiment at Fermilab~\cite{Abi:2021gix,Albahri:2021kmg,Albahri:2021ixb,Albahri:2021mtf}, with a tension of $4.2\,\sigma$ compared to the SM prediction~\cite{Aoyama:2020ynm}, can be explained with a $Z^\prime$ boson heavier than the electroweak (EW) scale if it couples flavour violatingly to the second and third lepton generation~\cite{Foot:1994vd,Gninenko:2001hx, Murakami:2001cs, Baek:2001kca, Ma:2001md, Pospelov:2008zw, Heeck:2011wj, Davoudiasl:2012ig, Carone:2013uh, Harigaya:2013twa, Altmannshofer:2014cfa, Tomar:2014rya, Altmannshofer:2014pba, Lee:2014tba, Allanach:2015gkd, Heeck:2016xkh, Patra:2016shz,Altmannshofer:2016brv,Iguro:2020rby}.
\smallskip

At first sight, new neutral gauge bosons coupling only to leptons and not to quarks might appear artificial. But as already in the SM gluons couple only to quarks and not to leptons, it is actually an interesting possibility that for a $Z^\prime$ boson the situation could be reversed. For example, gauged abelian flavour symmetries in the lepton sector, such as $L_\mu-L_\tau$~\cite{He:1990pn,Foot:1990mn,He:1991qd}, can naturally generate the observed pattern of the PMNS matrix~\cite{Binetruy:1996cs,Bell:2000vh,Choubey:2004hn} and lead by definition to leptophilic $Z^\prime$ bosons, which, after the breaking of the symmetry, can also induce charged LFV processes such as $\tau\to3\mu$~\cite{Dutta:1994dx,Heeck:2011wj,Crivellin:2015mga} and $h\to \mu \tau$~\cite{Heeck:2014qea,Crivellin:2015mga,Altmannshofer:2016oaq}.
\smallskip

{Such scenarios are particularly interesting as within recent years several hints for the violation of lepton flavour universality (LFU) have been acquired. These include $\tau\to\mu\nu\bar\nu/\tau\to e\nu\bar\nu$, $\tau\to\mu\nu\bar\nu/\mu\to e\nu\bar\nu$~\cite{Aubert:2009qj,Amhis:2019ckw} and the Cabibbo angle anomaly~\cite{Belfatto:2019swo,Grossman:2019bzp,Shiells:2020fqp,Seng:2020wjq}, which can also be interpreted as a sign of LFUV~\cite{Coutinho:2019aiy,Crivellin:2020lzu,Coutinho:2020xhc,Capdevila:2020rrl,Crivellin:2020ebi,Kirk:2020wdk,Alok:2020jod,Crivellin:2020oup,Crivellin:2020klg,Crivellin:2021egp,Crivellin:2021njn,Crivellin:2021rbf}. Furthermore, even though in this case also (small) couplings} to bottom and strange quarks are necessary, $Z^\prime$ bosons are among the prime candidates for explaining the discrepancies between the SM predictions and data in $b\!\to\! s\ell^+\ell^-$ transitions~\cite{Buras:2013qja,Gauld:2013qba,Gauld:2013qja,Altmannshofer:2014cfa,Crivellin:2015mga,Crivellin:2015lwa,Niehoff:2015bfa,Carmona:2015ena,Falkowski:2015zwa,Celis:2015eqs,Celis:2015ara,Crivellin:2015era,Crivellin:2016ejn,GarciaGarcia:2016nvr,Altmannshofer:2016oaq,Faisel:2017glo,King:2017anf,Chiang:2017hlj,DiChiara:2017cjq,Ko:2017lzd,Sannino:2017utc,Falkowski:2018dsl,Benavides:2018rgh,Maji:2018gvz,Singirala:2018mio,Guadagnoli:2018ojc,Allanach:2018lvl,Duan:2018akc,King:2018fcg,Kohda:2018xbc,Dwivedi:2019uqd,Foldenauer:2019vgn,Ko:2019tts,Allanach:2019iiy,Altmannshofer:2019xda,Calibbi:2019lvs,Aebischer:2019blw,Crivellin:2020oup,Allanach:2020kss,Greljo:2021xmg}. Here, LHCb measurements~\cite{Aaij:2017vbb,Aaij:2019wad} indicate a deficit in muons with respect to electrons, i.e. LFUV with a combined significance of $\approx4\sigma$~\cite{Capdevila:2017bsm, Altmannshofer:2017yso, DAmico:2017mtc, Ciuchini:2017mik, Hiller:2017bzc, Geng:2017svp,Hurth:2017hxg,Alguero:2019ptt,Aebischer:2019mlg,Ciuchini:2019usw,Arbey:2019duh}. This is consistent with many other measurements involving the same current, in particular with angular observables~\cite{Matias:2012xw,Descotes-Genon:2013vna}, where data also shows a deficit in muonic channels~\cite{Aaij:2015oid,Aaij:2020nrf} such that the most up-to-date global analysis finds several NP scenarios to be preferred over the SM at the $5-6\sigma$ level~\cite{Alguero:2019ptt,Aebischer:2019mlg,Ciuchini:2019usw,Ciuchini:2020gvn,Altmannshofer:2021qrr}. In order to respect LHC bounds, it is again advantageous if the couplings to quarks are small, i.e. if the $Z^\prime$ is to a good approximation leptophilic, which can e.g. be achieved by generating the quark couplings effectively via heavy vector-like quarks~\cite{Altmannshofer:2014cfa,Crivellin:2015mga,Bobeth:2016llm}.  
\smallskip

Therefore, it is very interesting to explore the phenomenology of leptophilic $Z^\prime$ bosons. Even in the absence of quark couplings, such a $Z^\prime$ boson affects many observables with the most interesting being
\begin{itemize}
	\item $\ell\to\ell^\prime\gamma$ {decays},
	\item {Anomalous magnetic moments (AMMs) and electric dipole moments (EDMs) of charged leptons,}
	\item $\ell\to3\ell^\prime$ {decays (7 in total),}
	\item $Z\to \ell\ell^{(\prime)}$ decays,
	\item LFU violation in $\ell\to \ell^\prime\nu\bar\nu$
	\item Neutrino trident production
	\item LEP searches for contact interactions
\end{itemize}
Therefore, in order to fully explore the allowed/preferred parameter space of such a new physics
 scenario, the aim of this article is to perform a global fit to all available data
 in a number of scenarios for $Z^\prime$ couplings to leptons.
For this we will use the publicly available HEPfit code~\cite{deBlas:2019okz} which also allows
 us to perform a global fit taking into account many degrees of freedom at the same time.
\smallskip

The article is structured as follows: In the next section we will define our setup before we consider the relevant observables, and calculate the relevant $Z^\prime$ contributions to them in Sec.~\ref{observables}. We then perform our phenomenological analysis in Sec.~\ref{phenomenology} before concluding in Sec.~\ref{conclusions}. In Appendix~\ref{app:LEP2} we give some details on the LEP-II bounds and in Appendix~\ref{QEDP} we list the contributions of QED penguins
to LFV decays like $\ell\to 3 \ell^\prime$ and to $\mu\to e$ conversion in nuclei. In Appendix~\ref{app:C} we present additional scenarios for LFV couplings beyond those presented in the main text.
\medskip

\section{Setup}
\label{Setup}
We extend the SM by adding a heavy neutral gauge boson $Z'_0$ (i.e. with a mass above the electroweak symmetry breaking scale). Following Ref.~\cite{delAguila:2010mx,deBlas:2012qp} we supplement the SM Lagrangian by a part containing the kinetic terms and the mass terms of the $Z'_0$-field,
\begin{equation}
\mathcal{L}_{Z'_0}=-\frac{1}{4}Z'_{0,\mu\nu}Z_0^{'\mu\nu}+\frac{\mu_Z'^2}{2}Z'_{0\mu} Z'^\mu_0\,,\label{LZp}
\end{equation}
where $Z'_{0,\mu\nu}\equiv \partial_\mu Z'_{0\nu}-\partial_\nu Z'_{0\mu}$ is the field strength tensor associated to the $Z'_0$-field, and a part describing the interactions of the $Z'_0$-field with the SM fields, 
\begin{align}
\mathcal{L}_{Z'_0}^{int}&=g_{Z'}Z'_{0\mu} Z'^\mu_0\phi^\dagger \phi 
-i g_{Z'}^\phi Z'^\mu_0\phi^\dagger \overleftrightarrow{D}_\mu \phi \\
&+\overline{\ell}_i \left(g^{L}_{ij}\gamma_\mu P_L +g^{R}_{ij}\gamma_\mu P_R\right)\ell_j Z'^\mu_0 +\overline{\nu}_i g^{L}_{ij}\gamma_\mu P_L\nu_j Z'^\mu_0\,,\notag \label{Lint}
\end{align}
where $\overset{\leftrightarrow}{D}_{\mu}\; =\overset{\rightarrow}{D}_{\mu}- (\overset{\leftarrow}{D}_{\mu})^\dagger$ and
$g_{Z'}^\phi$ is real. Note that here the subscript $0$ refers to the fact that these are not mass, but rather interaction eigenstates. $g_{ij}^{L/R}$ are hermitian and due to {$\text{SU(2)}_L$} invariance the coupling to neutrinos is the same as to left-handed charged leptons\footnote{Here we neglected small active neutrino masses and therefore set the PMNS matrix to the unit matrix.}. $\phi$ is the SM Higgs $SU(2)_L$ doublet and we use
\begin{equation}
D_\mu=\partial_\mu + ig_2W^a_\mu T^a+ig_1YB_\mu \label{CovariantDerivative}\,,
\end{equation}
as the definition of the covariant derivative.
\smallskip

The coupling $g_{Z'}^\phi$ leads to mixing of the $Z'_0$-boson with the SM $Z$. The corresponding mass matrix in the interaction eigenbasis ($Z_0,\,Z_0'$) is then given by
\begin{equation}
\mathcal{M}^2=\begin{pmatrix}
m_{Z_0}^2 & -\frac{y}{c_W}\\
-\frac{y}{c_W}& M_{Z'_0}^2
\end{pmatrix}\,, \qquad    y\equiv \frac{v^2}{2}\,g_2 \,g_{Z'}^\phi
\end{equation}
with $m_{Z_0}^2=\frac{v^2}{4}\left(g_1^2+g_2^2\right)$ and $\frac{v}{\sqrt{2}}\approx 174\,$TeV. To order $\frac{v^2}{m_{Z_0'}^2}$ the eigenvalues are 
\begin{align}
m_Z^2\simeq & \;m_{Z_0}^2-\frac{y^2}{c_W^2 M_{Z_0 '}^2}\equiv m_{Z_0}^2\left(1+\delta m_Z^2\right)\,,\label{MZcorr}\\
M_{Z'}^2\simeq & \;M_{Z_0'}^2+\frac{y^2}{c_W^2M_{Z_0'}^2}\,\label{MZpcorr}\,.
\end{align}
Hence the corrections to the mass of the SM $Z_0$ can only be destructive. The mass eigenstates $Z^{(\prime)}$ can then be expressed as
\begin{equation}
\left( {\begin{array}{*{20}{c}}
	{Z^\prime} \\	Z
	\end{array}} \right) = \left( {\begin{array}{*{20}{c}}
	{Z_{0}^{\prime} \cos \xi \; - Z_{0} \sin \xi}\\
	{Z_{0}^{\prime} \sin \xi + Z_{0} \cos \xi }
	\end{array}} \right)
\label{eq:mixing}
\end{equation}
where 
\begin{equation}
\sin \xi\simeq \frac{y}{c_W M_{Z_0'}^2}
\label{sinalpha}
\end{equation}
describes the $Z_0-Z'_0$ mixing. Note that only the relative phase between $\sin \xi$ and $g^{L,R}_{ij}$ is physical. Therefore, one can assume one of the diagonal couplings $g^{L,R}_{ii}$ to be positive without loss of generality. We write the interactions of the SM $Z$ with fermions as 
\begin{align}
\mathcal{L}_{Zff}=&\;\overline{\ell}_i \gamma_\mu \left(\Delta_{ij}^{\ell L} P_L+\Delta_{ij}^{\ell R} P_R\right) \ell_j Z^\mu +\overline{\nu}_i \gamma_\mu \Delta_{ij}^{\nu L} P_L\nu_j Z^\mu \notag\\
&+\overline{u}_k\,\gamma_\mu \left(g_{\rm SM}^{uL}P_L+g_{\rm SM}^{uR}P_R\right)u_k\,Z^\mu+\overline{d}_l\,\gamma_\mu \left(g_{\rm SM}^{dL}P_L+g_{\rm SM}^{dR}P_R\right)d_l\,Z^\mu\,,
\end{align}
with $i,j = e,\mu,\tau$, $k=u,c,t$, $l=d,s,b$ and
\begin{equation}
\begin{aligned}
\Delta_{ij}^{\ell L,R} \simeq &\sin \xi \, g_{ij}^{L,R}+\,g_{\rm SM}^{\ell L,R}\delta_{ij}\,, \qquad
\Delta_{ij}^{\nu L} \simeq \sin \xi \, g_{ij}^{L}+\,g_{\rm SM}^{\nu L}\delta_{ij}\,,
\end{aligned}
\label{Deltas}
\end{equation}
where $g_{\rm SM}^{\ell,\nu \,L,R}$ are the SM $Z {\overline{\ell}\ell}$ and $Z\overline{\nu}\nu$ couplings given by 
\begin{align}
\begin{aligned}
g_{\rm SM}^{\nu L} &=-\frac{e}{2s_W c_W}\,,\\
g_{\rm SM}^{\ell L} &=-\frac{e}{2s_W c_W}\left(-1+2 s_W^2\right)\,,\qquad
g_{\rm SM}^{\ell R} =-\frac{e \,s_W}{c_W}\,,\\
g_{\rm SM}^{uL}&=-\frac{e}{s_W c_W}\left(\frac{1}{2}-\frac{2}{3}s_W^2\right)\,,\qquad
g _{\rm SM}^{uR}=\frac{2}{3}\frac{e\,s_W}{c_W}\,,\\
g _{\rm SM}^{dL}&=-\frac{e}{s_W c_W}\left(-\frac{1}{2}+\frac{1}{3}s_W^2\right)\,,\qquad
g _{\rm SM}^{dR}=-\frac{1}{3}\frac{e \,s_W}{c_W}\,,
\label{eq:SMcouplings}
\end{aligned}
\end{align}
{with $e=g_1 g_2/\sqrt{g_1^2+g_2^2} = g_1 c_W = g_2 s_W$ being} the electric charge.
\smallskip

$Z^\prime$ scenarios are in general subject to gauge anomalies, which are often assumed to be canceled by additional heavy fields at a higher scale \cite{Langacker:2008yv,Alonso:2018bcg,Smolkovic:2019jow}. This is what we will do in the present paper. {Gauge anomaly cancellation in $Z'$ models was discussed recently in Refs.~\cite{Ellis:2017nrp,DAmico:2017mtc,Aebischer:2019blw,Altmannshofer:2019xda}.}
\medskip

\section{{Basic Formulae for Observables}}
\label{observables}

\subsection{Lepton Flavor Universality}
\begin{figure*}[t!]
\centering
        \includegraphics[width=0.5\textwidth]{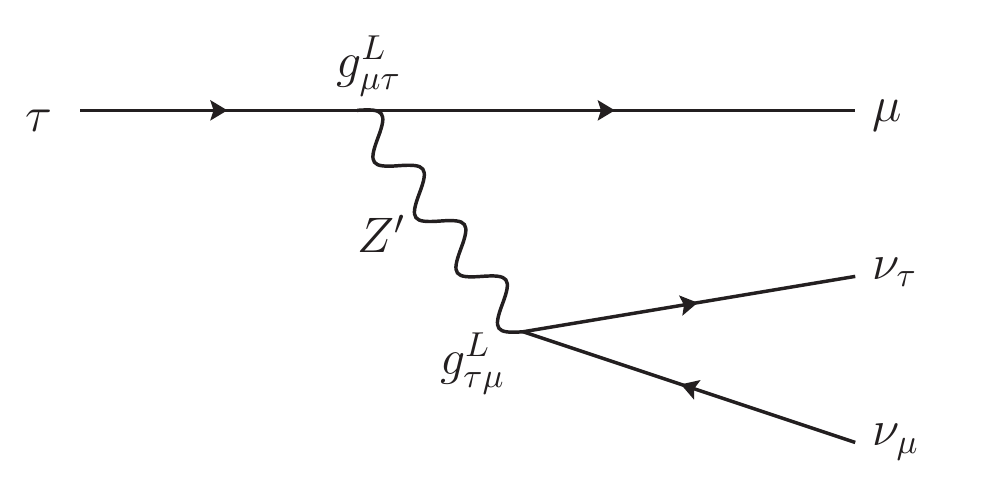} 
        \hspace{10mm}
            \includegraphics[width=0.41\textwidth]{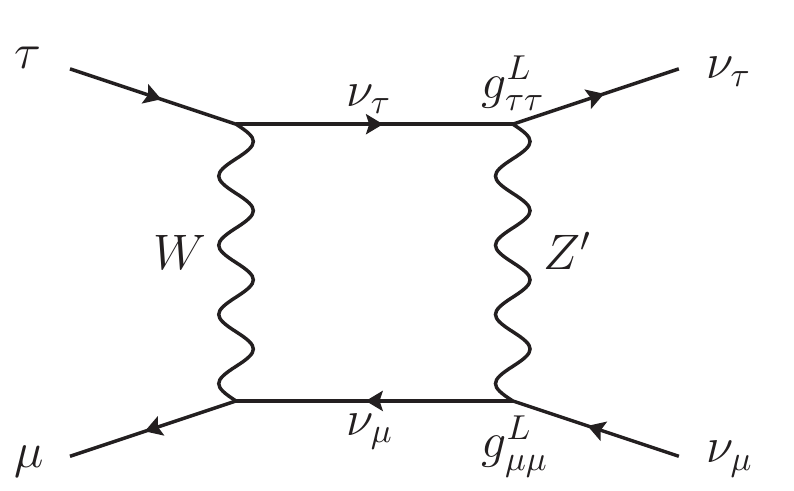}
	\caption{Feynman diagrams illustrating the leading $Z'$-contributions to the process
 $\tau\to \mu \nu_\tau \bar\nu_\mu$. In the presence of flavour off-diagonal
 $Z'$-couplings, the process arises already at tree-level (see left diagram), whereas
 in the case of flavour-diagonal couplings, corrections to
 $\tau\to \mu \nu_\tau \bar\nu_\mu$ are induced by four boxes, one of which is shown on the right.\label{FeynmanDiagrams}}
\end{figure*}

With this setup we are in a position to calculate the effects in the relevant observables. Flavour off-diagonal couplings already generate at tree-level $\ell\to\ell^\prime\nu\bar\nu$  amplitudes which interfere with the SM one, while flavour diagonal couplings can only do this via box diagrams (see Fig.~\ref{FeynmanDiagrams}). Neglecting non-interfering contributions we thus have
\begin{align}
R\left[\tau \to \mu\right]=&\frac{\mathcal{A}(\tau \to \mu \,\nu_\tau \overline{\nu}_\mu)}{\mathcal{A}(\tau \to \mu \,\nu_\tau \overline{\nu}_\mu)_{SM}}
=1+2\frac{|g_{\mu \tau}^L|^2}{g_2^2}\frac{m_W^2}{M_{Z^{\prime}}^2}
-\frac{3}{8\pi^2}\,g^L_{\mu\mu}\,g^L_{\tau\tau} \frac{\ln \left(\frac{m_W^2}{M_{Z^{\prime}}^2}\right)}{1-\frac{M_{Z^{\prime}}^2}{m_W^2}}\,,
\label{taumununu}
\end{align}
where $\cal A$ denotes the amplitude. Analogous expressions for $\tau \to e \,\nu_\tau \overline{\nu}_e$ and $\mu \to e \,\nu_\mu \overline{\nu}_u$ follow by a straightforward exchange of indices. This has to be compared to the experimental results~\cite{Amhis:2019ckw}
\begin{align}
\label{eq:LFUratios}
	\begin{split}
\left.\frac{\mathcal{A}\left[\tau  \to \mu \nu \bar \nu \right]}{\mathcal{A}\left[\mu  \to e\nu \bar \nu \right]}\right|_{\rm EXP} &= 1.0029 \pm 0.0014\,,\\
\left.\frac{\mathcal{A}\left[\tau  \to \mu \nu \bar \nu \right]}{\mathcal{A}\left[\tau  \to e\nu \bar \nu \right]}\right|_{\rm EXP} &= 1.0018 \pm 0.0014\,,\\
\left.\frac{\mathcal{A}\left[\tau  \to e\nu \bar \nu \right]}{\mathcal{A}\left[\mu  \to e\nu \bar \nu \right]}\right|_{\rm EXP} &= 1.0010 \pm 0.0014\,,
\end{split}
\end{align}
with the correlation matrix~\cite{Amhis:2019ckw}
\begin{align}
\begin{pmatrix}
1.00  & 0.49 & 0.51\\
0.49  & 1.00 & -0.49\text{\footnotemark} \\
0.51  & -0.49  & 1.00
\end{pmatrix}\,.	
\end{align}
\footnotetext{The HFLAV collaboration reports -0.50, however for the practical usage we choose -0.49 to have a positive semi-definite correlation matrix.}
\smallskip

Furthermore, the effect in $\mu \to e \,\nu_\mu \overline{\nu}_e$ is related to a modification of the Fermi constant which enters not only electroweak precision observables (to be discussed later) but also in the determination of $V_{ud}$ from beta decays, in particular super-allowed beta decays, which allow for the most precise determination of $V_{ud}$. Here a tension with kaon, tau and $D$ decays has been observed, whose significance depends strongly on the radiative corrections applied to $\beta$ decays~\cite{Marciano:2005ec,Seng:2018yzq,Seng:2018qru,Gorchtein:2018fxl,Czarnecki:2019mwq,Seng:2020wjq,Hayen:2020cxh,Hardy:2020qwl}, but also on the treatment of tensions between $K_{\ell 2}$ and $K_{\ell 3}$ 
decays~\cite{Moulson:2017ive} and on the bounds from $\tau$ decays~\cite{Amhis:2019ckw}, see Ref.~\cite{Crivellin:2020lzu} for more details. In the end, quoting a significance of $3\sigma$ should provide a realistic representation of the current situation, and for definiteness we will thus use the estimate of the first-row CKM unitarity violation from Ref.~\cite{Zyla:2020zbs} 
\begin{align}
\big|V_{ud}\big|^2+\big|V_{us}\big|^2+\big|V_{ub}\big|^2
= 0.9985(5).
\label{1throw}
\end{align}
In addition, note that there is also a deficit in the first-column CKM unitarity relation~\cite{Zyla:2020zbs}
\begin{equation}
\big|V_{ud}\big|^2+\big|V_{cd}\big|^2+\big|V_{td}\big|^2 = 0.9970(18),
\end{equation}
less significant than Eq.~\eqref{1throw}, but suggesting that if the deficits were due to NP, they would likely be related to $\beta$ decays. This unitarity deficit constitutes the so-called Cabibbo Angle Anomaly {(CAA)} and could be alleviated by our NP effect given by
\begin{equation}
R\left( {\mu  \to e} \right)=0.00075\pm0.00025,
\end{equation}
with {$R(\mu\to e)$ defined in Eq.~(\ref{taumununu}) with properly changed flavour indices.}
Note that in these tests of LFU we can neglect the modifications of the $W\ell\nu$ vertices since these would be loop-induced and thus suppressed by a factor $ m_\ell ^2/M_Z^{\prime 2}$.
\medskip

\begin{boldmath}
\subsection{Lepton Flavor Violation in $\ell_j \to \ell_i \gamma$}
\end{boldmath}
Defining the effective Hamiltonian by
\begin{align}
\mathcal{H}_{NP}=c_L^{ij} \;\bar{\ell_i}\,\sigma_{\mu \nu}P_L\, \ell_j\, F^{\mu \nu}+\mathrm{h.c.}\, ,
\label{Hmuegamma}
\end{align}
we find
\begin{align}
c_L^{ij}=&\frac{e}{48\pi^2 M_{Z^{\prime}}^2}\sum_{k}
\bigg(m_j \, g_{ik}^R \, g_{kj}^R -3\, m_k \, g_{ik}^R\,  g_{kj}^L + m_i\, g_{ik}^L \, g_{kj}^L \bigg)\,
\label{cLmuegamma}
\end{align}
and $c_R^{ij}$, which can be obtained from $c_L^{ij}$ by interchanging $L$ and $R$. We find then
 the branching ratio\footnote{The coefficients $c_L^{ij}$ in the case of $Z^\prime$ contribution
 can be obtained from its contribution through the chromomagnetic penguin to $b\to s\gamma$ decay
 calculated in Ref.~\cite{Buras:2011zb}. Using formulae in that paper and adjusting the couplings
 to the case at hand we obtain (at leading order in $m_\ell/m_{Z^\prime}$ ) consistent results
 with the generic formula of Ref.~\cite{Crivellin:2018qmi}. However, our result for the branching
 ratio is a factor $1/2$ smaller than the result of Ref.~\cite{Lindner:2016bgg}. }
\begin{align}
\mathrm{Br}\left[\ell_j \to \ell_i \gamma\right]=\frac{m_j^3}{4\pi  \Gamma_j}\left(|c_L^{ij}|^2 +|c_R^{ij}|^2\right).\label{rad}
\end{align}
%
The current experimental limits for lepton flavour violation processes at $90\%$ C.L. are~\cite{Bertl:2006up,Aubert:2009ag,TheMEG:2016wtm}:
\begin{align}\modified{\label{gammabounds}}
\begin{split}
{\rm Br}\!\left[\mu\rightarrow e\gamma\right] &\leq 4.2\times10^{-13}\,,\\
{\rm Br}\!\left[\tau\rightarrow \mu\gamma\right] &\leq 4.4\times10^{-8} \,,\\
{\rm Br}\!\left[\tau\rightarrow e\gamma\right] &\leq 3.3\times10^{-8} \,.
\end{split}
\end{align}
Improvements of approximately one order of magnitude for tau decays can be achieved at BELLE II~\cite{Kou:2018nap} and MeG II will further increase the sensitivity for $\mu\to e\gamma$~\cite{Baldini:2018nnn}.
\medskip

\subsection{Anomalous Magnetic Moments and Electric Dipole Moments}

Using the coefficients in \eq{cLmuegamma} for the flavour conserving case, we obtain for the NP contributions to anomalous magnetic moments~$\Delta a_i$ and the electric dipole moments~$d_i$,
\begin{align}
\begin{split}
\Delta a_i=&-\frac{4\,m_i}{e}{\rm Re}\left[c_R^{ii}\right]\,,\\
d_i=&-2\,{\rm Im}\left[c_R^{ii}\right]\,.
\label{AMM}
\end{split}
\end{align}
These expressions have to be compared with experimental bounds
\begin{align*}
\Delta a_e^{\rm Cs}&=a_e^{\rm exp} -a_e^{\rm SM, Cs}=-0.88(36)\times 10^{-12}\,,\\
\Delta a_e^{\rm Rb}&=a_e^{\rm exp} -a_e^{\rm SM, Rb}=0.48(30)\times 10^{-12}\,,\\
\Delta a_\mu&=a_\mu^{\rm exp} -a_\mu^{\rm SM}=251(59)\times 10^{-11}\,,\\
d_e &<1.1\times 10^{-29} \; \text{e cm}\,,\\
d_\mu &<-0.1(0.9)\times 10^{-19} \; \text{e cm}\,,\\
-0.22 \times 10^{-16}\; \text{e cm} &< {\rm Re}(d_\tau) < 0.45 \times 10^{-16}\; \text{e cm}\,,\\
-0.25 \times  10^{-16}\; \text{e cm} &< {\rm Im}(d_\tau) < 0.08 \times 10^{-16}\; \text{e cm}\,.
\end{align*}
Here the value of $\Delta a_e$ extracted from the corresponding measurement~\cite{Hanneke:2008tm}
 and the SM prediction~\cite{Aoyama:2017uqe,Laporta:2017okg}, using the fine structure constant
 from $^{133}$Cs~\cite{Parker:2018vye}, is incompatible with the determination using $\alpha$
 from $^{87}$Rb~\cite{Morel:2020dww}. Therefore, we quoted both values and we will also
 distinguish between these cases in our numerical analysis (where relevant). 
$\Delta a_\mu$ is extracted from the measurement of Ref.~\cite{Bennett:2006fi,Mohr:2015ccw} and from the recent results from the Fermilab Muon $g-2$ experiment~\cite{Abi:2021gix,Albahri:2021kmg,Albahri:2021ixb,Albahri:2021mtf}. The theory consensus is taken from Ref.~\cite{Aoyama:2020ynm}\footnote{This theory consensus does not
 include the determination of hadronic vacuum polarization (HVP) from the
 Budapest-Marseilles-Wuppertal lattice collaboration~\cite{Borsanyi:2020mff} which would decrease
 the tension in $\Delta a_\mu$ but differs from HVP determined via 
$e^+e^- \to {\rm hadrons}$~\cite{Hoferichter:2019gzf,Davier:2019can,Colangelo:2018mtw,Ananthanarayan:2018nyx,Davier:2017zfy,Keshavarzi:2018mgv,Davier:2019can,Keshavarzi:2019abf} 
and would increase the tensions in the global EW fit~\cite{Crivellin:2020zul,Keshavarzi:2020bfy}.}
 while the bound on $d_\mu$ originates from Ref.~\cite{Bennett:2008dy} and $d_e$ from
 Refs.~\cite{Baron:2013eja,Andreev:2018ayy}. Note that while currently the bounds from tau
 leptons~\cite{Inami:2002ah,Abdallah:2003xd,Eidelman:2007sb} and the muon EDM are not
 constraining, the latter could be significantly improved by a dedicated experiment proposed at
 PSI~\cite{Adelmann:2021udj}.

\subsection{Lepton Flavor Violation in {$\ell\to 3\ell'$}}
Three body decays to charged leptons receive the following contributions:
\begin{enumerate}
\item  Tree-level $Z^\prime$ and $Z$ exchanges (in the presence of $Z-Z^{\prime}$ mixing).  The latter contributions to processes with just one flavour transition, such as $\tau ^-\to\mu ^-e^+e^-$ or $\tau \to 3\mu$, are proportional to $\sin\xi$ for the $Z$-$Z^\prime$ interference and proportional to $\sin^2\xi$ for the $Z$ contributions alone. 
 The processes with two flavour transitions, such as $\tau ^-\to\mu ^-e^+\mu^-$, are suppressed by higher powers of $\sin\xi$ and neglected in the following. However, for the $Z'$-mediated tree-level contributions we include the possibility that both vertices are flavour changing.
\item  {One-loop effects in dipole operators (as defined in Eq.~(\ref{cLmuegamma})) entering via on-shell photon. There contributions again only affect decays with same flavour $\ell^+\ell^-$ pairs in the final states. }
\item  {One-loop contributions generated through the mixing of tree-level induced 4-lepton operators  into the operators}
\begin{align}
\begin{aligned}
\mathcal{O}_{e\mu}^{LL}&=(\bar e\gamma_\mu P_L\mu)(\bar\ell\gamma^\mu P_L\ell), \quad \mathcal{O}_{e\tau}^{LL}=(\bar e\gamma_\mu P_L\tau)(\bar\ell\gamma^\mu P_L\ell),\\
\mathcal{O}_{\mu\tau}^{LL}&=(\bar \mu\gamma_\mu P_L\tau)(\bar\ell\gamma^\mu P_L\ell),\quad
\mathcal{O}_{e\mu}^{RL}=(\bar e\gamma_\mu P_R\mu)(\bar\ell\gamma^\mu P_L\ell), \\
 \mathcal{O}_{e\tau}^{RL}&=(\bar e\gamma_\mu P_R\tau)(\bar\ell\gamma^\mu P_L\ell),\quad \mathcal{O}_{\mu\tau}^{RL}=(\bar \mu\gamma_\mu P_R\tau)(\bar\ell\gamma^\mu P_L\ell),
 \end{aligned}\label{4-lepton-operators}
\end{align}
and the corresponding ones with $L$ and $R$ interchanged. The details of this mixing through the
 QED penguin diagrams are discussed in Appendix~\ref{QEDP}. There we list the results
 for the Wilson coefficients of the operators in question which, due to the vectorial nature of 
the photon coupling, satisfy the following relations 
  \begin{equation}
    C_{e\mu}^{LL}{\Big|_{\text{QED}}}= C_{e\mu}^{LR}{\Big|_{\text{QED}}},\qquad
 C_{e\mu}^{RL}{\Big|_{\text{QED}}}= C_{e\mu}^{RR}{\Big|_{\text{QED}}},
  \end{equation}
 with analogous relations for the remaining coefficients.
\end{enumerate}
\smallskip
Note that the formulae for the coefficients $C_{ij}^{AB}$ in Appendix~\ref{QEDP} 
carry no $\ell\ell$ indices (contrary in $C_{ij,\ell\ell}^{AB}$) and consequently apply to all operators. 
The difference will, however, be present in the parameters $X^{AB}_{ij,kl}$ 
which are sensitive to the choice of $kl$.
\begin{figure}%
\centering 
\subfloat[Penguin]{\includegraphics[width=0.45\textwidth]{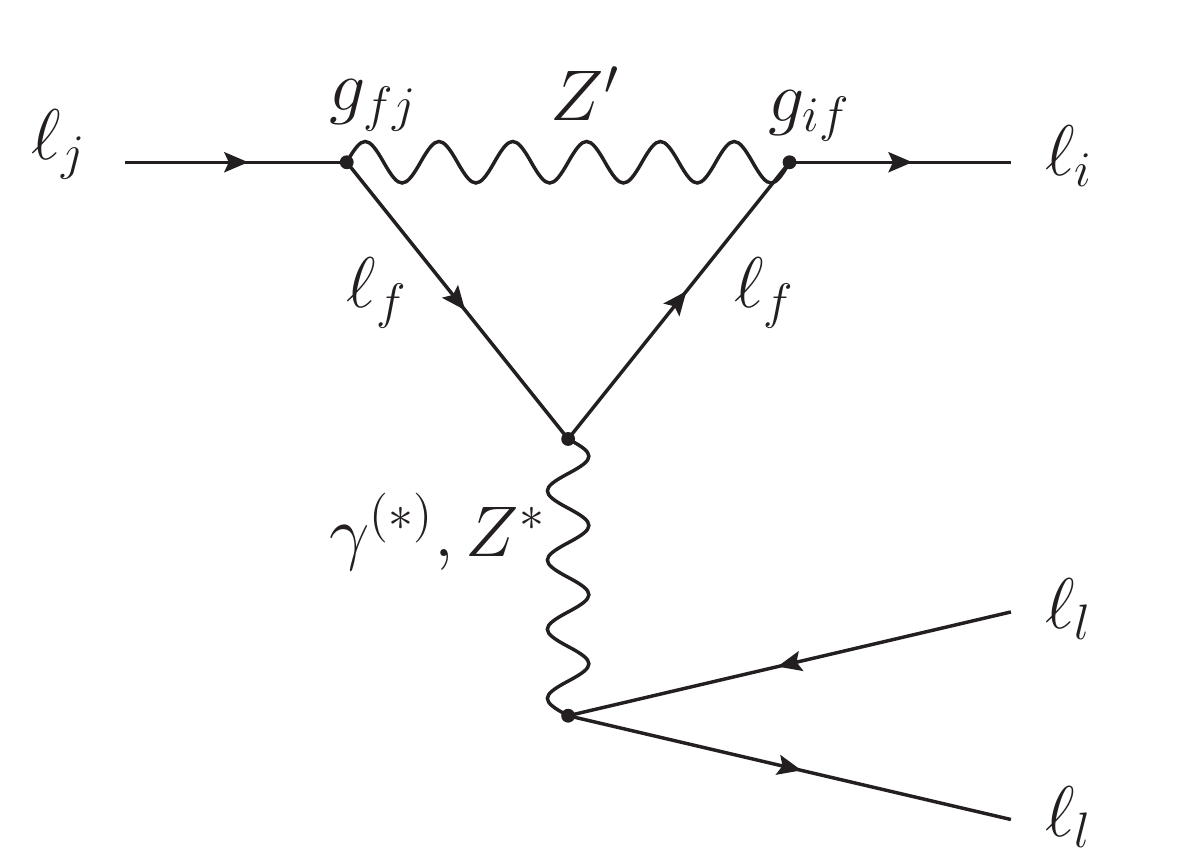} }
\qquad
\subfloat[Crossed penguin]{{\includegraphics[width=0.45\textwidth]{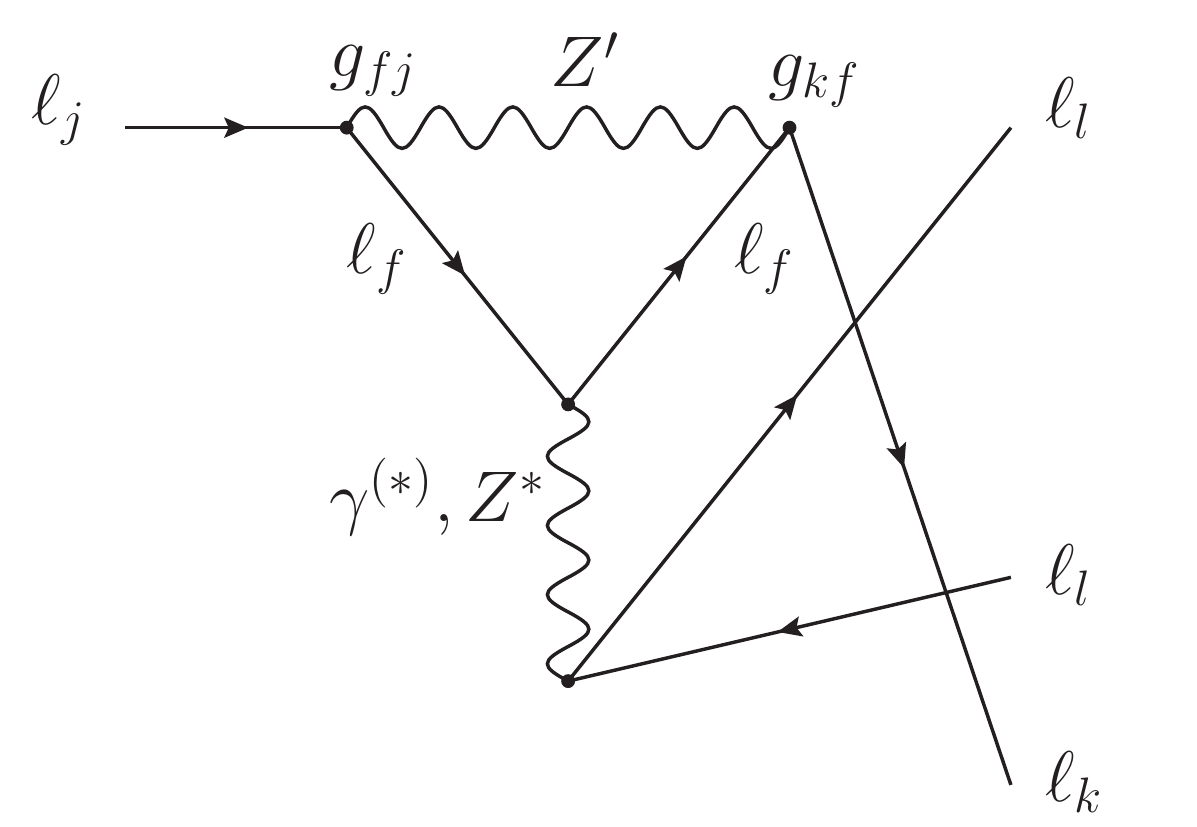} }}
\caption{Feynman diagrams showing the $Z$- and the photon-penguin contributions to $\ell_j\to \ell_i \, \overline{\ell}_l\, \ell_k -$processes.}%
 \label{l_lll_full_diags}%
\end{figure}
The latter are defined by
\begin{align}\label{Xij}
X^{AB}_{ij,kl} \equiv  & \,C_{ij}^{AB} \,\delta_{kl}
+ \frac{\Delta_{ij}^A\,\Delta_{kl}^B}{m_Z^2}\,\delta_{kl}
+ \frac{g_{ij}^A\; g_{kl}^B}{M_{Z'}^2}.
\end{align}
where $A,B=L,R$, which combines the contributions from tree-level $Z^\prime$ and $Z$ exchanges, and from QED penguin contributions. 
{The couplings $\Delta_{ij}^A$ are defined in \eq{Deltas}.}
We then find the following branching ratios for flavour-violating $\tau$ decays:
\begin{align}
{\rm Br}\left[\tau^\mp\to e^\mp\mu^\pm\mu^\mp\right]=
-&\frac{e^2\, m_\tau^3}{48\, \pi ^3\,\Gamma_\tau} 
\left(\left| c_L^{e\tau}\right|^2+\left|c_R^{e\tau}\right|^2\right) \left(\ln\left(\frac{m_\mu^2}{m_\tau^2}\right)+3\right)\\
+&\frac{m_\tau^5}{1536\, \pi ^3 \,\Gamma_\tau }
\left(\left| X^{LL}_{e\tau, \mu\mu} + X^{LL}_{\mu\tau, e\mu}\right|^2 +\left|X^{LR}_{e\tau, \mu\mu} \right|^2 + \left|X^{LR}_{\mu \tau ,e\mu}\right|^2 + L\leftrightarrow R\right)
\notag\\
+&\frac{e \,m_\tau{}^4}{192\, \pi ^3 \,\Gamma_\tau}\left({\rm Re}\left[c_R^{e\tau*}\left(X^{LL}_{e\tau,\mu\mu}+X^{LL}_{\mu\tau,e\mu}+X^{LR}_{e\tau, \mu\mu}+X^{LR}_{\mu \tau, e\mu}\right)\right]
+L\leftrightarrow R\right)\,,
\notag \\
{\rm Br}\left[\tau \to 3\mu\right]=&-\frac{e^2\, m_\tau^3 }{192\, \pi ^3\,\Gamma_\tau}\left(|c_L^{\mu\tau}|^2+|c_R^{\mu\tau}|^2\right) \left(4 \ln \left(\frac{m_\mu^2}{m_\tau^2}\right)+11\right)\notag\\
&+\frac{m_\tau^5 }{1536 \,\pi ^3\, \Gamma_\tau}
\left(2 \left|X_{\mu\tau, \mu\mu}^{LL}\right|^2+\left|X_{\mu\tau, \mu\mu}^{LR}\right|^2
+L\leftrightarrow R\right)\\
&+\frac{e\, m_\tau^4}{192\, \pi ^3 \,\Gamma_\tau }
\Big({\rm Re}\left[c_R^{\mu\tau *} \left(2\, X_{\mu\tau, \mu\mu}^{LL}+X_{\mu\tau,\mu\mu}^{LR}\right)\right]+L\leftrightarrow R\Big)\,,
\notag \\
{\rm Br}\left[\tau^\mp\to \mu^\mp e^\pm \mu^\mp\right]=&\frac{m_\tau^5}{1536\,\pi^3\Gamma_\tau}
\left(2 \,\left|X^{LL}_{\mu\tau,\mu e}\right|^2 + \left|X^{LR}_{\mu\tau,\mu e}\right|^2+L\leftrightarrow R\right)\,.
\label{lto3l}
\end{align}
{The expression for $\tau\to3e$ is obtained from $\tau\to3\mu$ by replacing $\mu$ by $e$, the one for  $\mu\to3e$ from $\tau\to3\mu$ by replacing index  $\tau$ by $\mu$ and $\mu$ by $e$. Finally the expression for  $\tau^\mp \rightarrow e^\mp \mu^\pm e^\mp$ is obtained from last formula by interchanging $\mu$ and $e$. }
\smallskip
The experimental bounds~\cite{Bellgardt:1987du,Hayasaka:2010np,Lees:2010ez,Aaij:2014azz,Amhis:2019ckw}
at $90\%$ C.L.  are:
\begin{align}\modified{\label{3bounds}}
\begin{aligned}
\operatorname{Br}\left[\mu^- \rightarrow e^- e^+ e^-\right]&\leq 1.0 \times 10^{-12}\,,\\
\operatorname{Br}\left[\tau^- \rightarrow e^- e^+ e^-\right]&\leq 1.4 \times 10^{-8}\,,\\
\operatorname{Br}\left[\tau^- \rightarrow e^- \mu^+ \mu\right] &\leq 1.6 \times 10^{-8}\,,\\
\operatorname{Br}\left[\tau^- \rightarrow \mu^- e^+ \mu^-\right] &\leq 9.8 \times 10^{-9}\,,\\
\operatorname{Br}\left[\tau^- \rightarrow \mu^- e^+ e^-\right] &\leq 1.1 \times 10^{-8}\,,\\
\operatorname{Br}\left[\tau^- \rightarrow e^- \mu^+ e^-\right]&\leq 8.4 \times 10^{-9}\,,\\
\operatorname{Br}\left[\tau^- \rightarrow \mu^- \mu^+ \mu^-\right] &\leq 1.1 \times 10^{-8}\,.
\end{aligned}
\end{align}
Here we can expect future improvements in $\tau$ decays by BELLE II~\cite{Kou:2018nap} (and also LHCb~\cite{Cerri:2018ypt}) and for $\mu\to3e$ by Mu3e~\cite{Blondel:2013ia,Perrevoort:2016nuv}.
\medskip

\subsection{{$\mu\to e$ conversion}}

We define
\begin{align}
\mathcal{L}_{\text{eff}}=\!\!\!\sum_{q=u,d}\!\left(C_{e\mu,qq}^{LL}\,O_{e\mu,qq}^{LL}\,+\,C_{e\mu,qq}^{LR}\,O_{e\mu,qq}^{LR}\right)+\left(L\leftrightarrow R\right)+\mathrm{h.c.}\,,
\end{align}
with
\begin{align}
\begin{aligned}
O_{e\mu,qq}^{LL}&=(\bar{e}\gamma^\mu P_L\mu)(\bar{q}\gamma_\mu P_L q)\,,\qquad O_{e\mu,qq}^{RL}&=(\bar{e}\gamma^\mu P_R\mu)(\bar{q}\gamma_\mu P_L q)\,,\\
O_{e\mu,qq}^{LR}&=(\bar{e}\gamma^\mu P_L\mu)(\bar{q}\gamma_\mu P_R q)\,,\quad\quad O_{e\mu,qq}^{RR}&=(\bar{e}\gamma^\mu P_R\mu)(\bar{q}\gamma_\mu P_R q)\,.
\end{aligned}
\end{align}
In the presence of $Z-Z'$ mixing, the flavour off-diagonal $Z'$-couplings lead to $\mu\to e$ conversion already at tree-level
\begin{align}
\begin{aligned}
C_{e\mu,qq}^{AB}&=\frac{\sin \xi \; g_{e \mu}^{A}}{m_Z^2} \;g_{\rm SM}^{q B}\Big|_{Z-Z'} \,, 
\end{aligned}
\end{align}
with the $\sin \xi$ given in \eq{eq:mixing} and $A,B=L,R$. In addition, for small $Z_0-Z_0'$
 mixing, the mixing {of} four-lepton operators into $O_{qq}^{V,LL}$ and $O_{qq}^{V,LR}$ can be
 relevant and is obtained analogously to the off-shell photon effects in $\ell\to3 \ell^\prime$ decays
 (see Appendix~\ref{QEDP}).   The standard renormalization group evolution is then performed from
 scale	$M_{Z^\prime}$ down to $m_\mu$, taking into account that the $\tau$-lepton is integrated out
 at $m_\tau$.  Taking into account that for operators with three electrons or three muons two
 different Wick contractions exist, which leads to a relative factor of 2 w.r.t the
 $\tau$-leptons case, and only considering the contributions of the hidden 
operators\footnote{For the contributions of the visible operators we refer to
 Eq.~(\ref{visible_mue_conv}) of Appendix \ref{AppC} where this naming of operators is explained.
}, we find 
\begin{align}
\begin{split}
  C_{e\mu,qq}^{LL}&=\frac{e^2 Q_q}{16\pi^2}\frac{2}{3M_{Z'}^{2}}\bigg(\!g_{e\mu}^L(2g_{\mu\mu}^L+g_{\mu\mu}^R) \ln\!\left(\!\frac{M_{Z'}^2}{m_\mu^2}\!\right)
  +\left( g_{e\tau}^L g_{\tau\mu}^{L}+g_{e\mu}^L (g_{\tau\tau}^L+g_{\tau\tau}^R)\right) \ln\!\left(\!\frac{M_{Z'}^2}{m_\tau^2}\!\right)\!\!\bigg) \Big|_{\text{QED}},
  \\
 C_{e\mu,qq}^{RL}&=\frac{e^2 Q_q}{16\pi^2}\frac{2}{3M_{Z'}^{2}}\bigg(\!g_{e\mu}^R(2g_{\mu\mu}^R+g_{\mu\mu}^L)
    \ln\!\left(\!\frac{M_{Z'}^2}{m_\mu^2}\!\right)
    + \left(g_{e\tau}^R g_{\tau\mu}^{R}+g_{e\mu}^R (g_{\tau\tau}^L+g_{\tau\tau}^R)\right)\ln\!\left(\!\frac{M_{Z'}^2}{m_\tau^2}\!\right)\!\!\bigg) \Big|_{\text{QED}},
\end{split}\label{hidden_mue_conv}
\end{align} 
and
\begin{equation}
    C_{e\mu,qq}^{LL}\Big|_{\text{QED}}=C_{e\mu,qq}^{LR}\Big|_{\text{QED}}\,,\qquad C_{e\mu,qq}^{RL}\Big|_{\text{QED}}=C_{e\mu,qq}^{RR}\Big|_{\text{QED}}\,,
\end{equation}
where $Q_q$ is the electric charge of the quarks ($Q_u=+\frac{2}{3},\;Q_d=-\frac{1}{3}$).
\smallskip

The transition rate $\Gamma_{\mu\to e}^N\equiv \Gamma(\mu N\to eN)$ is given by 
\begin{align}
\Gamma_{\mu\to e}^N =&\frac{m_\mu^5}{4} \Bigg \vert \frac{c_L^{e\mu}}{m_\mu}D_N+4\sum_{q=u,d}\left(C_{e\mu,qq}^{RL}+C_{e\mu,qq}^{RR}\right)\left(f_{Vp}^{(q)}V_N^p+ f_{Vn}^{(q)}V_N^n\right)
\Bigg\vert^2{+}(L\leftrightarrow R)\,.
\end{align}
The quantities $D_N$ and $V_{p/n}^N$ are related to the overlap integrals between the lepton wave functions and the nucleon densities, and thus depend on the nature of the target $N$. We use the numerical values \cite{Kitano:2002mt}
\begin{align}
D_{\rm Au}&=0.189,\hspace{1cm} 
V_{\text{Au}}^p=0.0974\,,\hspace{1cm}
V_{\text{Au}}^n=0.146\,.
\end{align}
The nucleon vector form factors are the same as the ones measured in elastic electron-hadron scattering, i.e.
\begin{align}
f_{Vp}^{(u)}=2,\quad f_{Vn}^{(u)}=1,\quad f_{Vp}^{(d)}=1,\quad f_{Vn}^{(d)}=2\,.
\end{align}
Finally, the branching ratio of $\mu\to e$ conversion is defined as the transition rate divided by the $\mu$ capture rate:
\begin{align}
\mathrm{Br}\left[\mu \to e\right]=\frac{\Gamma^{\rm conv}}{\Gamma^{\rm capt}}\,,
\end{align}
 and for the latter we use \cite{Suzuki:1987jf}
\begin{align}
\Gamma_{\rm Au}^{\rm capt}=8.7\times 10^{-18}\; \text{GeV}\,.
\end{align}
The experimental limit on $\mu\to e$ conversion from SINDRUM II is~\cite{Bertl:2006up}
\begin{align}
\frac{\Gamma_{\rm Au}^\text{conv}}{\Gamma_{\rm Au}^\text{capt}}&<7.0\times 10^{-13}\,.
\end{align}
{It is expected to be improved by three orders of magnitude by COMET and Mu2e collaborations in the coming years~\cite{Baldini:2018uhj}.}
\medskip

\subsection{Electroweak Precision Observables}

The EW sector of the SM has been tested with a very high precision at LEP and Tevatron~\cite{Schael:2013ita,ALEPH:2005ab}. Since it can be parametrised by only three Lagrangian parameters, we choose the set with the smallest experimental error consisting of the Fermi constant ($G_F=1.1663787(6)\times10^{-5}\,{\rm GeV}^{-2}$~\cite{Tanabashi:2018oca}), the mass of the Z boson ($m_Z=91.1875(21)$~\cite{ALEPH:2005ab}) and the fine structure constant ($\alpha_{em}=7.2973525664(17)\times10^{-3}$~\cite{Tanabashi:2018oca}). \\
In our model, the Lagrangian values for $G_F$ and $m_Z$ are shifted with respect to their measurements. In particular, the effect in $\mu \to e\nu \bar \nu $ leads to the following relation
\begin{align}
\label{eq:GF}
\dfrac{G_F}{G_F^{\mathcal L}} &=1+2\,\frac{| g_{e\mu}^L|^2}{g_2^2}\frac{m_W^2}{M_{Z'}^2}\equiv 1+\delta G_F\,,
\end{align}
while the Z boson mass is modified via \eq{MZcorr}. Moreover, taking into account the tree-level effects in \eq{Deltas} and the loop effects in Ref.~\cite{Altmannshofer:2014cfa,Haisch:2011up}, we have the following modified $Z$ couplings to leptons
\begin{equation}
\begin{aligned}
\Delta _{ij}^{\ell L}(q^2=m_Z^2) &= g_{{\rm{SM}}}^{\ell L}\left( {{\delta _{ij}} + \sin \xi \frac{{g_{ij}^L}}{{g_{{\rm{SM}}}^{\ell L}}} + \sum\limits_k {\frac{{g_{ik}^Lg_{kj}^L}}{{{{(4\pi )}^2}}}} {K_F}\left( {\frac{{m_Z^2}}{{M_{Z'}^2}}} \right)} \right){\kern 1pt} ,
\\
\Delta _{ij}^{\nu L}(q^2=m_Z^2) &= g_{{\rm{SM}}}^{\nu L}\left( {{\delta _{ij}} + \sin \xi \frac{{g_{ij}^L}}{{g_{{\rm{SM}}}^{\nu L}}} + \sum\limits_k {\frac{{g_{ik}^Lg_{kj}^L}}{{{{(4\pi )}^2}}}} {K_F}\left( {\frac{{m_Z^2}}{{M_{Z'}^2}}} \right)} \right){\kern 1pt} ,
\\
\Delta _{ij}^{\ell R}(q^2=m_Z^2) &= g_{{\rm{SM}}}^{\ell R}\left( {{\delta _{ij}} + \sin \xi \frac{{g_{ij}^R}}{{g_{{\rm{SM}}}^{\ell R}}} + \sum\limits_k {\frac{{g_{ik}^Rg_{kj}^R}}{{{{(4\pi )}^2}}}} {K_F}\left( {\frac{{m_Z^2}}{{M_{Z'}^2}}} \right)} \right){\kern 1pt} ,
\end{aligned}
\label{Deltas_with_loops}
\end{equation}
with
\begin{align}
{K_F}\left( x \right)& =  - \frac{{2{{(x + 1)}^2}({\rm{L}}{{\rm{i}}_2}( - x) + \ln (x)\ln (x + 1))}}{{{x^2}}} 
- \frac{{7x + 4}}{{2x}} + \frac{{(3x + 2)\ln (x)}}{x}.
\end{align}
For the numerical analysis, we implemented the EW observables shown in Table~\ref{tab:EWCC} in HEPfit~\cite{deBlas:2019okz} taking into account the modifications induced by \eq{MZcorr}, \eq{eq:GF} and \eq{Deltas_with_loops}. In addition, the Higgs mass ($M_H = 125.16 \pm 0.13$ GeV~\cite{Aaboud:2018wps,CMS:2019drq}), the top mass ($m_t = 172.80 \pm 0.40$ GeV~\cite{TevatronElectroweakWorkingGroup:2016lid,Aaboud:2018zbu,Sirunyan:2018mlv}), the strong coupling constant ($\alpha_s(M_Z) = 0.1181\pm 0.0011$~\cite{Tanabashi:2018oca}) and the hadronic contribution to the running of $\alpha_{em}$ ($\Delta\alpha_{\rm had}=276.1(11) \times 10^{-4}$~\cite{Tanabashi:2018oca}) have been used as input parameters, since they enter EW observables indirectly via loop effects.

\begin{table}[t]
\centering
	\begin{tabular}{l | r } \hline
		Observable & Experimental value \\
		\hline
		$m_W\,[\text{GeV}]$  & $80.379(12)$  \\
		$\Gamma_W\,[\text{GeV}]$  & $2.085(42)$ \\
		$\text{BR}(W\to \text{had})$  & $0.6741(27)$ \\
		$\text{BR}(W\to \text{lep})$  & $0.1086(9)$ \\
		$\text{sin}^2\alpha_{\rm eff,\, e}^{\rm CDF}$ & $0.23248(52)$  \\
		$\text{sin}^2\alpha_{\rm eff,\, e}^{\rm D0}$ & $0.23146(47)$  \\
		$\text{sin}^2\alpha_{\rm eff,\, \mu}^{\rm CDF}$ & $0.2315(10)$  \\
		$\text{sin}^2\alpha_{\rm eff,\, \mu}^{\rm CMS}$ & $0.2287(32)$  \\
		$\text{sin}^2\alpha_{\rm eff,\, \mu}^{\rm LHCb}$ & $0.2314(11)$  \\
		$P_{\tau}^{\rm pol}$  &$0.1465(33)$  \\
		$A_{e}$  &$0.1516(21)$ \\
		$A_{\mu}$  &$0.142(15)$ \\
		$A_{\tau}$  &$0.136(15)$ \\		
		$\Gamma_Z\,[\text{GeV}]$  &$2.4952(23)$  \\
		$\sigma_h^{0}\,[\text{nb}]$  &$41.541(37)$ \\
		$R^0_{e}$  &$20.804(50)$ \\
		$R^0_{\mu}$  &$20.785(33)$ \\
		$R^0_{\tau}$  &$20.764(45)$ \\
		$A_{\rm FB}^{0,e}$ &$0.0145(25)$  \\
		$A_{\rm FB}^{0,\mu}$ &$0.0169(13)$  \\
		$A_{\rm FB}^{0,\tau}$ &$0.0188(17)$  \\
		$R_{b}^{0}$  &$0.21629(66)$  \\
		$R_{c}^{0}$  &$0.1721(30)$  \\
		$A_{\rm FB}^{0,b}$  &$0.0992(16)$   \\
		$A_{\rm FB}^{0,c}$  &$0.0707(35)$  \\
		$A_{b}$  &$0.923(20)$  \\
		$A_{c}$  &$0.670(27)$  \\
	\end{tabular}
	\caption{ Electroweak observables~\cite{ALEPH:2005ab,Tanabashi:2018oca} used in our fit which are calculated by HEPfit~\cite{deBlas:2019okz} using $m_Z^L$, $\alpha$ and $G_F$ as input.}
	\label{tab:EWCC}
\end{table}
\medskip

\subsection{{$Z\to \ell \ell^{\prime}$}}
In the presence of $Z-Z^{\prime}$ mixing, we obtain a tree-level contribution to $Z\to \ell_i \ell_j,\,i\neq j$, leading to the branching ratios
\begin{align}
\begin{aligned}
{\rm{Br}}\left[ {Z \to {\ell_i}\bar \ell_j } \right] &= \frac{1}{{24\pi }}\frac{{{m_Z}}}{{{\Gamma _Z}}}\left( {|\Delta _{ij}^{\ell L}(q^2=m_Z^2){|^2} + |\Delta _{ij}^{\ell R}(q^2=m_Z^2){|^2}} \right)\, ,\\
{\rm{Br}}\left[ {Z \to {\nu _i}\bar \nu _j } \right] &= \frac{1}{{24\pi }}\frac{{{m_Z}}}{{{\Gamma _Z}}}|\Delta _{ij}^{\nu L}(q^2=m_Z^2){|^2}\, ,
\end{aligned}
\end{align}
with $\Gamma_Z=2.4952\pm 0.0023$~GeV~\cite{Zyla:2020zbs}. We compare these results to the ATLAS and LEP measurements given in Table~\ref{Zll_constraints}.
\begin{table}[h!]
\begin{tabular}{|r l l|}
\hline
$\mathrm{Br}\left[Z\to e^\pm\tau^\mp\right]$    & $=(-0.1\pm 3.5\text{(stat)}\pm 2.3\text{(syst)})\times 10^{-6}$ & ATLAS: \cite{Aad:2020gkd} \\
 $\mathrm{Br}\left[Z\to e^\pm\tau^\mp\right]$   & $<9.8\times 10^{-6}$                                            & LEP (OPAL): \cite{Akers:1995gz,Zyla:2020zbs}\\
$\mathrm{Br}\left[Z\to \mu^\pm\tau^\mp\right]$  & $=(4.3\pm 2.8\text{(stat)}\pm 1.6\text{(syst)})\times 10^{-6}$  & ATLAS: \cite{Aad:2020gkd} \\
 $\mathrm{Br}\left[Z\to \mu^\pm\tau^\mp\right]$ & $<1.2\times 10^{-5}$                                           & LEP (DELPHI): \cite{Abreu:1996mj,Zyla:2020zbs}\\
$\mathrm{Br}\left[Z\to e^\pm \mu^\mp\right]$    & $<7.5\times 10^{-7}\; (95\%\,{\rm CL})$                         & ATLAS: \cite{Aad:2014bca,Zyla:2020zbs}\\
\hline
\end{tabular}
\caption{Experimental bounds on $Z\to \ell\ell'$.}\label{Zll_constraints}
\end{table}
\medskip

\subsection{Neutrino Trident Production}

Neutrino trident production can be used to constrain couplings of muons to muon neutrinos~\cite{Altmannshofer:2014cfa,Altmannshofer:2014pba}. Generalizing the formula of Ref.~\cite{Altmannshofer:2014pba} to the case of chiral $Z^\prime$ couplings, we find
	\begin{equation}
	\dfrac{{{\sigma_{SM + NP}}}}{{{\sigma _{SM}}}} = 1 + 8\dfrac{{\left( {1 + 4s_W^2} \right)\dfrac{{g_{22}^L\left( {g_{22}^L + g_{22}^R} \right)}}{{g_2^2}}\dfrac{{m_W^2}}{{M_{Z'}^2}} - \dfrac{{g_{22}^L{{\left( {g_{22}^R - g_{22}^L} \right)}^2}}}{{g_2^2}}\dfrac{{m_W^2}}{{M_{Z'}^2}}}}{{{{\left( {1 + 4s_W^2} \right)}^2} + 1}}\,.
	\end{equation}
This ratio is bounded by the weighted average
\begin{equation}
\sigma_{\rm exp } / \sigma_{\mathrm{SM}}=0.83 \pm 0.18\,,
\end{equation}
obtained from averaging the 
CHARM-II~\cite{Geiregat:1990gz}, CCFR~\cite{Mishra:1991bv} and NuTeV results~\cite{Adams:1998yf}.
\medskip

\subsection{LEP-II bounds}

LEP-II set stringent bounds on 4-lepton operators from $e^+ e^- \to \ell^+ \ell^-$ (with $\ell = e, \mu, \tau$)~\cite{Schael:2013ita} for specific chiralities.
Some of our more general scenarios cannot be matched to the 4-lepton operators as given in Ref.~\cite{Schael:2013ita}. For these cases we derive the constraints in Appendix~\ref{app:LEP2} and provide the formula which we implemented in HEPfit.
\medskip

\section{Phenomenological Analysis}\label{phenomenology}

In our phenomenological analysis we perform a global fit taking into account all observables discussed in the last section. This includes EW precision observables, as implemented in the HEPfit distribution~\cite{deBlas:2019okz}.
To include the other observables discussed previously, we added them to the HEPfit code such that we can perform a Bayesian statistical analysis using the Markov Chain Monte Carlo (MCMC) determination of posteriors of the Bayesian Analysis Toolkit (\texttt{BAT})~\cite{Caldwell:2008fw}. 
\smallskip

With this setup we can now consider several different scenarios. For our numerical analysis we fix $M_{Z^\prime}=1\,$TeV unless stated otherwise and assume real couplings. Note that despite small logarithmic corrections, the results we obtain scale like $g^2/M_{Z^\prime}^2$. {However, for the loop-induced modifications of $Z\ell\ell$ couplings in the scenario which aims at an explanation of the data on the anomalous magnetic moment of the muon, these logarithmic corrections to the $g^2/M_{Z^\prime}^2$ scaling can indeed be relevant.}

\subsection{Lepton Flavour Universality}\label{LFU}

Here we consider four scenarios which respect lepton flavour universality:
\begin{enumerate}
	\item Left-handed couplings: $g^L_{ii}=g^L$
	\item Right-handed couplings: $g^R_{ii}=g^R$
	\item Vectorial couplings: $g^L_{ii}=g^R_{ii}=g^V$
	\item Generic chiral couplings: $g^L_{ii}=g^L$, $g^R_{ii}=g^R$ 
\end{enumerate}

\begin{figure}[t]
	\centering
	\includegraphics[width=1.05\textwidth]{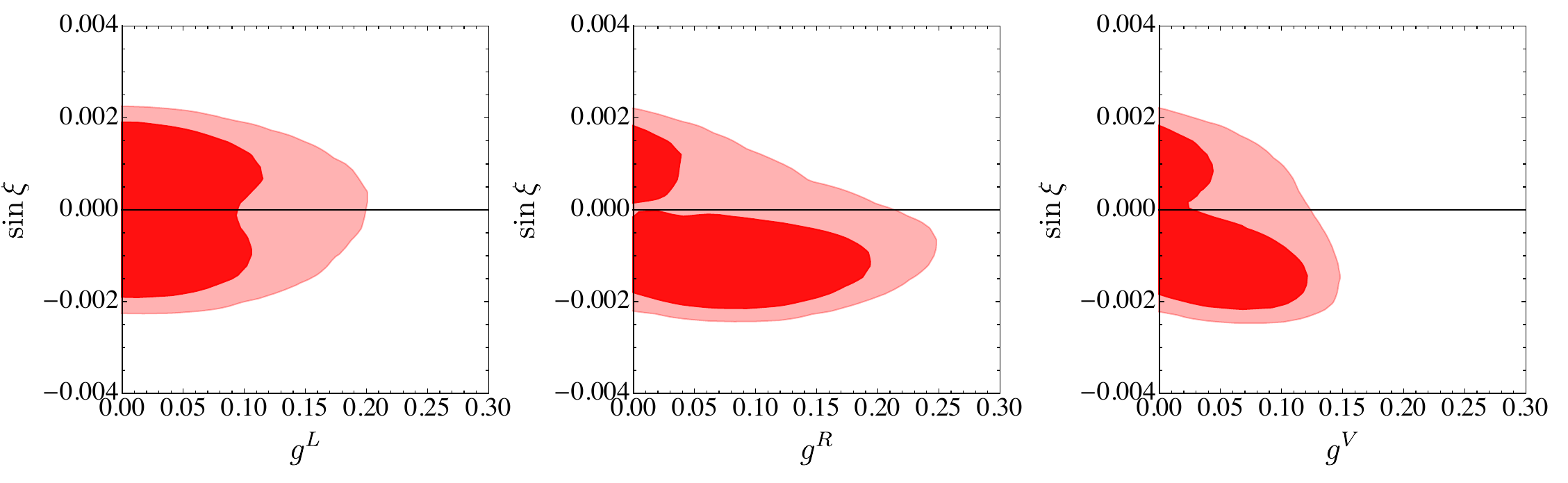} \\
	\caption{{68\% and 95\% CL regions for the three LFU cases $g^L$, $g^R$ and $g^V$. Note that due to the preference for a slightly lower $W$ mass than predicted in the SM, the origin in the $g^R$ scenario is not within the 68\% CL region. The $Z-Z^\prime$ mixing angle is bounded to be $\xi\lessapprox 0.02$ and $g^{L,R}\lessapprox 0.2$, while $g^{V}\lessapprox 0.1$ for $M_{Z^\prime}=1\,$TeV. } \label{LFUplot}}
\end{figure}

\begin{figure}[h!]
	\centering
	\includegraphics[width=0.7\textwidth]{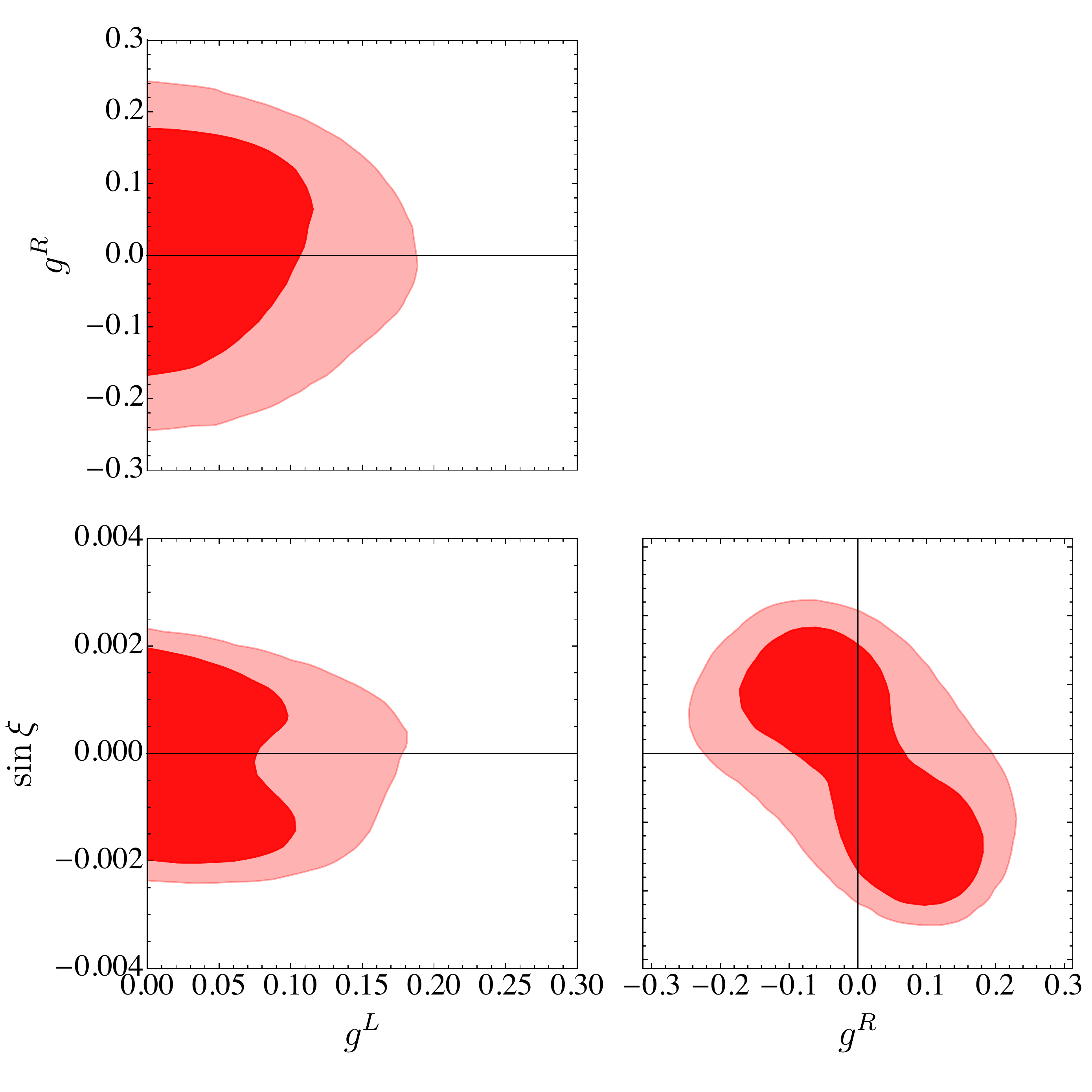} \\
	\caption{{68\% and 95\% CL regions for the three-dimensional LFU fit, where $g_L$, $g_R$ and $\xi$ are free parameters for $M_{Z^\prime}=1\,$TeV.  }\label{LFU_gLgR_plot}}
\end{figure}

Each of the first three LFU scenarios is two dimensional, with the coupling and the $Z$-$Z^\prime$ mixing angle $\xi$ being free parameters, and shown in Fig.~\ref{LFUplot}. We can see that for a $Z^\prime$ mass of 1~TeV, the couplings should be smaller than $\approx 0.2$ and the $Z-Z^\prime$ mixing is bounded to be less than $\approx0.002$ at 95\%~CL. In the case of $g_L$ and  $g_R$ being independent of each other, shown in  Fig.~\ref{LFU_gLgR_plot}, there is a mild preference for a non-zero mixing angle at the 68\%~CL, which is due to the slight tension in the $W$ mass prediction within the EW fit. However, neither the tension in {$\tau\to\mu\nu\bar\nu/\tau\to e\nu\bar\nu$ nor in the first row CKM unitarity or $(g-2)_\mu$ can be explained in these LFU setups.}

\begin{figure}[h!]
	\centering
	\includegraphics[width=1.05\textwidth]{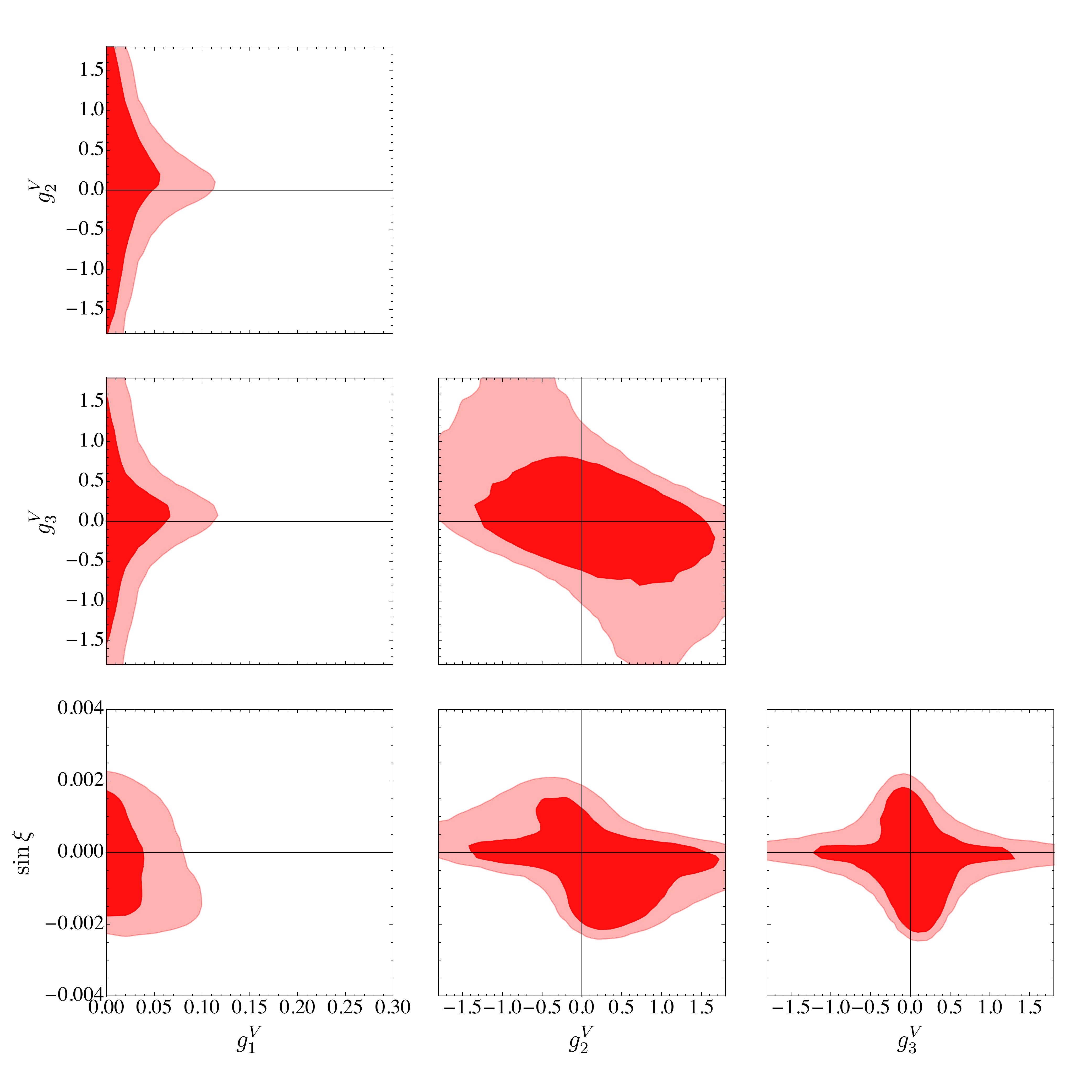} \\
	\caption{68\% and 95\% CL regions for the LFUV case with vectorial couplings $g^{L}_{ii}=g^{R}_{ii}=g^V_{i}$. \label{LFUV_V_plot}}
\end{figure}

\subsection{Lepton Flavour Universality Violation}\label{LFUV}

\begin{figure}[h!]
	\centering
	\includegraphics[width=1.05\textwidth]{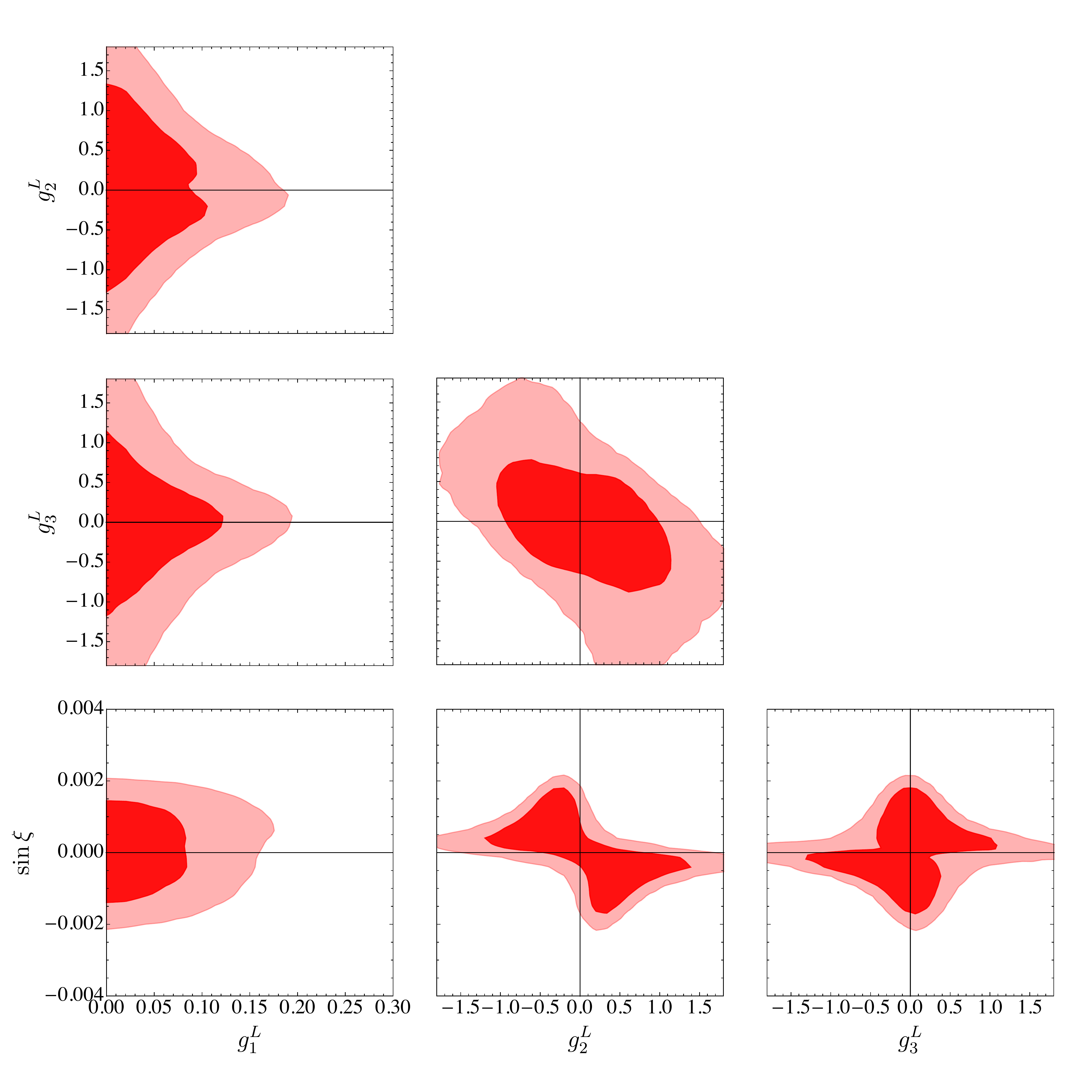} \\
	\caption{68\% and 95\% CL regions for the LFUV case with
          {left-handed} couplings $g^L_{i}=g^L_{ii}$. \label{LFUV_L_plot}}
\end{figure}

\begin{figure}[h!]
	\centering
	\includegraphics[width=1.05\textwidth]{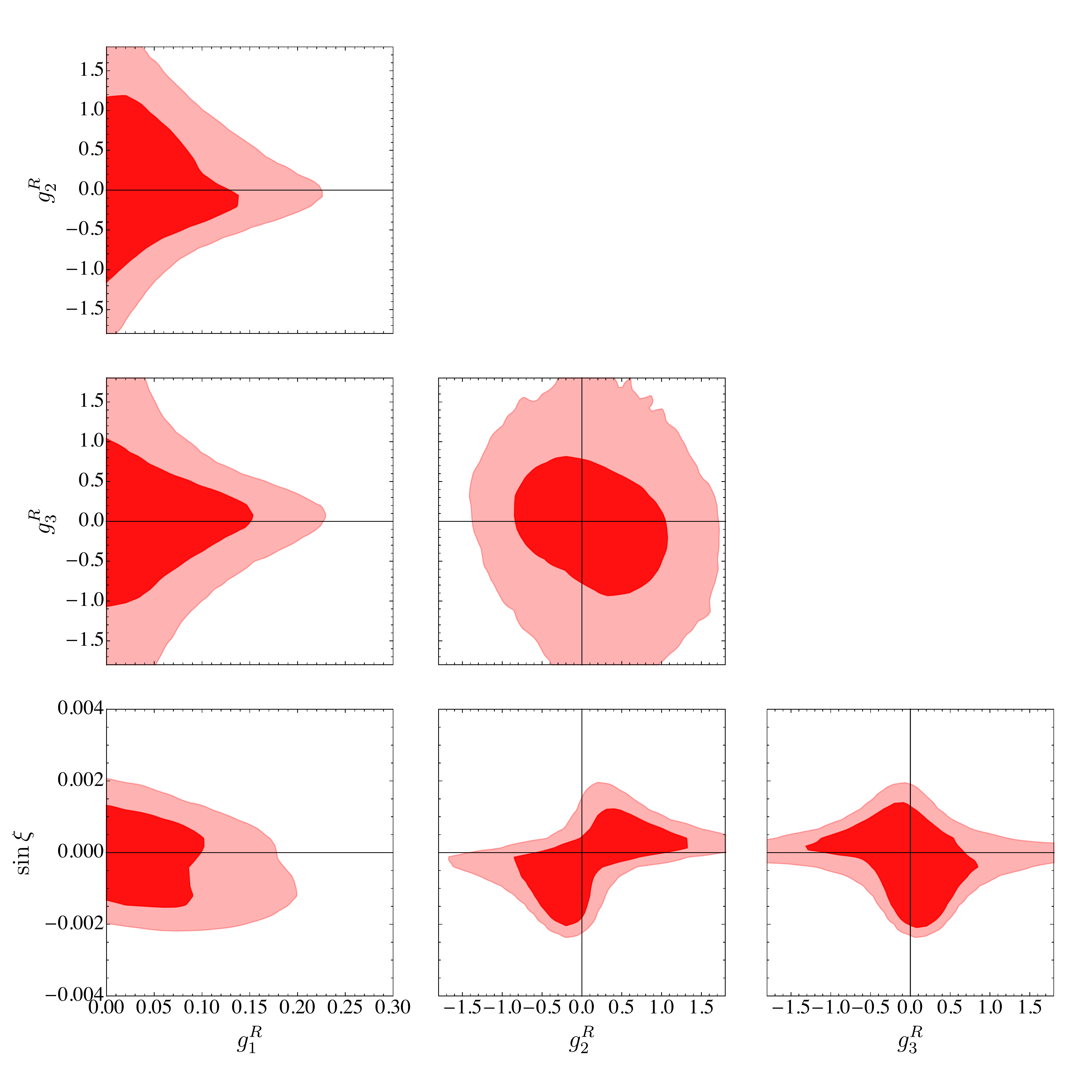} \\
	\caption{68\% and 95\% CL regions for the LFUV case with {right-handed}
 couplings $g^R_{i}=g^R_{ii}$. \label{LFUV_R_plot}}
\end{figure}

Here we study the case in which the couplings that are diagonal in flavour space but not proportional to the unit matrix:
\begin{enumerate}
	\item Vectorial couplings: $g^L_{ii}=g^R_{ii}=g^V_{i}$
	\item Left-handed couplings: $g^L_{ii}=g_{i}^L$
	\item Right-handed couplings: $g^R_{ii}=g_{i}^R$
\end{enumerate}
These scenarios are shown in Figs.~\ref{LFUV_V_plot}-\ref{LFUV_R_plot}.
The couplings to electrons are very well constrained and can be at most of the order of 0.2 due to the LEP bounds on 4-electron contact interactions. The bounds on muon and tau  couplings are less stringent and therefore can be as large as 2 for larger $M_{Z^\prime}=1\,$TeV. This is also due to the  fact that $\tau\to\mu\nu\bar\nu$ prefers larger couplings to muons and taus which enter via the box contributions. However, in this case effects in the EW fit are generated as well, such that no significant preference over the SM fit can be achieved.

\subsection{Lepton Flavour Violation}\label{LFV}
\begin{figure}
\centering
\includegraphics[scale=1]{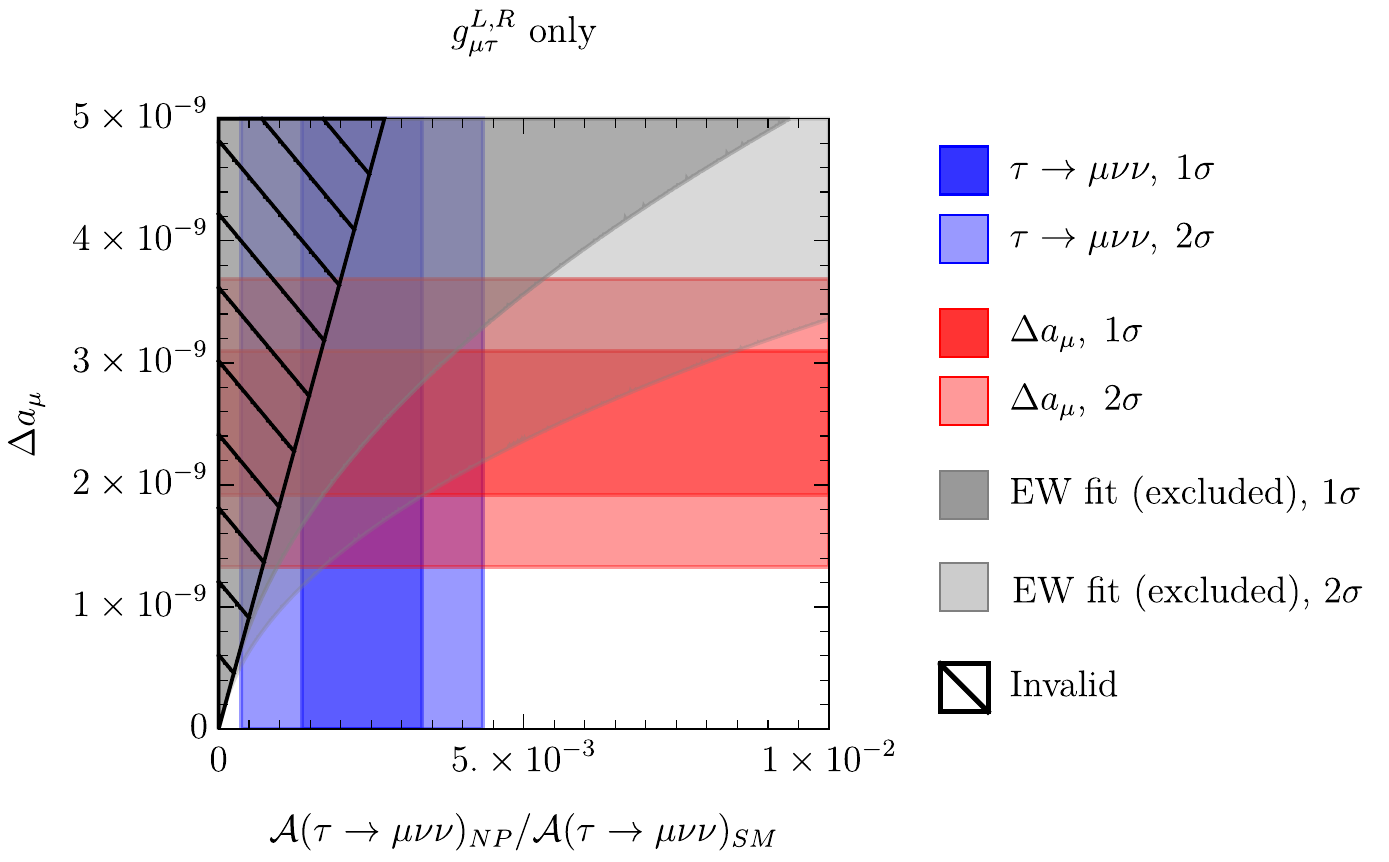}
\caption{Correlations between $(g-2)_\mu$ and {${\mathcal A}(\tau\to\mu\nu\bar\nu)_{\rm NP}/{\mathcal A}(\tau\to\mu\nu\bar\nu)_{\rm SM}$} in the scenario with zero $Z-Z'$-mixing ($\sin\xi=0$), where all $Z'$ couplings to leptons are set to zero apart from $g_{\mu\tau}^{L,R}\neq 0$. The colored regions are preferred by the anomalous magnetic moment of the muon and by LFUV in tau decays, while the gray region is excluded by electroweak data for 
$M_{Z'}=1\,$TeV. Note that for lighter $Z^\prime$ masses EW precision observables would be less constraining. The hatched region is excluded in this setup in the sense that points within it cannot be reached in this setup.}\label{correlations}
\end{figure}

Here we study the following scenarios for the couplings:
\begin{enumerate}
	\item Flavour violating tau-muon couplings: $g^R_{23}$ and $g^L_{23}$
	\item $L_\mu-L_\tau$ symmetry with left-handed rotations
	\item Vectorial couplings: $g^L_{ij}=g^R_{ij}=g^V_{ij}$
	\item Left-handed couplings: $g^L_{ij}$
	\item Right-handed couplings: $g^R_{ij}$
\end{enumerate}
Scenario 1, which assumes that only $g_{23}^L$ and $g_{23}^R$ are non-zero, allows us to find interesting correlations as shown in Fig.~\ref{correlations}. Here one can see that it is possible to explain $\tau\to\mu\nu\bar\nu$ and $(g-2)_\mu$ simultaneously, predicting observable effects in $Z\to\mu\bar\mu$ and $Z\to\tau\bar\tau$ for $M_{Z^\prime}=1\,$TeV. However, due to the logarithmic corrections involved here, the effects in $Z\to\mu\bar\mu$ and $Z\to\tau\bar\tau$ become weaker for smaller masses. These effects will allow in the future to conquer the parameter space (assuming an explanation of $(g-2)_\mu$), as direct searches and EW precision constraints test complementary regions in parameter space.
\smallskip

Next, let us consider case 2 with a (broken) $L_\mu-L_\tau$ symmetry. This means that the coupling matrix takes the form
	\begin{equation}
	g_{ij}^V = \left( {\begin{array}{*{20}{c}}
		0&0&0\\
		0&g&0\\
		0&0&{ - g}
		\end{array}} \right)
     \end{equation}
in the interaction basis. Now we assume that $L_\mu-L_\tau$ is broken by the charged lepton Yukawa couplings in the left-handed sector such that, after EW symmetry breaking,
	\begin{equation}
	g_{ij}^L = {\left( {\left( {\begin{array}{*{20}{c}}
				1&0&0\\
				0&{\cos {\beta _{23}}}&{  \sin {\beta _{23}}}\\
				0&{-\sin {\beta _{23}}}&{\cos {\beta _{23}}}
				\end{array}} \right) \cdot \left( {\begin{array}{*{20}{c}}
				0&0&0\\
				0&g&0\\
				0&0&{ - g}
				\end{array}} \right)\cdot\left( {\begin{array}{*{20}{c}}
				1&0&0\\
				0&{\cos {\beta _{23}}}&{-\sin {\beta _{23}}}\\
				0&{  \sin {\beta _{23}}}&{\cos {\beta _{23}}}
				\end{array}} \right)} \right)_{ij}}    \,,
 \end{equation}
for $2$--$3$ rotations. The analogous formula for $1$--$2$ rotations follows straightforwardly.
\smallskip

\begin{figure}[h!]
	\centering
	\includegraphics[width=0.7\textwidth]{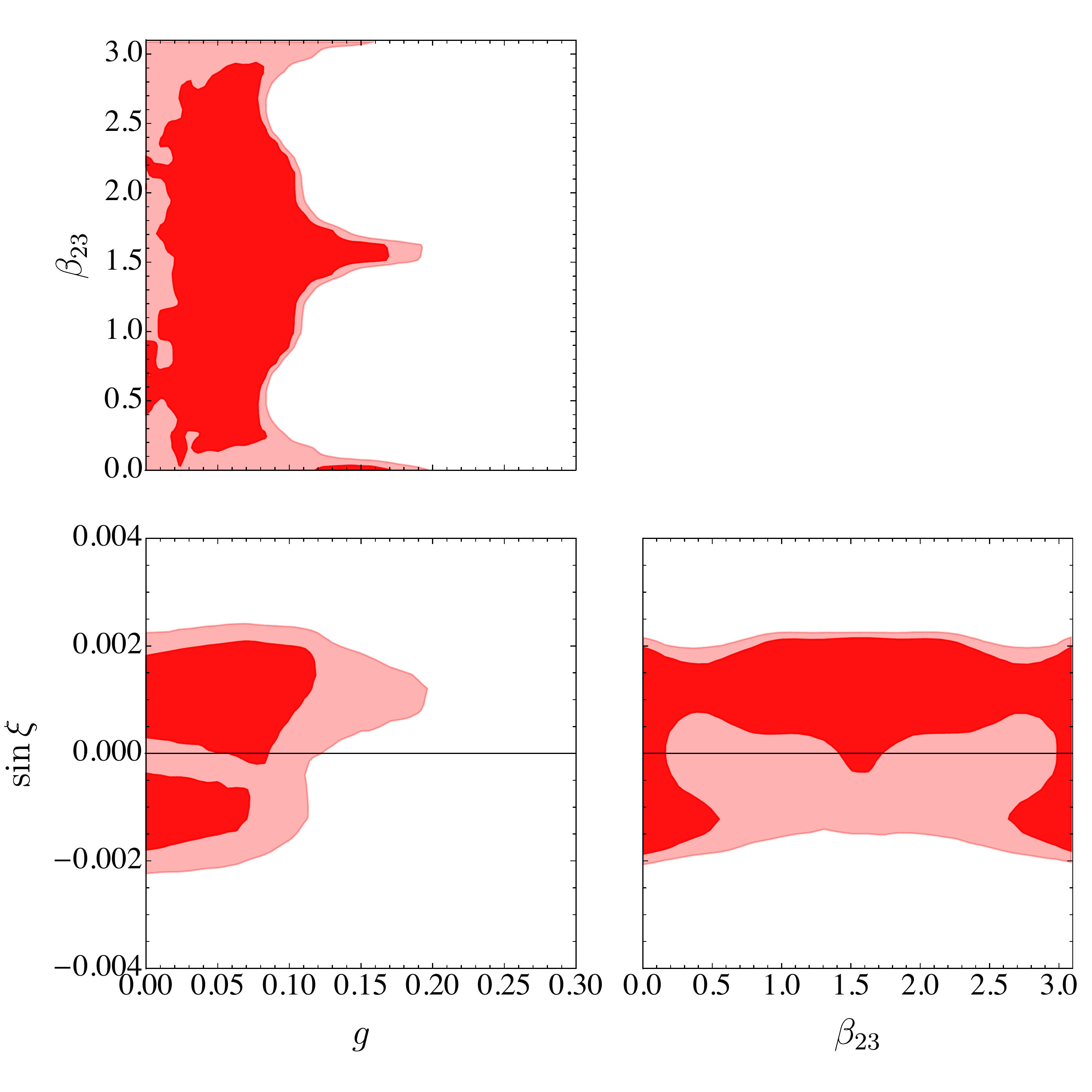} \\
	\caption{68\% and 95\% C.L. regions for the $L_\mu-L_\tau$ coupling $g$, the rotation angle $\beta_{23}$ and the mixing angle $\sin \xi$. \label{LmuLtau23}}
\end{figure}

\begin{figure}[h!]
	\centering
	\includegraphics[width=0.7\textwidth]{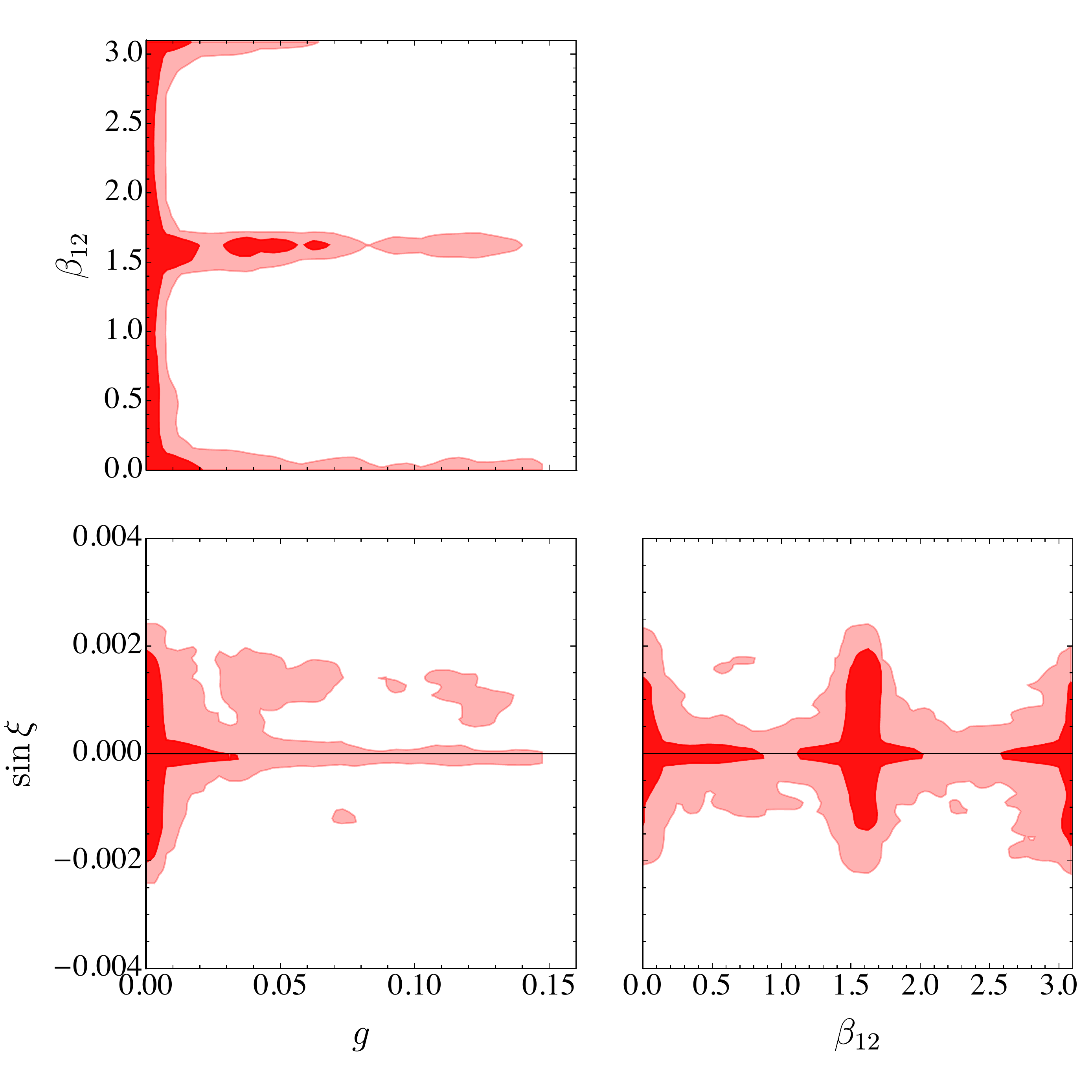} \\
	\caption{68\% and 95\% C.L. regions for the $L_\mu-L_\tau$ coupling $g$, the rotation angle $\beta_{12}$ and the mixing angle $\sin \xi$. \label{LmuLtau12}}
\end{figure}

\begin{figure}[h!]
	\centering
	\includegraphics[scale=1]{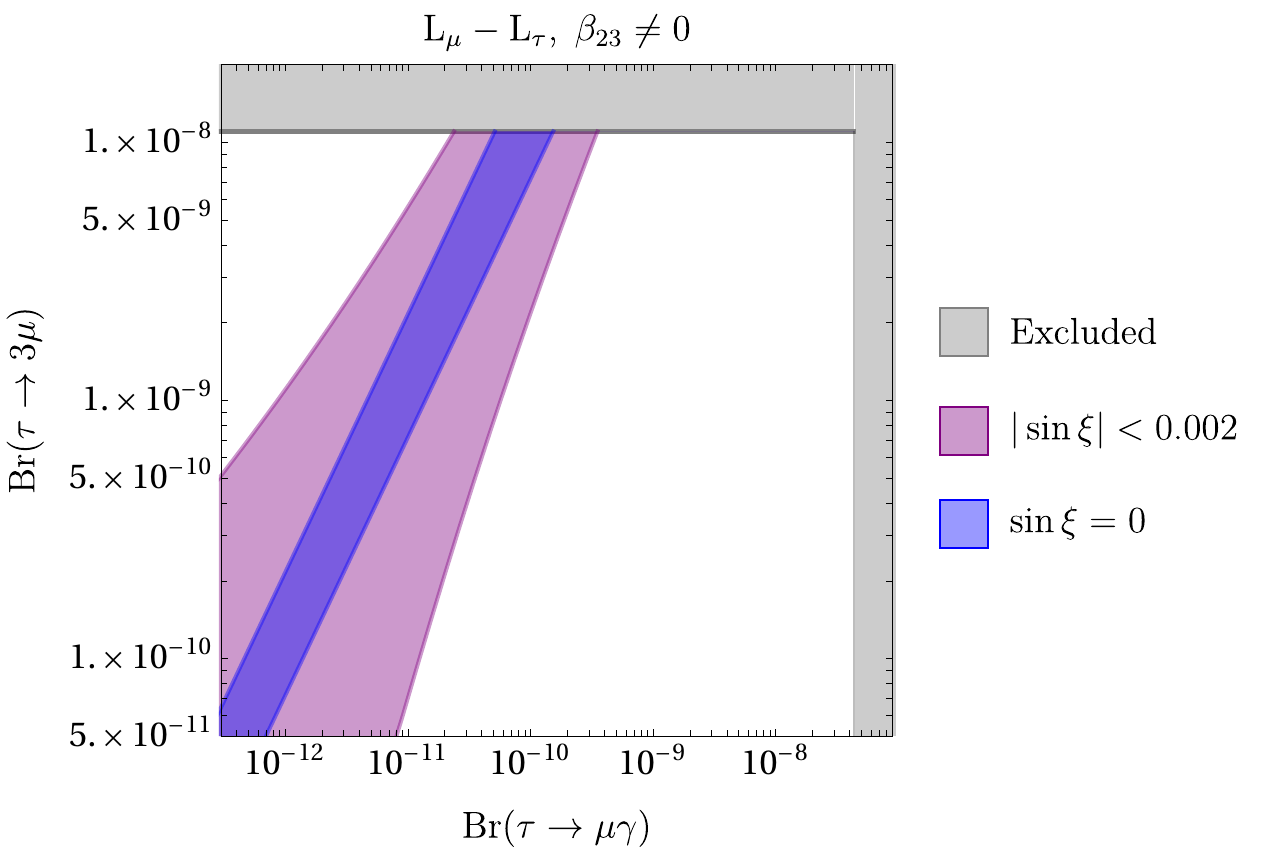} \\
	\caption{Correlations between Br($\tau \to \mu \mu \mu$) and Br($\tau \to \mu \gamma$) for different values of $\sin \xi$ within the $L_\mu-L_\tau$ scenario with left-handed 2-3 rotations ($\beta_{23}\neq 0$). The colored regions are allowed within this setup. \label{Correlation23}}
\end{figure}

\begin{figure}[h!]
	\centering
	\includegraphics[scale=1]{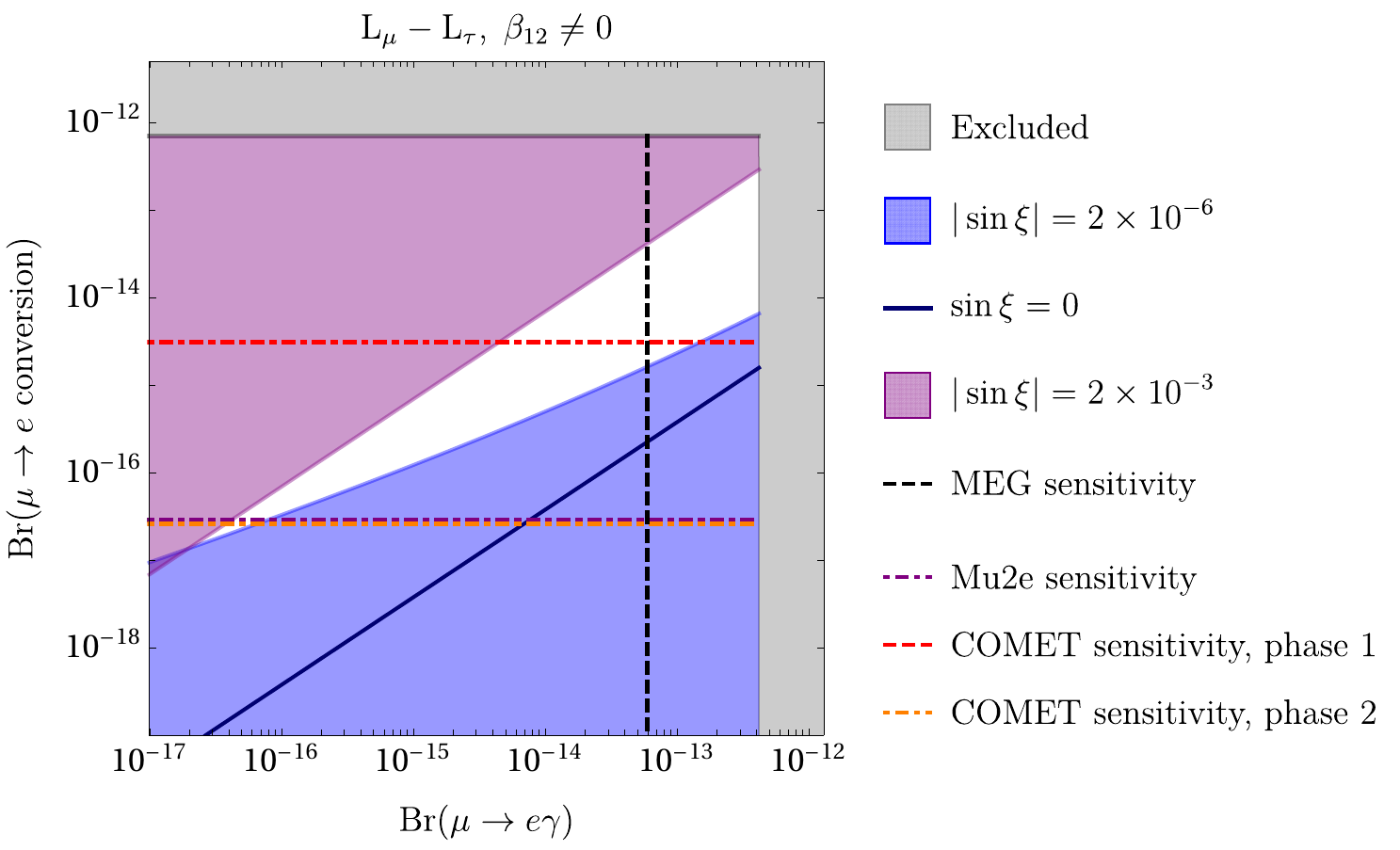}
	\\
		\includegraphics[scale=1]{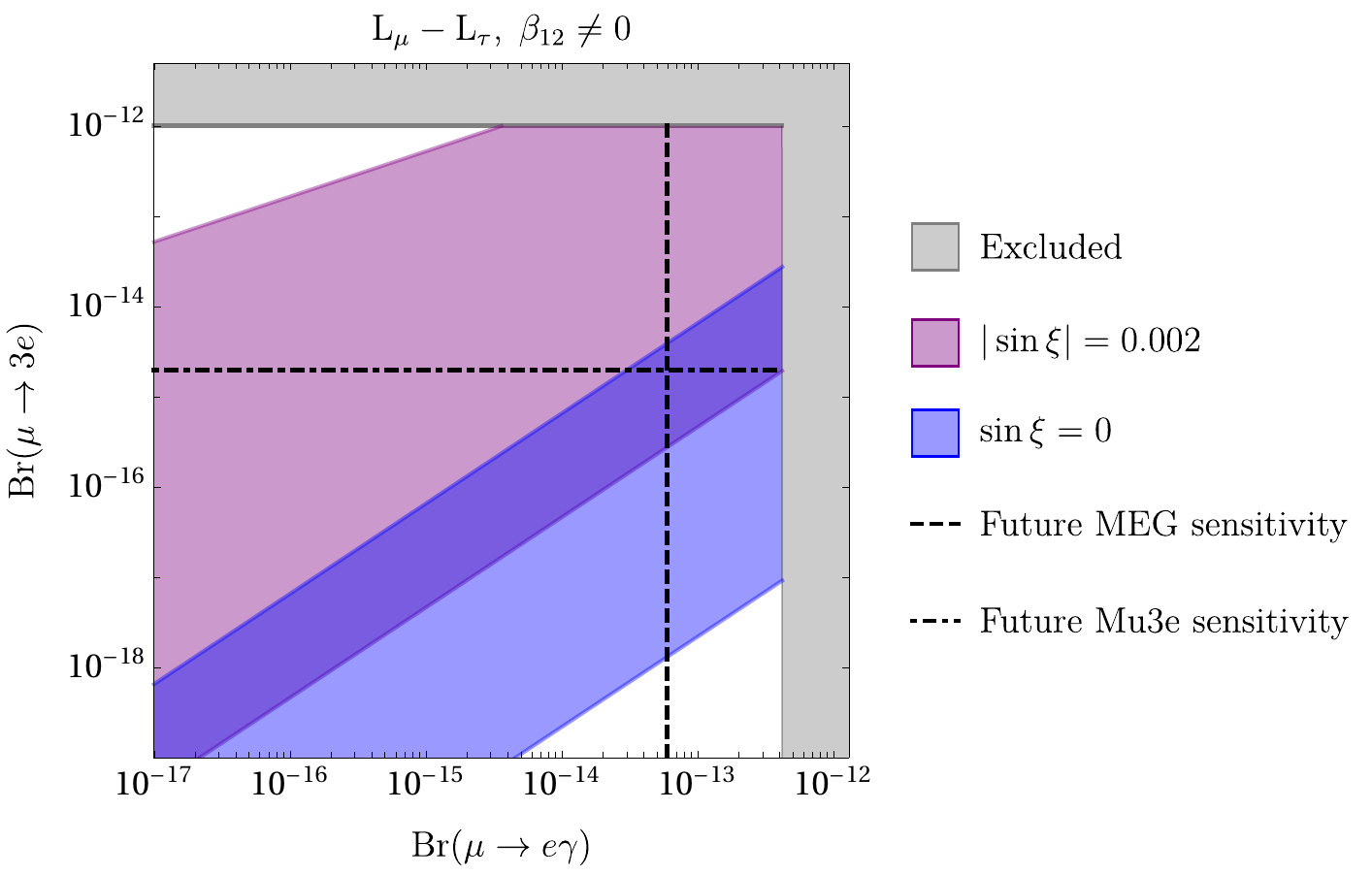}
	\caption{Correlations between different $\mu\to e$ transitions in the $L_\mu-L_\tau$ scenario with left-handed 1-2 rotations ($\beta_{12}\neq 0$). \label{Correlation12}}
\end{figure}

For this setup we show in  Figs.~\ref{LmuLtau23} and \ref{LmuLtau12} the 68\% and 95\% CL regions for the coupling $g$ and the rotation angle $\beta_{23}$ and $\beta_{12}$, respectively. In Fig.~\ref{Correlation23} we show the correlations between 
Br[$\tau \to \mu \gamma$] and Br[$\tau \to \mu \mu \mu$], which display that in this scenario the present experimental upper bound on  Br[$\tau \to \mu \mu \mu$]
can easily be saturated, while Br[$\tau \to \mu \gamma$] is orders of magnitude below the present current bound. Therefore, finding in the coming years Br[$\tau \to \mu \gamma$] at the level of $10^{-9}$ would rule out the $L_\mu-L_\tau$ scenario.
Similarly for $\mu\to e$ transitions: Fig.~\ref{Correlation12} demonstrates the importance of $Z-Z^\prime$ mixing and thereby the role of $Z$ in the enhancement of $\mu\to 3e$ and $\mu\to e$ conversion. However, in this case, in contrast to $\rm Br[\tau \to \mu \gamma]$, the branching ratio for $\rm Br[\mu \to e \gamma]$ can easily saturate the present experimental upper bound. MEG, Mu2e, Mu2e and COMET will constrain the allowed space in these plots in the coming years significantly.  
\medskip

We discuss the more general cases 3, 4 and 5, with seven free parameters each, in Appendix~\ref{app:C}.

\section{Conclusions}
\label{conclusions}
{In this paper we have performed a global fit to leptophilic $Z^\prime$ models with the goal to obtain bounds on the $Z^\prime$ couplings to leptons in multi-dimensional scenarios. In our global analysis we took into account a large number of observables,} including $\ell\to\ell^\prime\nu\bar\nu$ decays, anomalous magnetic moments of charged leptons, $\ell\to\ell^\prime\gamma$, $\ell\to3\ell^\prime$ decays,
$\mu\to e$ conversion, electroweak precision observables, lepton flavour violating $Z$ decays, neutrino trident production and LEP searches for four-lepton contact interactions.
\smallskip

Properly extending the  HEPfit code \cite{deBlas:2019okz} by implementing these observables, and performing a Bayesian statistical
 analysis, we obtained bounds on the $Z^\prime$ couplings
in a number of generic scenarios, as listed in Section~\ref{phenomenology}. The results are presented in Figs.~\ref{LFUV_V_plot}-\ref{LFV_R_plot}.
The plots in these figures are self-explanatory but the main message is that the couplings involving electrons are much more strongly bounded than those involving muons or tau leptons. These results should turn out to be useful for building models where the patterns for the couplings are governed by flavour symmetries. 
\smallskip

In more detail, we find that in the LFU scenario neither the tension in $\tau\to\mu\nu\bar\nu/\tau\to e\nu\bar\nu$ nor in the first row CKM unitarity or in $(g-2)_\mu$ can be explained.  In the LFUV scenario the couplings to electrons are very well constrained and can be at most of the order of 0.2 due to the LEP bounds on 4-electron contact interactions. The bounds on muon and tau  couplings are less stringent and therefore can be as large as 2 for larger $M_{Z^\prime}=1\,$TeV.
We then studied more specific scenarios with constrained patterns for the couplings. Here we found that if only $g^L_{23}$ and $g^R_{23}$ are non-zero, the anomaly in the anomalous magnetic moment of the muon and the hint for LFUV in leptonic tau decays can be explained, with interesting predictions for EW precision observables. Furthermore, in scenarios with a $L_\mu-L_\tau$ symmetry we were able to correlate $\tau\to3\mu$ to $\tau\to\mu\gamma$ (and similarly for $\mu\to e$ transitions). These correlations can be used to test this setup with future experiments.
\medskip

\acknowledgments
{We would like to thank Dimitri Bourilkov for useful discussions concerning the LEP bounds.
The work of A.C., C.A.M, F.K. and M.M. is supported by a Professorship Grant (PP00P2\_176884) of the Swiss National Science Foundation. A.C. thanks CERN for the support via the scientific associate program. A.J.B acknowledges financial support
from the Excellence Cluster ORIGINS,
funded by the Deutsche Forschungsgemeinschaft (DFG, German Research Foundation), 
{Excellence Strategy, EXC-2094, 390783311.}}
\medskip

\appendix

\section{LEP-II constraints for general coupling structure}
\label{app:LEP2}

LEP-II measured with high precision the total cross section $\sigma_{\rm TOT} =  \sigma_F+\sigma_B$ and the Forward-Backward assymetry $A_{FB} = (\sigma_F-\sigma_B)/\sigma$ for the process $e^+ e^- \to \ell^+ \ell^-$ (with $\ell = e, \mu, \tau$) at various $\sqrt{s}$  between 130 and 207 GeV (see Tables~3.4 and 3.8 - 3.12 from Ref.~\cite{Schael:2013ita}). One can extract bounds on our model parameters by computing the BSM contribution to $\sigma_F\pm\sigma_B$. Here we work within an EFT aproach with a Lagrangian defined as $\mathcal{L}= \mathcal{L}_{\rm SM}+\frac{1}{v^2}C_iO_i$ and we fierzed the operators giving an effect at the dimension six level into the basis:
\begin{align}
	\begin{split}
	\big[O_{\ell\ell}\big]_{11jj} &= \big(\ell_1\gamma_{\mu}P_L\ell_1 \big) \big(\ell_j\gamma^{\mu}P_L\ell_j \big)\,,\\
	\big[O_{ee}\big]_{11jj} &= \big(\ell_1\gamma_{\mu}P_R\ell_1 \big) \big(\ell_j\gamma^{\mu}P_R\ell_j \big)\,,\\
	\big[O_{\ell e}\big]_{11jj} &= \big(\ell_1\gamma_{\mu}P_L\ell_1 \big) \big(\ell_j\gamma^{\mu}P_R\ell_j \big)\,,\\
	\big[O_{\ell e}\big]_{jj11} &= \big(\ell_j\gamma_{\mu}P_L\ell_j \big) \big(\ell_1\gamma^{\mu}P_R \ell_1 \big)\,.\\
	\end{split}
	\label{LEPope}
\end{align}
 Assuming that the $Z-Z^\prime$ mixing effects are sub-leading, and for vanishing lepton masses, the BSM contributions to $\sigma_{\rm TOT}$ and $A_{FB}$ for $\ell = \mu$ are given by:
\begin{align}
\begin{split}
\delta\left(\sigma_{F}+\sigma_{B}\right) = \frac{1}{24 \pi v^{2}}\Bigg\{& e^{2}\big(\big[c_{\ell \ell}\big]_{1122}+ \big[c_{e e}\big]_{1122}+\big[c_{\ell e}\big]_{1122}
+\big[c_{\ell e}\big]_{2211}\big)\\
&+\frac{s\big(g_{2}^{2}+g_{1}^{2}\big)}{s-m_{Z}^{2}}\Bigg[\left(-\frac{1}{2}+s_W^2\right)^{2} \, \big[c_{\ell \ell}\big]_{1122}+s_W^4\big[c_{e e}\big]_{1122}\\
&+\left(-\frac{1}{2}+s_W^2\right) \, s_W^2\left(\big[c_{\ell e}\big]_{1122}+\big[c_{\ell e}\big]_{2211}\right)\Bigg]\Bigg\}\,,
\end{split}
\end{align} 
\begin{align}
	\begin{split}
 \delta\left(\sigma_{F}-\sigma_{B}\right) = \frac{1}{32 \pi v^{2}}\Bigg\{& e^{2}\left(\big[c_{\ell \ell}\big]_{1122}+\big[c_{e e}\big]_{1122}-\big[c_{\ell e}\big]_{1122}-\big[c_{\ell \ell}\big]_{2211}\right) \\
&+\frac{s\big(g_{2}^{2}+g_{1}^{2}\big)}{s-m_{Z}^{2}}\Bigg[\left(-\frac{1}{2}+s_W^2\right)^{2} \, \big[c_{\ell \ell}\big]_{1122}+s_W^4\big[c_{e e}\big]_{1122}\\
&-\left(-\frac{1}{2}+s_W^2\right) \,s_W^2\left(\big[c_{\ell e}\big]_{1122}+\big[c_{\ell e}\big]_{2211}\right)\Bigg]\Bigg\}\,.
\end{split}
\end{align} 
For the $\tau^+\tau^- $ channel we have the same {expressions} with the index exchange
 $2 \leftrightarrow 3$. 
 
 On the other hand for $e^+ e^- \to e^+ e^-$ Ref.~\cite{Schael:2013ita}
 gives also the angular differential cross section, 
measured at different $\sqrt{s}$ and $\cos\theta$ ranges (see Tables~3.8 - 3.12 in Ref.~\cite{Schael:2013ita}). In this case the BSM contribution to the differential cross section is given by: 
\begin{align}
	\begin{split}
 \delta\frac{{\rm d}\sigma}{{\rm d}\cos\theta}  = \frac{1}{8 \pi s} \frac{1}{v^{2}}\Bigg\{&2u^{2}\Bigg[e^{2}\left(\big[c_{\ell \ell}\big]_{1111}+\left[c_{e e}\right]_{1111}\right)\left(\frac{1}{s}+\frac{1}{t}\right)\\
&+\left(g_{2}^{2}+g_{1}^{2}\right)\left(\frac{1}{s-m_{Z}^{2}}+\frac{1}{t-m_{Z}^{2}}\right) \left(\left(\frac{1}{2}+s_W^2\right)^2\left[c_{\ell \ell}\right]_{1111}+s_W^4 \left[c_{e e}\right]_{1111}\right) \Bigg] \\
&+t^{2}\left[\left[c_{\ell e}\right]_{1111} \frac{e^{2}}{s}+\left[c_{\ell e}\right]_{1111} \frac{\left(g_{2}^{2}+g_{1}^{2}\right) (\frac{1}{2}+s_W^2) s_W^2} {s-m_{Z}^{2}}\right] \\
& +s^{2}\left[\left[c_{\ell e}\right]_{1111} \frac{e^{2}}{t}+\left[c_{\ell e}\right]_{1111} \frac{\left(g_{2}^{2}+g_{1}^{2}\right) (\frac{1}{2}+s_W^2) s_W^2}{t-m_{Z}^{2}}\right]\Bigg\}\,,
\end{split}
\end{align} 
where $t=-\frac{s}{2}\left(1-\cos\theta\right)$ and $u=-\frac{s}{2}\left(1+\cos\theta\right)$.
\smallskip

The relations between our model parameters and the SMEFT at dimension six level are as follows:
\begin{align}
\begin{split}
\big[c_{\ell\ell}\big]_{1122} &= -\frac{v^2}{M_{Z^\prime}^2}\,  \left(g^L_{ee} g^L_{\mu\mu}+g^L_{e\mu} g^L_{\mu e}\right), \\ 
\big[c_{\ell e}\big]_{1122} &= -\frac{ v^2}{M_{Z^\prime}^2} \left(g^L_{ee} g^R_{\mu \mu}+ \,g^R_{\mu e} g^L_{e \mu} \right), \\
\big[c_{\ell e}\big]_{2211} &= -\frac{ v^2}{M_{Z^\prime}^2} \left( g^R_{ee} g^L_{\mu\mu}+g^L_{\mu e} g^R_{e \mu} \right)\,, \\
\big[c_{e e}\big]_{1122} &= -\frac{ v^2}{M_{Z^\prime}^2} \left(g^R_{ee} g^R_{\mu \mu}+g^R_{\mu e} g^R_{e \mu}\right) \,,\\
\big[c_{\ell\ell}\big]_{1111} &= -\frac{ v^2}{2M_{Z^\prime}^2} g^L_{ee} g^L_{ee}\,, \\
\big[c_{e e}\big]_{1111} &= -\frac{v^2}{2M_{Z^\prime}^2}  g^R_{ee} g^R_{ee}\,, \\
\big[c_{\ell e}\big]_{1111} &= -\frac{ v^2}{M_{Z^\prime}^2} g^R_{ee} g^L_{\mu\mu}\,. \\
\end{split}
\label{eq:matching}
\end{align}
and similarly for $2 \leftrightarrow 3$.
\medskip

\section{QED Penguin Contributions}\label{QEDP}
\subsection{Hidden Operators}

\begin{figure}
\centering 
\subfloat[]{{\includegraphics[width=0.45\textwidth]{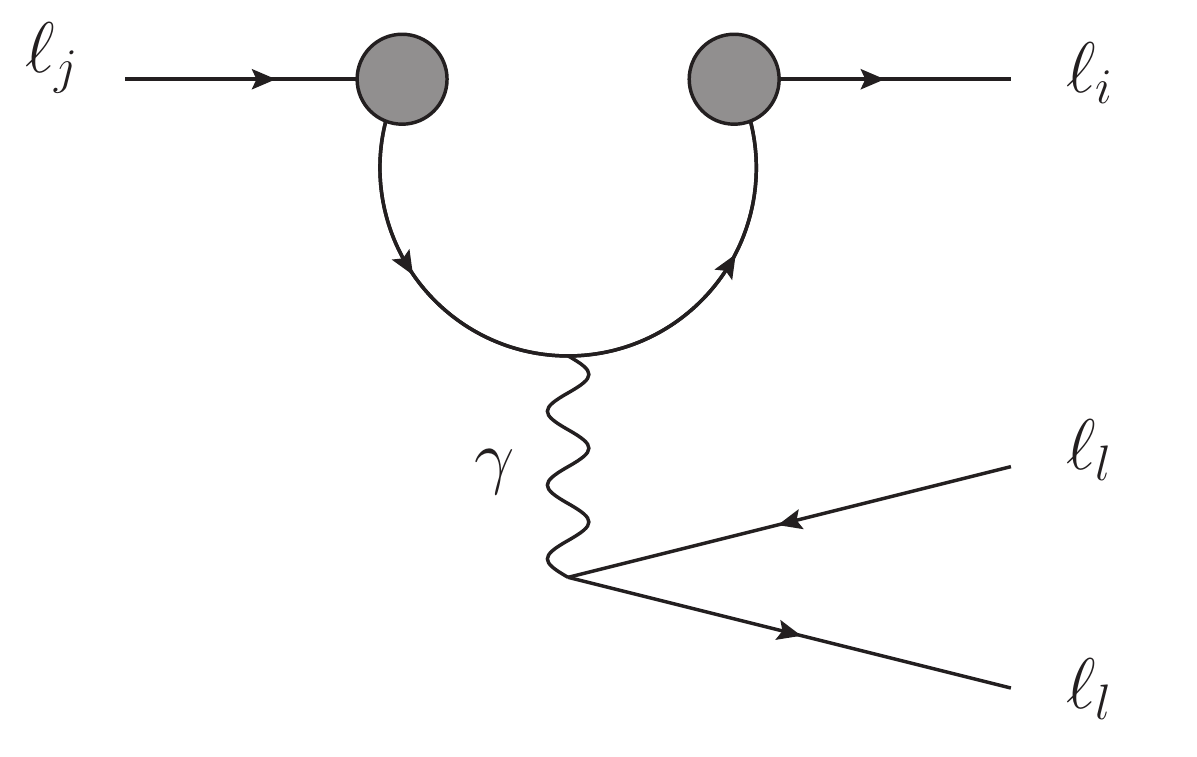} }}
\qquad\subfloat[]{{\includegraphics[width=0.45\textwidth]{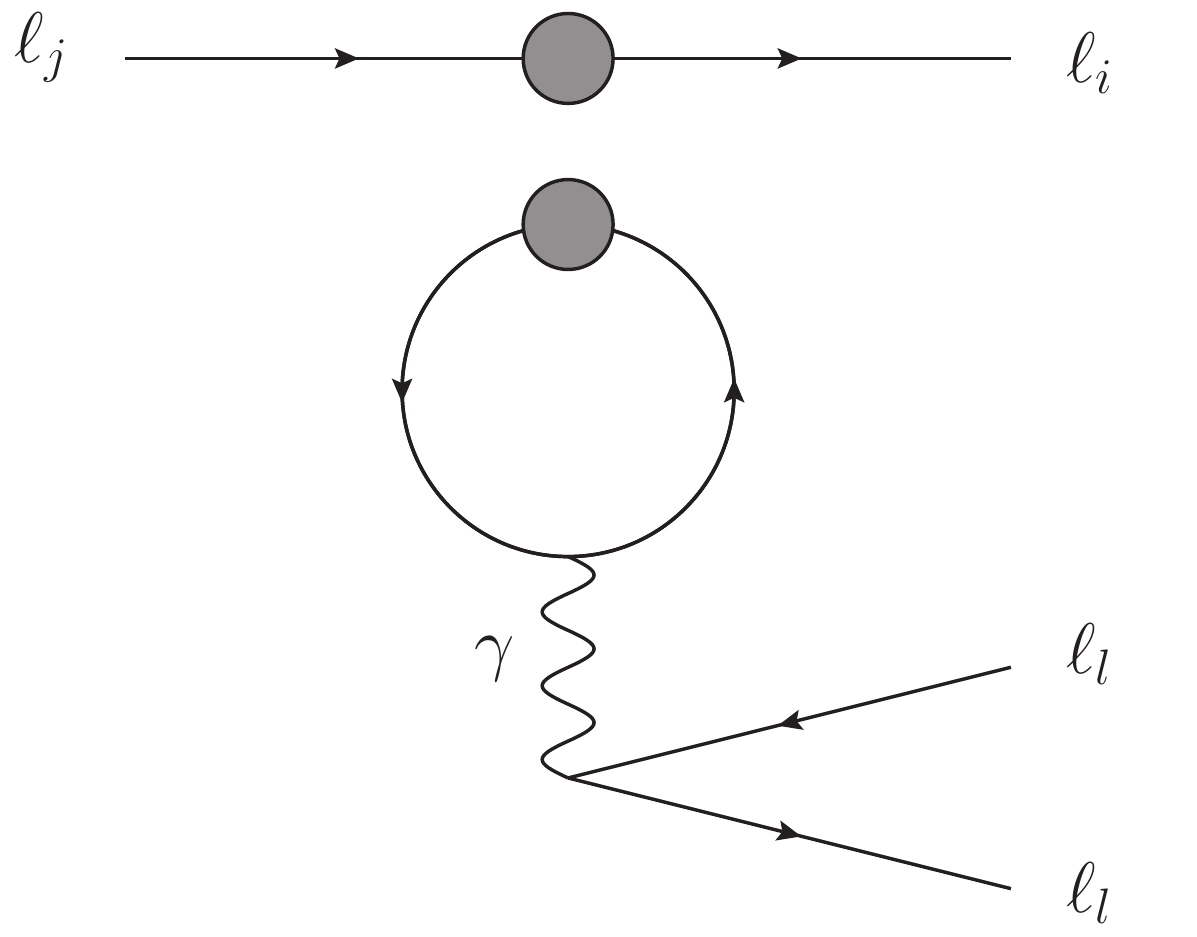} }}
\caption{Feynman diagrams with two operator insertions corresponding to different Wick contractions in the EFT where the $Z^\prime$ is integrated out. }%
 \label{l_lll_eft_photon_diags}%
\end{figure}

The following operators contained in Eq.~(\ref{4-lepton-operators})
\begin{equation}\label{RNEWOPE1}
  \mathcal{O}_{e\mu,\mu\mu}^{AB},\qquad \mathcal{O}_{e\mu,\tau\tau}^{AB},\qquad
  \mathcal{O}_{e\tau,\tau\mu}^{AB},\qquad \mathcal{O}_{e\tau,\tau\tau}^{AB},\qquad  \mathcal{O}_{\mu\tau,\tau\tau}^{AB},
\end{equation}
with $AB=LL,LR,RL,RR$, do not contribute to the flavour changing processes considered by us at tree level and could thus be considered as {\em hidden operators}. However, they contribute to these processes through QED penguin diagrams (as depicted in Fig.~\ref{l_lll_eft_photon_diags}) ~\cite{Crivellin:2017rmk,Pruna:2014asa,Aebischer:2017gaw}.
\smallskip

In the formal language, these diagrams generate mixing of the operators in \eq{RNEWOPE1}  into operators contributing already
at tree-level to flavour changing processes. In this appendix
we present the results for this additional effect.
\smallskip

This mixing can be found by  means of standard methods \cite{Buras:2020xsm}. That is by calculating the relevant one-loop anomalous dimensions obtained by inserting the operators in 
\eq{RNEWOPE1} into one-loop
 off-shell photon penguin diagrams~\cite{Crivellin:2017rmk,Pruna:2014asa,Aebischer:2017gaw}. As shown in Fig.~\ref{l_lll_eft_photon_diags}, there are two possible operator insertions which result in the same contributions to the Wilson coefficients, unless they vanish. In the cases at hand one should note that for the inserted $LL$ and $RR$ operators with three electrons or three muons both penguin topologies contribute, which brings in a factor of two for this mixing relative to the remaining operator insertions.
 However, in the case of $RL$ and $LR$ operators, only the insertions into diagram (b) in Fig.~\ref{l_lll_eft_photon_diags} contribute, and this implies that the operator $ \mathcal{O}_{e\tau,\tau\mu}^{V,RL}$ cannot contribute to this mixing.

 The standard renormalization group evolution is then performed from the scale $M_{Z^\prime}$ down to $m_\tau$ or $m_\mu$ for $\tau$ decays and $\mu$ decay, respectively. In the latter case one takes into acount that the $\tau$-lepton
  is integrated out at $m_\tau$, which implies two different logarithms in the final results. Keeping only the leading logarithms, we find the following results for the additional contributions to the Wilson coefficients of the operators in \eq{4-lepton-operators}. 
\begin{align}\label{hidden}
  C_{e\mu}^{LL}&=\frac{-e^2}{16\pi^2M_{Z'}^{2}}\, \frac{2}{3} \, \bigg(
  g_{e\mu}^L(2g_{\mu\mu}^L+g_{\mu\mu}^R)\ln\!\left(\frac{M_{Z'}^2}{m_\mu^2}\right)
  \!+\!\left( g_{e\tau}^L g_{\tau\mu}^L+g_{e\mu}^L (g_{\tau\tau}^L+g_{\tau\tau}^R)\right) \ln\!\left(\frac{M_{Z'}^2}{m_\tau^2}\right)\!\!\bigg) ,\nonumber
  \\
 C_{e\mu}^{RL}&=\frac{-e^2}{16\pi^2{M_{Z'}^{2}}}\, \frac{2}{3} \,\bigg(g_{e\mu}^R(2g_{\mu\mu}^R+g_{\mu\mu}^L)
    \ln\!\left(\frac{M_{Z'}^2}{m_\mu^2}\right)
    \!+\! \left(g_{e\tau}^R g_{\tau\mu}^R+g_{e\mu}^R (g_{\tau\tau}^L+g_{\tau\tau}^L)\right)\ln\!\left(\frac{M_{Z'}^2}{m_\tau^2}\right)\!\!\bigg),\nonumber
    \\
  C_{e\tau}^{LL}&=\frac{-e^2}{16\pi^2M_{Z'}^{2}}\, \frac{2}{3} \,
  g_{e\tau}^L(2g_{\tau\tau}^L+g_{\tau\tau}^R)
   \ln\!\left(\frac{M_{Z'}^2}{m_\tau^2}\right),\nonumber
  \\
 C_{e\tau}^{RL}&=\frac{-e^2}{16\pi^2M_{Z'}^{2}}\, \frac{2}{3} \,g_{e\tau}^R(g_{\tau\tau}^L+2g_{\tau\tau}^R)
    \ln\!\left(\frac{M_{Z'}^2}{m_\tau^2}\right),\nonumber
    \\
  C_{\mu\tau}^{LL}&=\frac{-e^2}{16\pi^2M_{Z'}^{2}}\, \frac{2}{3} \,
 g_{\mu\tau}^L(2g_{\tau\tau}^L+g_{\tau\tau}^R)
   \ln\!\left(\frac{M_{Z'}^2}{m_\tau^2}\right),\nonumber
  \\
 C_{\mu\tau}^{RL}&=\frac{-e^2}{16\pi^2M_{Z'}^{2}}\, \frac{2}{3} \,
g_{\mu\tau}^R(g_{\tau\tau}^L+2g_{\tau\tau}^R)
   \ln\!\left(\frac{M_{Z'}^2}{m_\tau^2}\right).
\end{align}
The vectorial nature of the photon implies $C_{ij}^{LR}=C_{ij}^{LL}$ and $C_{ij}^{RR}=C_{ij}^{RL}$. It should be noted that these results apply universally to all operators with $j=k$ independent of the lepton flavour, as seen in the first term on the r.h.s in (\ref{Xij}).

The Wilson coefficients of the operators contributing already at tree-level
can also be affected by inserting them in penguin and one-loop
current-current operators.  But since these are already
constrained by tree-level processes and loop-effects turn out to be subleading.
Yet, for completeness we present expressions for these effects in the next appendix.
\subsection{Running of Visible Operators}\label{AppC}
\begin{figure}
\centering 
\subfloat[]{{\includegraphics[scale=.9]{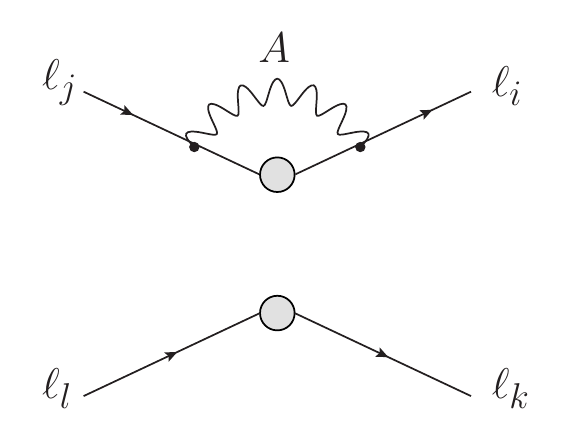}}}
\qquad\subfloat[]{{\includegraphics[scale=.9]{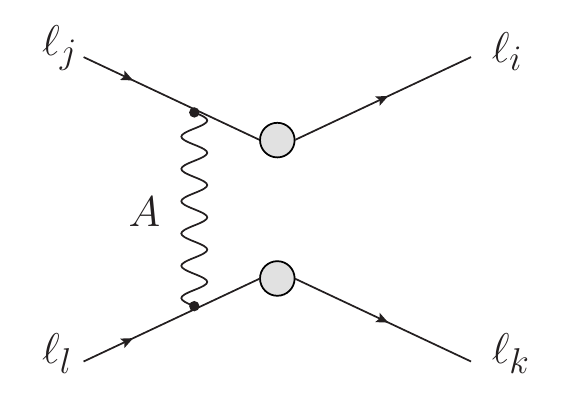}}}
\qquad\subfloat[]{{\includegraphics[scale=.9]{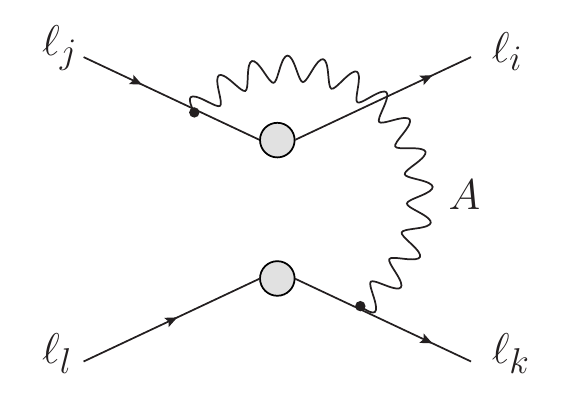}}}
\caption{1-loop QED corrections generating mixing among four-lepton operators.}
\label{qed_corr_diags}
\end{figure}
So far we have only included the contributions from hidden operators mixing into
visible operators. But there are also contributions from the mixing of
visible operators into visible operators, both through QED penguins and insertions in current-current topologies. The QED penguin diagrams are again given in
Fig.~\ref{l_lll_eft_photon_diags}, the current-current topologies in Fig.~\ref{qed_corr_diags}.
\smallskip

The QED penguin contributions imply shifts in the coefficients of the operators in (\ref{hidden}). We find 
\begin{align}
  \Delta C_{e\mu}^{LL}&=\frac{-e^2}{16\pi^2M_{Z'}^{2}}\, \frac{2}{3} \,
  g_{e\mu}^L\left(2g_{ee}^L+g_{ee}^R\right)\ln\left(\frac{M_{Z'}^2}{m_\mu^2}\right)
  ,
  \\
\Delta  C_{e\mu}^{RL}&=\frac{-e^2}{16\pi^2{M_{Z'}^{2}}}\, \frac{2}{3} \,g_{e\mu}^R\left(g_{ee}^L+2g_{ee}^R\right)
    \ln\left(\frac{M_{Z'}^2}{m_\mu^2}\right)\nonumber
    ,\\
  \Delta C_{e\tau}^{LL}&=\frac{-e^2}{16\pi^2M_{Z'}^{2}}\, \frac{2}{3} \,\bigg(
  \,g_{e\tau}^L\,\left(2\,g_{ee}^L+\,g_{ee}^R\right)
  +\left( g_{e\mu}^L\, g_{\mu\tau}^L+g_{e\tau}^L \,\left(g_{\mu\mu}^L+g_{\mu\mu}^R\right)\right)\bigg)
   \ln\!\left(\frac{M_{Z'}^2}{m_\tau^2}\right),\nonumber
  \\
\Delta C_{e\tau}^{RL}&=\frac{-e^2}{16\pi^2M_{Z'}^{2}}\, \frac{2}{3} \,\bigg(g_{e\tau}^R\,\left(2\,g_{ee}^R +\,g_{ee}^L\right)
    + \left(g_{e\mu}^R \,g_{\mu\tau}^R+g_{e\tau}^R\, \left(g_{\mu\mu}^R+g_{\mu\mu}^L\right)\right)\,\bigg)
    \ln\!\left(\frac{M_{Z'}^2}{m_\tau^2}\right),\nonumber
    \\
  \Delta C_{\mu\tau}^{LL}&=\frac{-e^2}{16\pi^2M_{Z'}^{2}}\, \frac{2}{3} \, \bigg(
 \,g_{\mu\tau}^L\,\left(2\,g_{\mu\mu}^L+\,g_{\mu\mu}^R \right)
  +\left( g_{\mu e }^L\, g_{e\tau }^L+g_{\mu\tau}^L \left(g_{ee}^L+\,g_{ee}^R\right)\right) \bigg)
   \ln\!\left(\frac{M_{Z'}^2}{m_\tau^2}\right),\nonumber
  \\
 \Delta C_{\mu\tau}^{RL}&=\frac{-e^2}{16\pi^2M_{Z'}^{2}}\, \frac{2}{3} \,
\bigg(g_{\mu\tau}^R\,(2\,g_{\mu\mu}^R + g_{\mu\mu}^L)\, 
    + \left(g_{\mu e}^R \,g_{e\tau}^R+g_{\mu\tau}^R\, \left(g_{ee}^R+g_{ee}^L\right)\right)\bigg)
   \ln\!\left(\frac{M_{Z'}^2}{m_\tau^2}\right),\nonumber
\end{align}
and for $RR$ and $LR$ with $L$ and $R$ interchanged.
\smallskip

On the other hand, the current-current contributions  modify the
coefficients $X^{AB}_{ij,kl}$ in (\ref{Xij}). We find
\begin{equation}
\Delta  X^{LL}_{ij,kl} =\frac{-e^2}{16\pi^2M_{Z'}^{2}} \, 6 \, 
  g_{ij}^Lg_{kl}^L\ln\!\left(\frac{M_{Z'}^2}{m_r^2}\right),
\end{equation}
\begin{equation}
\Delta  X^{RL}_{ij,kl} =\frac{e^2}{16\pi^2M_{Z'}^{2}} \,6 \,
  g_{ij}^Rg_{kl}^L\ln\!\left(\frac{M_{Z'}^2}{m_r^2}\right),
\end{equation}
where $m_r=m_\tau$ and $m_r=m_\mu$ for $\tau$ and $\mu$ decays, respectively.
For $RR$- and $LR$-coefficients, $L$ and $R$ should be interchanged.

The visible operators also lead to contributions to $\mu\to e$ conversion. The resulting shifts in the Wilson coefficients of Eq.~(\ref{hidden_mue_conv}) are given by
\begin{align}
\begin{split}
	\Delta C_{e\mu,qq}^{LL}&=\frac{e^2Q_q}{16\pi^2}\frac{2}{3}\frac{1}{M_{Z'}^{2}}g_{e\mu}^L\big(2\,g_{ee}^L
	+g_{ee}^R\big) \log\left(\frac{M_{Z'}^2}{m_\mu^2}\right),
\\
	\Delta C_{e\mu,qq}^{RL}&=\frac{e^2Q_q}{16\pi^2}\frac{2}{3}\frac{1}{M_{Z'}^{2}}g_{e\mu}^R\big(2\,g_{ee}^R
	+ g_{ee}^L \big)	\log\left(\frac{M_{Z'}^2}{m_\mu^2}\right).
\end{split}\label{visible_mue_conv}
\end{align}
where, as in Eq.~(\ref{hidden_mue_conv}), $Q_q$ stands for the electric charge of the quarks.

\section{General LFV scenarios}
\label{app:C}

\begin{figure}[h!]
	\centering
	\includegraphics[width=1.05\textwidth]{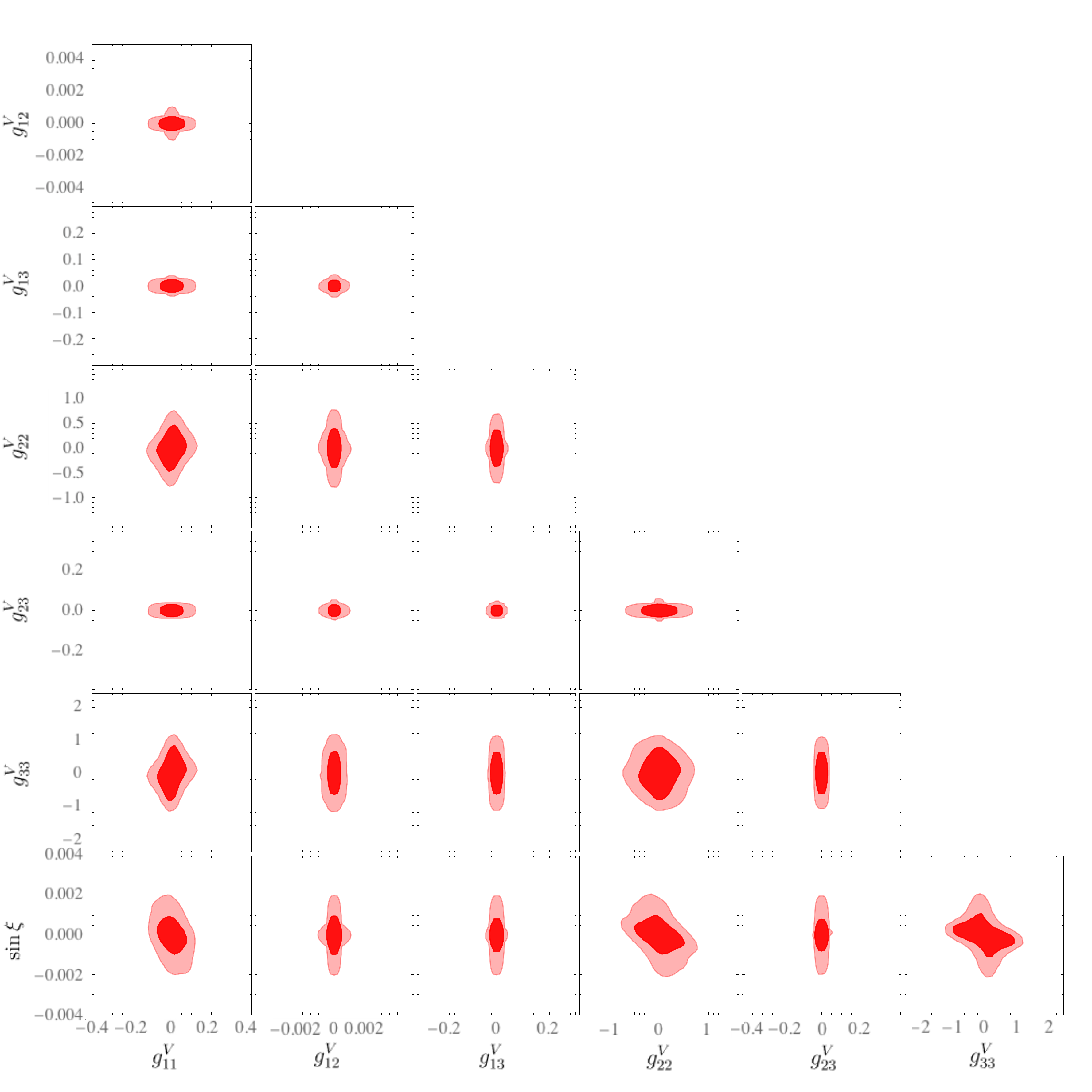} \\
	\caption{68\% and 95\% CL regions for $g^{V}_{ij}=g^{L}_{ij}=g^{R}_{ij}$ and $\sin\xi$. \label{LFV_V_plot}}
\end{figure}

\begin{figure}[h!]
	\centering
	\includegraphics[width=1.05\textwidth]{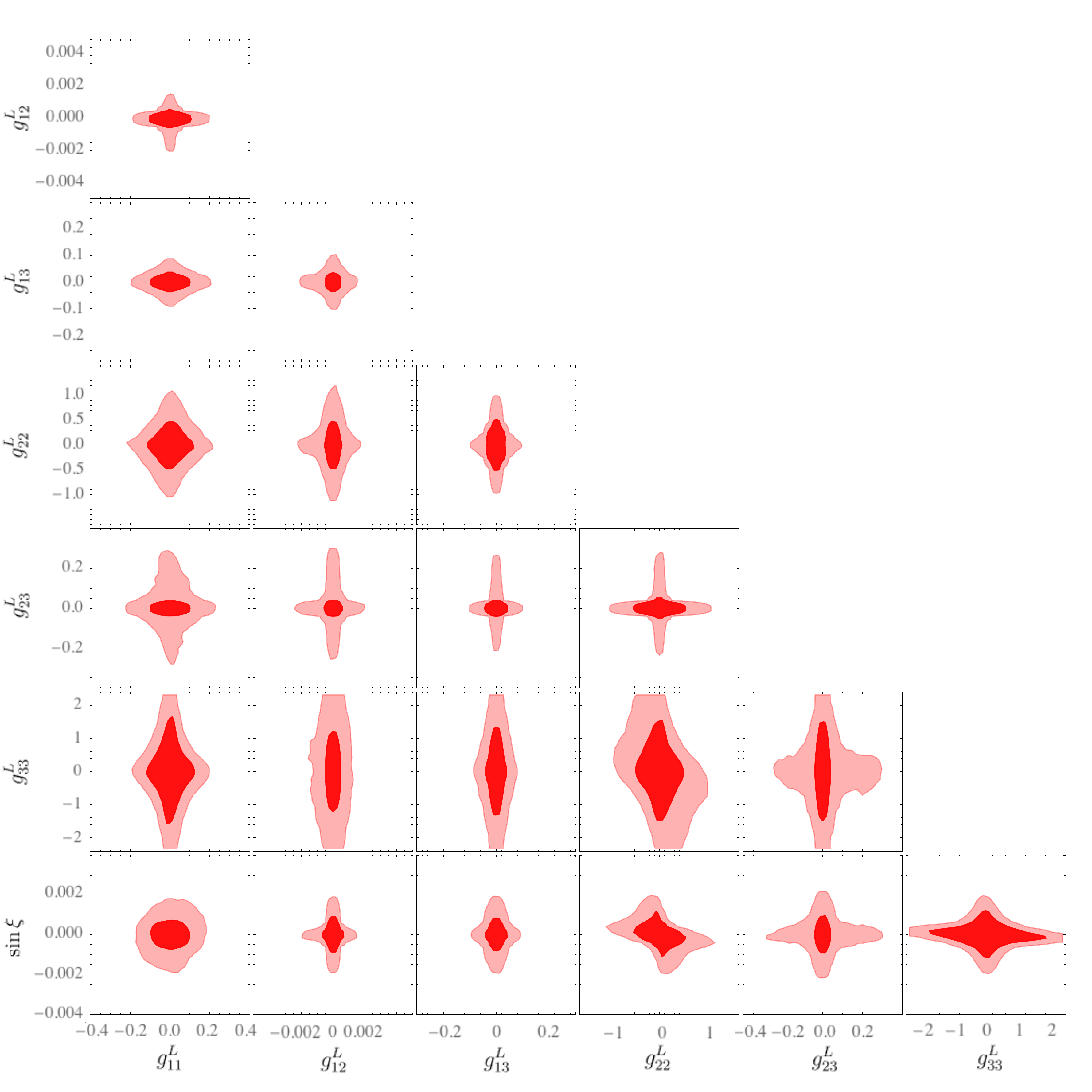} \\
	\caption{68\% and 95\% CL regions for the LFV case where only $g^{L}_{ij}$ is non-zero. \label{LFV_L_plot}}
\end{figure}

\begin{figure}[h!]
	\centering
	\includegraphics[width=1.05\textwidth]{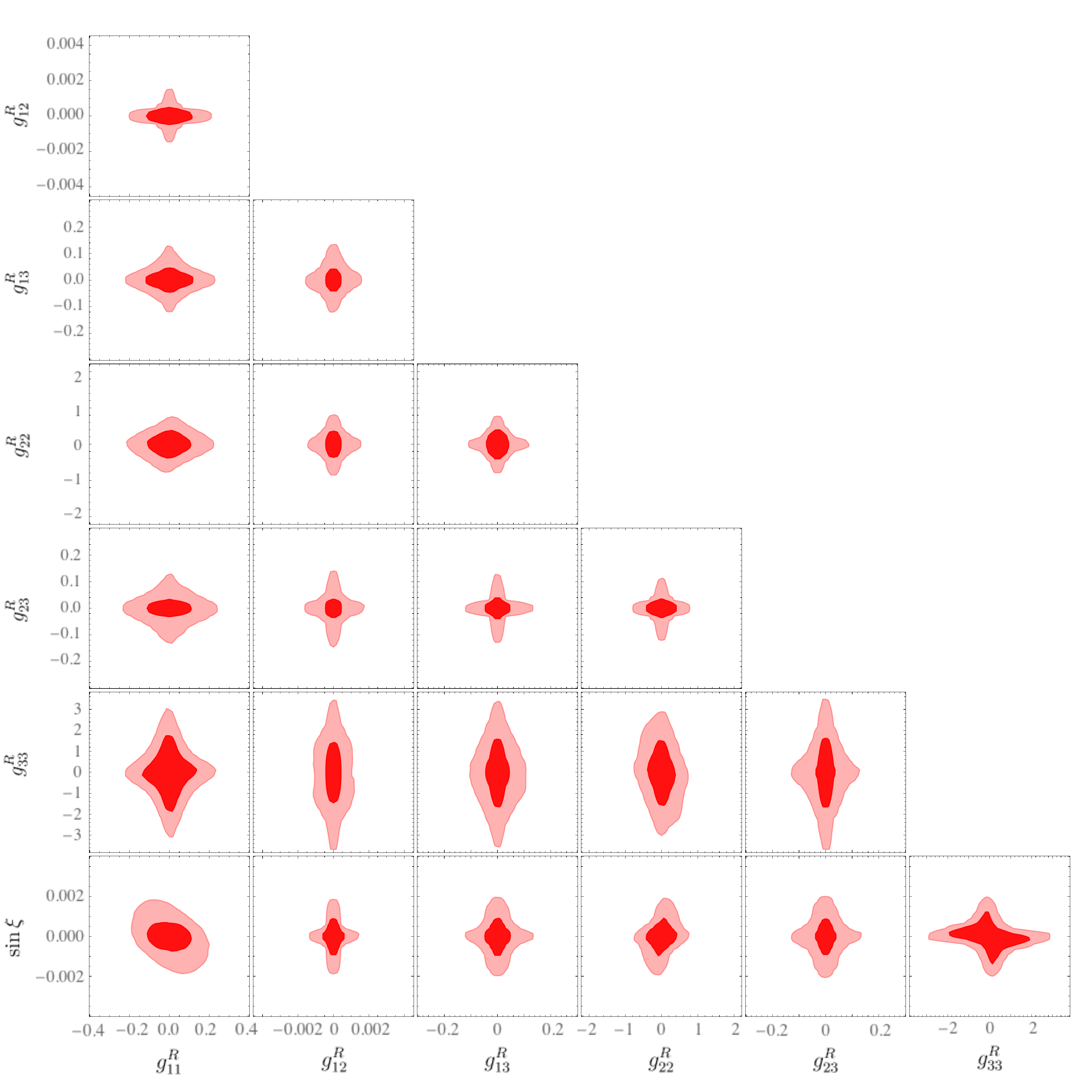} \\
	\caption{68\% and 95\% CL regions for the LFV case where only $g^{R}_{ij}$ is non-zero. \label{LFV_R_plot}}
\end{figure}

Here we study the general cases 4, 5 and 6 from Section~\ref{LFV}. Case 4, where $g^L_{ij}=g^R_{ij}=g^V_{ij}$, is shown in Fig~\ref{LFV_V_plot}. Similarly to previous cases in Sections~\ref{LFU} and~\ref{LFUV}, all couplings involving the first generation are strongly constrained. Also the flavour violating ones are strictly bounded while the flavour conserving ones involving muons and taus can be of order unity. Similarly, we present the results for the cases 5 and 6, where only $g^L_{ij}$ and $g^R_{ij}$ are non-zero, as shown in Fig.~\ref{LFV_L_plot} and Fig.~\ref{LFV_R_plot}, respectively.  Evidently, in these cases the couplings are allowed to be larger than for the case of vectorial couplings.
\smallskip

\clearpage

\bibliographystyle{JHEP}
\bibliography{bibliography}

\providecommand{\href}[2]{#2}\begingroup\raggedright\begin{thebibliography}{100}

\bibitem{Aad:2012tfa}
{\scshape ATLAS} collaboration, \emph{{Observation of a new particle in the
  search for the Standard Model Higgs boson with the ATLAS detector at the
  LHC}}, \href{https://doi.org/10.1016/j.physletb.2012.08.020}{\emph{Phys.
  Lett. B} {\bfseries 716} (2012) 1}
  [\href{https://arxiv.org/abs/1207.7214}{{\ttfamily 1207.7214}}].

\bibitem{Chatrchyan:2012xdj}
{\scshape CMS} collaboration, \emph{{Observation of a New Boson at a Mass of
  125 GeV with the CMS Experiment at the LHC}},
  \href{https://doi.org/10.1016/j.physletb.2012.08.021}{\emph{Phys. Lett. B}
  {\bfseries 716} (2012) 30} [\href{https://arxiv.org/abs/1207.7235}{{\ttfamily
  1207.7235}}].

\bibitem{Aaboud:2017yvp}
{\scshape ATLAS} collaboration, \emph{{Search for new phenomena in dijet events
  using 37 fb$^{-1}$ of $pp$ collision data collected at $\sqrt{s}=$13 TeV with
  the ATLAS detector}},
  \href{https://doi.org/10.1103/PhysRevD.96.052004}{\emph{Phys. Rev. D}
  {\bfseries 96} (2017) 052004}
  [\href{https://arxiv.org/abs/1703.09127}{{\ttfamily 1703.09127}}].

\bibitem{Sirunyan:2018xlo}
{\scshape CMS} collaboration, \emph{{Search for narrow and broad dijet
  resonances in proton-proton collisions at $ \sqrt{s}=13 $ TeV and constraints
  on dark matter mediators and other new particles}},
  \href{https://doi.org/10.1007/JHEP08(2018)130}{\emph{JHEP} {\bfseries 08}
  (2018) 130} [\href{https://arxiv.org/abs/1806.00843}{{\ttfamily
  1806.00843}}].

\bibitem{Aad:2019fac}
{\scshape ATLAS} collaboration, \emph{{Search for high-mass dilepton resonances
  using 139 fb$^{-1}$ of $pp$ collision data collected at $\sqrt{s}=$13 TeV
  with the ATLAS detector}},
  \href{https://doi.org/10.1016/j.physletb.2019.07.016}{\emph{Phys. Lett. B}
  {\bfseries 796} (2019) 68}
  [\href{https://arxiv.org/abs/1903.06248}{{\ttfamily 1903.06248}}].

\bibitem{CMS:2019tbu}
{\scshape CMS} collaboration, \emph{{Search for a narrow resonance in high-mass
  dilepton final states in proton-proton collisions using
  140$~\mathrm{fb}^{-1}$ of data at $\sqrt{s}=13~\mathrm{TeV}$}}, .

\bibitem{Sirunyan:2021khd}
{\scshape CMS} collaboration, \emph{{Search for resonant and nonresonant new
  phenomena in high-mass dilepton final states at $\sqrt{s} = $ 13 TeV}},
  \href{https://arxiv.org/abs/2103.02708}{{\ttfamily 2103.02708}}.

\bibitem{Langacker:2000ju}
P.~Langacker and M.~Plumacher, \emph{{Flavor changing effects in theories with
  a heavy $Z^\prime$ boson with family nonuniversal couplings}},
  \href{https://doi.org/10.1103/PhysRevD.62.013006}{\emph{Phys. Rev. D}
  {\bfseries 62} (2000) 013006}
  [\href{https://arxiv.org/abs/hep-ph/0001204}{{\ttfamily hep-ph/0001204}}].

\bibitem{Langacker:2008yv}
P.~Langacker, \emph{{The Physics of Heavy $Z^\prime$ Gauge Bosons}},
  \href{https://doi.org/10.1103/RevModPhys.81.1199}{\emph{Rev. Mod. Phys.}
  {\bfseries 81} (2009) 1199}
  [\href{https://arxiv.org/abs/0801.1345}{{\ttfamily 0801.1345}}].

\bibitem{Buras:2012jb}
A.~J. Buras, F.~De~Fazio and J.~Girrbach, \emph{{The Anatomy of $Z^\prime$ and
  Z with Flavour Changing Neutral Currents in the Flavour Precision Era}},
  \href{https://doi.org/10.1007/JHEP02(2013)116}{\emph{JHEP} {\bfseries 02}
  (2013) 116} [\href{https://arxiv.org/abs/1211.1896}{{\ttfamily 1211.1896}}].

\bibitem{Altmannshofer:2015mqa}
W.~Altmannshofer and I.~Yavin, \emph{{Predictions for lepton flavor
  universality violation in rare B decays in models with gauged $L_\mu -
  L_\tau$}}, \href{https://doi.org/10.1103/PhysRevD.92.075022}{\emph{Phys. Rev.
  D} {\bfseries 92} (2015) 075022}
  [\href{https://arxiv.org/abs/1508.07009}{{\ttfamily 1508.07009}}].

\bibitem{Crivellin:2015era}
A.~Crivellin, L.~Hofer, J.~Matias, U.~Nierste, S.~Pokorski and J.~Rosiek,
  \emph{{Lepton-flavour violating $B$ decays in generic $Z'$ models}},
  \href{https://doi.org/10.1103/PhysRevD.92.054013}{\emph{Phys. Rev. D}
  {\bfseries 92} (2015) 054013}
  [\href{https://arxiv.org/abs/1504.07928}{{\ttfamily 1504.07928}}].

\bibitem{Allanach:2019mfl}
B.~C. Allanach, J.~M. Butterworth and T.~Corbett, \emph{{Collider constraints
  on $Z^\prime$ models for neutral current B-anomalies}},
  \href{https://doi.org/10.1007/JHEP08(2019)106}{\emph{JHEP} {\bfseries 08}
  (2019) 106} [\href{https://arxiv.org/abs/1904.10954}{{\ttfamily
  1904.10954}}].

\bibitem{Buras:2020xsm}
A.~Buras, \emph{{Gauge Theory of Weak Decays}}. Cambridge University Press, 6,
  2020, \href{https://doi.org/10.1017/9781139524100}{10.1017/9781139524100}.

\bibitem{Schael:2013ita}
{\scshape ALEPH, DELPHI, L3, OPAL, LEP Electroweak} collaboration,
  \emph{{Electroweak Measurements in Electron-Positron Collisions at
  W-Boson-Pair Energies at LEP}},
  \href{https://doi.org/10.1016/j.physrep.2013.07.004}{\emph{Phys. Rept.}
  {\bfseries 532} (2013) 119}
  [\href{https://arxiv.org/abs/1302.3415}{{\ttfamily 1302.3415}}].

\bibitem{Bennett:2006fi}
{\scshape Muon g-2} collaboration, \emph{{Final Report of the Muon E821
  Anomalous Magnetic Moment Measurement at BNL}},
  \href{https://doi.org/10.1103/PhysRevD.73.072003}{\emph{Phys. Rev. D}
  {\bfseries 73} (2006) 072003}
  [\href{https://arxiv.org/abs/hep-ex/0602035}{{\ttfamily hep-ex/0602035}}].

\bibitem{Mohr:2015ccw}
P.~J. Mohr, D.~B. Newell and B.~N. Taylor, \emph{{CODATA Recommended Values of
  the Fundamental Physical Constants: 2014}},
  \href{https://doi.org/10.1103/RevModPhys.88.035009}{\emph{Rev. Mod. Phys.}
  {\bfseries 88} (2016) 035009}
  [\href{https://arxiv.org/abs/1507.07956}{{\ttfamily 1507.07956}}].

\bibitem{Abi:2021gix}
{\scshape Muon g-2} collaboration, \emph{{Measurement of the Positive Muon
  Anomalous Magnetic Moment to 0.46 ppm}},
  \href{https://doi.org/10.1103/PhysRevLett.126.141801}{\emph{Phys. Rev. Lett.}
  {\bfseries 126} (2021) 141801}
  [\href{https://arxiv.org/abs/2104.03281}{{\ttfamily 2104.03281}}].

\bibitem{Albahri:2021kmg}
{\scshape Muon g-2} collaboration, \emph{{Magnetic Field Measurement and
  Analysis for the Muon g-2 Experiment at Fermilab}},
  \href{https://doi.org/10.1103/PhysRevA.103.042208}{\emph{Phys. Rev. A}
  {\bfseries 103} (2021) 042208}
  [\href{https://arxiv.org/abs/2104.03201}{{\ttfamily 2104.03201}}].

\bibitem{Albahri:2021ixb}
{\scshape Muon g-2} collaboration, \emph{{Measurement of the anomalous
  precession frequency of the muon in the Fermilab Muon g-2 experiment}},
  \href{https://doi.org/10.1103/PhysRevD.103.072002}{\emph{Phys. Rev. D}
  {\bfseries 103} (2021) 072002}
  [\href{https://arxiv.org/abs/2104.03247}{{\ttfamily 2104.03247}}].

\bibitem{Albahri:2021mtf}
{\scshape Muon g-2} collaboration, \emph{{Beam dynamics corrections to the
  Run-1 measurement of the muon anomalous magnetic moment at Fermilab}},
  \href{https://arxiv.org/abs/2104.03240}{{\ttfamily 2104.03240}}.

\bibitem{Aoyama:2020ynm}
T.~Aoyama et~al., \emph{{The anomalous magnetic moment of the muon in the
  Standard Model}},
  \href{https://doi.org/10.1016/j.physrep.2020.07.006}{\emph{Phys. Rept.}
  {\bfseries 887} (2020) 1} [\href{https://arxiv.org/abs/2006.04822}{{\ttfamily
  2006.04822}}].

\bibitem{Foot:1994vd}
R.~Foot, X.~G. He, H.~Lew and R.~R. Volkas, \emph{{Model for a light $Z^\prime$
  boson}}, \href{https://doi.org/10.1103/PhysRevD.50.4571}{\emph{Phys. Rev. D}
  {\bfseries 50} (1994) 4571}
  [\href{https://arxiv.org/abs/hep-ph/9401250}{{\ttfamily hep-ph/9401250}}].

\bibitem{Gninenko:2001hx}
S.~N. Gninenko and N.~V. Krasnikov, \emph{{The Muon anomalous magnetic moment
  and a new light gauge boson}},
  \href{https://doi.org/10.1016/S0370-2693(01)00693-1}{\emph{Phys. Lett. B}
  {\bfseries 513} (2001) 119}
  [\href{https://arxiv.org/abs/hep-ph/0102222}{{\ttfamily hep-ph/0102222}}].

\bibitem{Murakami:2001cs}
B.~Murakami, \emph{{The Impact of lepton flavor violating $Z^\prime$ bosons on
  muon g-2 and other muon observables}},
  \href{https://doi.org/10.1103/PhysRevD.65.055003}{\emph{Phys. Rev. D}
  {\bfseries 65} (2002) 055003}
  [\href{https://arxiv.org/abs/hep-ph/0110095}{{\ttfamily hep-ph/0110095}}].

\bibitem{Baek:2001kca}
S.~Baek, N.~G. Deshpande, X.~G. He and P.~Ko, \emph{{Muon anomalous g-2 and
  gauged L(muon) - L(tau) models}},
  \href{https://doi.org/10.1103/PhysRevD.64.055006}{\emph{Phys. Rev. D}
  {\bfseries 64} (2001) 055006}
  [\href{https://arxiv.org/abs/hep-ph/0104141}{{\ttfamily hep-ph/0104141}}].

\bibitem{Ma:2001md}
E.~Ma, D.~P. Roy and S.~Roy, \emph{{Gauged $L_\mu - L_\tau$ with large muon
  anomalous magnetic moment and the bimaximal mixing of neutrinos}},
  \href{https://doi.org/10.1016/S0370-2693(01)01428-9}{\emph{Phys. Lett. B}
  {\bfseries 525} (2002) 101}
  [\href{https://arxiv.org/abs/hep-ph/0110146}{{\ttfamily hep-ph/0110146}}].

\bibitem{Pospelov:2008zw}
M.~Pospelov, \emph{{Secluded U(1) below the weak scale}},
  \href{https://doi.org/10.1103/PhysRevD.80.095002}{\emph{Phys. Rev. D}
  {\bfseries 80} (2009) 095002}
  [\href{https://arxiv.org/abs/0811.1030}{{\ttfamily 0811.1030}}].

\bibitem{Heeck:2011wj}
J.~Heeck and W.~Rodejohann, \emph{{Gauged $L_\mu - L_\tau$ Symmetry at the
  Electroweak Scale}},
  \href{https://doi.org/10.1103/PhysRevD.84.075007}{\emph{Phys. Rev. D}
  {\bfseries 84} (2011) 075007}
  [\href{https://arxiv.org/abs/1107.5238}{{\ttfamily 1107.5238}}].

\bibitem{Davoudiasl:2012ig}
H.~Davoudiasl, H.-S. Lee and W.~J. Marciano, \emph{{Dark Side of Higgs Diphoton
  Decays and Muon g-2}},
  \href{https://doi.org/10.1103/PhysRevD.86.095009}{\emph{Phys. Rev. D}
  {\bfseries 86} (2012) 095009}
  [\href{https://arxiv.org/abs/1208.2973}{{\ttfamily 1208.2973}}].

\bibitem{Carone:2013uh}
C.~D. Carone, \emph{{Flavor-Nonuniversal Dark Gauge Bosons and the Muon g-2}},
  \href{https://doi.org/10.1016/j.physletb.2013.03.011}{\emph{Phys. Lett. B}
  {\bfseries 721} (2013) 118}
  [\href{https://arxiv.org/abs/1301.2027}{{\ttfamily 1301.2027}}].

\bibitem{Harigaya:2013twa}
K.~Harigaya, T.~Igari, M.~M. Nojiri, M.~Takeuchi and K.~Tobe, \emph{{Muon g-2
  and LHC phenomenology in the $L_\mu-L_\tau$ gauge symmetric model}},
  \href{https://doi.org/10.1007/JHEP03(2014)105}{\emph{JHEP} {\bfseries 03}
  (2014) 105} [\href{https://arxiv.org/abs/1311.0870}{{\ttfamily 1311.0870}}].

\bibitem{Altmannshofer:2014cfa}
W.~Altmannshofer, S.~Gori, M.~Pospelov and I.~Yavin, \emph{{Quark flavor
  transitions in $L_\mu-L_\tau$ models}},
  \href{https://doi.org/10.1103/PhysRevD.89.095033}{\emph{Phys. Rev. D}
  {\bfseries 89} (2014) 095033}
  [\href{https://arxiv.org/abs/1403.1269}{{\ttfamily 1403.1269}}].

\bibitem{Tomar:2014rya}
G.~Tomar and S.~Mohanty, \emph{{Muon anomalous magnetic moment and positron
  excess at AMS-02 in a gauged horizontal symmetric model}},
  \href{https://doi.org/10.1007/JHEP11(2014)133}{\emph{JHEP} {\bfseries 11}
  (2014) 133} [\href{https://arxiv.org/abs/1403.6301}{{\ttfamily 1403.6301}}].

\bibitem{Altmannshofer:2014pba}
W.~Altmannshofer, S.~Gori, M.~Pospelov and I.~Yavin, \emph{{Neutrino Trident
  Production: A Powerful Probe of New Physics with Neutrino Beams}},
  \href{https://doi.org/10.1103/PhysRevLett.113.091801}{\emph{Phys. Rev. Lett.}
  {\bfseries 113} (2014) 091801}
  [\href{https://arxiv.org/abs/1406.2332}{{\ttfamily 1406.2332}}].

\bibitem{Lee:2014tba}
H.-S. Lee, \emph{{Muon g\ensuremath{-}2 anomaly and dark leptonic gauge
  boson}}, \href{https://doi.org/10.1103/PhysRevD.90.091702}{\emph{Phys. Rev.
  D} {\bfseries 90} (2014) 091702}
  [\href{https://arxiv.org/abs/1408.4256}{{\ttfamily 1408.4256}}].

\bibitem{Allanach:2015gkd}
B.~Allanach, F.~S. Queiroz, A.~Strumia and S.~Sun, \emph{{$Z^\prime$ models for
  the LHCb and $g-2$ muon anomalies}},
  \href{https://doi.org/10.1103/PhysRevD.93.055045}{\emph{Phys. Rev. D}
  {\bfseries 93} (2016) 055045}
  [\href{https://arxiv.org/abs/1511.07447}{{\ttfamily 1511.07447}}].

\bibitem{Heeck:2016xkh}
J.~Heeck, \emph{{Lepton flavor violation with light vector bosons}},
  \href{https://doi.org/10.1016/j.physletb.2016.05.007}{\emph{Phys. Lett. B}
  {\bfseries 758} (2016) 101}
  [\href{https://arxiv.org/abs/1602.03810}{{\ttfamily 1602.03810}}].

\bibitem{Patra:2016shz}
S.~Patra, S.~Rao, N.~Sahoo and N.~Sahu, \emph{{Gauged $U(1)_{L_\mu - L_\tau}$
  model in light of muon $g-2$ anomaly, neutrino mass and dark matter
  phenomenology}},
  \href{https://doi.org/10.1016/j.nuclphysb.2017.02.010}{\emph{Nucl. Phys. B}
  {\bfseries 917} (2017) 317}
  [\href{https://arxiv.org/abs/1607.04046}{{\ttfamily 1607.04046}}].

\bibitem{Altmannshofer:2016brv}
W.~Altmannshofer, C.-Y. Chen, P.~S. Bhupal~Dev and A.~Soni, \emph{{Lepton
  flavor violating $Z^\prime$ explanation of the muon anomalous magnetic
  moment}}, \href{https://doi.org/10.1016/j.physletb.2016.09.046}{\emph{Phys.
  Lett. B} {\bfseries 762} (2016) 389}
  [\href{https://arxiv.org/abs/1607.06832}{{\ttfamily 1607.06832}}].

\bibitem{Iguro:2020rby}
S.~Iguro, Y.~Omura and M.~Takeuchi, \emph{{Probing $\mu\tau$ flavor-violating
  solutions for the muon $g-2$ anomaly at Belle II}},
  \href{https://doi.org/10.1007/JHEP09(2020)144}{\emph{JHEP} {\bfseries 09}
  (2020) 144} [\href{https://arxiv.org/abs/2002.12728}{{\ttfamily
  2002.12728}}].

\bibitem{He:1990pn}
X.~G. He, G.~C. Joshi, H.~Lew and R.~R. Volkas, \emph{{New $Z^\prime$
  phenomenology}}, \href{https://doi.org/10.1103/PhysRevD.43.R22}{\emph{Phys.
  Rev. D} {\bfseries 43} (1991) 22}.

\bibitem{Foot:1990mn}
R.~Foot, \emph{{New Physics From Electric Charge Quantization?}},
  \href{https://doi.org/10.1142/S0217732391000543}{\emph{Mod. Phys. Lett. A}
  {\bfseries 6} (1991) 527}.

\bibitem{He:1991qd}
X.-G. He, G.~C. Joshi, H.~Lew and R.~R. Volkas, \emph{{Simplest $Z^\prime$
  model}}, \href{https://doi.org/10.1103/PhysRevD.44.2118}{\emph{Phys. Rev. D}
  {\bfseries 44} (1991) 2118}.

\bibitem{Binetruy:1996cs}
P.~Binetruy, S.~Lavignac, S.~T. Petcov and P.~Ramond, \emph{{Quasidegenerate
  neutrinos from an Abelian family symmetry}},
  \href{https://doi.org/10.1016/S0550-3213(97)00211-3}{\emph{Nucl. Phys. B}
  {\bfseries 496} (1997) 3}
  [\href{https://arxiv.org/abs/hep-ph/9610481}{{\ttfamily hep-ph/9610481}}].

\bibitem{Bell:2000vh}
N.~F. Bell and R.~R. Volkas, \emph{{Bottom up model for maximal
  $\nu_\mu-\nu_\tau$ mixing}},
  \href{https://doi.org/10.1103/PhysRevD.63.013006}{\emph{Phys. Rev. D}
  {\bfseries 63} (2001) 013006}
  [\href{https://arxiv.org/abs/hep-ph/0008177}{{\ttfamily hep-ph/0008177}}].

\bibitem{Choubey:2004hn}
S.~Choubey and W.~Rodejohann, \emph{{A Flavor symmetry for quasi-degenerate
  neutrinos: $L\mu-L_\tau$}},
  \href{https://doi.org/10.1140/epjc/s2005-02133-1}{\emph{Eur. Phys. J. C}
  {\bfseries 40} (2005) 259}
  [\href{https://arxiv.org/abs/hep-ph/0411190}{{\ttfamily hep-ph/0411190}}].

\bibitem{Dutta:1994dx}
G.~Dutta, A.~S. Joshipura and K.~B. Vijaykumar, \emph{{Leptonic flavor
  violations in the presence of an extra Z}},
  \href{https://doi.org/10.1103/PhysRevD.50.2109}{\emph{Phys. Rev. D}
  {\bfseries 50} (1994) 2109}
  [\href{https://arxiv.org/abs/hep-ph/9405292}{{\ttfamily hep-ph/9405292}}].

\bibitem{Crivellin:2015mga}
A.~Crivellin, G.~D'Ambrosio and J.~Heeck, \emph{{Explaining
  $h\to\mu^\pm\tau^\mp$, $B\to K^* \mu^+\mu^-$ and $B\to K \mu^+\mu^-/B\to K
  e^+e^-$ in a two-Higgs-doublet model with gauged $L_\mu-L_\tau$}},
  \href{https://doi.org/10.1103/PhysRevLett.114.151801}{\emph{Phys. Rev. Lett.}
  {\bfseries 114} (2015) 151801}
  [\href{https://arxiv.org/abs/1501.00993}{{\ttfamily 1501.00993}}].

\bibitem{Heeck:2014qea}
J.~Heeck, M.~Holthausen, W.~Rodejohann and Y.~Shimizu, \emph{{Higgs
  \textrightarrow{}\ensuremath{\mu}\ensuremath{\tau} in Abelian and non-Abelian
  flavor symmetry models}},
  \href{https://doi.org/10.1016/j.nuclphysb.2015.04.025}{\emph{Nucl. Phys. B}
  {\bfseries 896} (2015) 281}
  [\href{https://arxiv.org/abs/1412.3671}{{\ttfamily 1412.3671}}].

\bibitem{Altmannshofer:2016oaq}
W.~Altmannshofer, M.~Carena and A.~Crivellin, \emph{{$L_\mu - L_\tau$ theory of
  Higgs flavor violation and $(g-2)_\mu$}},
  \href{https://doi.org/10.1103/PhysRevD.94.095026}{\emph{Phys. Rev. D}
  {\bfseries 94} (2016) 095026}
  [\href{https://arxiv.org/abs/1604.08221}{{\ttfamily 1604.08221}}].

\bibitem{Aubert:2009qj}
{\scshape BaBar} collaboration, \emph{{Measurements of Charged Current Lepton
  Universality and $|V_{us}|$ using Tau Lepton Decays to $e^- \bar{\nu}_e
  \nu_\tau$, $\mu^- \bar{\nu}_\mu \nu_\tau$, $\pi^- \nu_\tau$, and $K^-
  \nu_\tau$}},
  \href{https://doi.org/10.1103/PhysRevLett.105.051602}{\emph{Phys. Rev. Lett.}
  {\bfseries 105} (2010) 051602}
  [\href{https://arxiv.org/abs/0912.0242}{{\ttfamily 0912.0242}}].

\bibitem{Amhis:2019ckw}
{\scshape HFLAV} collaboration, \emph{{Averages of $b$-hadron, $c$-hadron, and
  $\tau$-lepton properties as of 2018}},
  \href{https://arxiv.org/abs/1909.12524}{{\ttfamily 1909.12524}}.

\bibitem{Belfatto:2019swo}
B.~Belfatto, R.~Beradze and Z.~Berezhiani, \emph{{The CKM unitarity problem: A
  trace of new physics at the TeV scale?}},
  \href{https://doi.org/10.1140/epjc/s10052-020-7691-6}{\emph{Eur. Phys. J. C}
  {\bfseries 80} (2020) 149}
  [\href{https://arxiv.org/abs/1906.02714}{{\ttfamily 1906.02714}}].

\bibitem{Grossman:2019bzp}
Y.~Grossman, E.~Passemar and S.~Schacht, \emph{{On the Statistical Treatment of
  the Cabibbo Angle Anomaly}},
  \href{https://doi.org/10.1007/JHEP07(2020)068}{\emph{JHEP} {\bfseries 07}
  (2020) 068} [\href{https://arxiv.org/abs/1911.07821}{{\ttfamily
  1911.07821}}].

\bibitem{Shiells:2020fqp}
K.~Shiells, P.~G. Blunden and W.~Melnitchouk, \emph{{Electroweak axial
  structure functions and improved extraction of the $V_{ud}$ CKM matrix
  element}},  \href{https://arxiv.org/abs/2012.01580}{{\ttfamily 2012.01580}}.

\bibitem{Seng:2020wjq}
C.-Y. Seng, X.~Feng, M.~Gorchtein and L.-C. Jin, \emph{{Joint lattice
  QCD\textendash{}dispersion theory analysis confirms the quark-mixing top-row
  unitarity deficit}},
  \href{https://doi.org/10.1103/PhysRevD.101.111301}{\emph{Phys. Rev. D}
  {\bfseries 101} (2020) 111301}
  [\href{https://arxiv.org/abs/2003.11264}{{\ttfamily 2003.11264}}].

\bibitem{Coutinho:2019aiy}
A.~M. Coutinho, A.~Crivellin and C.~A. Manzari, \emph{{Global Fit to Modified
  Neutrino Couplings and the Cabibbo-Angle Anomaly}},
  \href{https://doi.org/10.1103/PhysRevLett.125.071802}{\emph{Phys. Rev. Lett.}
  {\bfseries 125} (2020) 071802}
  [\href{https://arxiv.org/abs/1912.08823}{{\ttfamily 1912.08823}}].

\bibitem{Crivellin:2020lzu}
A.~Crivellin and M.~Hoferichter, \emph{{\ensuremath{\beta} Decays as Sensitive
  Probes of Lepton Flavor Universality}},
  \href{https://doi.org/10.1103/PhysRevLett.125.111801}{\emph{Phys. Rev. Lett.}
  {\bfseries 125} (2020) 111801}
  [\href{https://arxiv.org/abs/2002.07184}{{\ttfamily 2002.07184}}].

\bibitem{Coutinho:2020xhc}
C.~A. Manzari, A.~M. Coutinho and A.~Crivellin, \emph{{Modified lepton
  couplings and the Cabibbo-angle anomaly}},
  \href{https://doi.org/10.22323/1.382.0242}{\emph{PoS} {\bfseries LHCP2020}
  (2021) 242} [\href{https://arxiv.org/abs/2009.03877}{{\ttfamily
  2009.03877}}].

\bibitem{Capdevila:2020rrl}
B.~Capdevila, A.~Crivellin, C.~A. Manzari and M.~Montull, \emph{{Explaining
  $b\to s\ell^+\ell^-$ and the Cabibbo angle anomaly with a vector triplet}},
  \href{https://doi.org/10.1103/PhysRevD.103.015032}{\emph{Phys. Rev. D}
  {\bfseries 103} (2021) 015032}
  [\href{https://arxiv.org/abs/2005.13542}{{\ttfamily 2005.13542}}].

\bibitem{Crivellin:2020ebi}
A.~Crivellin, F.~Kirk, C.~A. Manzari and M.~Montull, \emph{{Global Electroweak
  Fit and Vector-Like Leptons in Light of the Cabibbo Angle Anomaly}},
  \href{https://arxiv.org/abs/2008.01113}{{\ttfamily 2008.01113}}.

\bibitem{Kirk:2020wdk}
M.~Kirk, \emph{{Cabibbo anomaly versus electroweak precision tests: An
  exploration of extensions of the standard model}},
  \href{https://doi.org/10.1103/PhysRevD.103.035004}{\emph{Phys. Rev. D}
  {\bfseries 103} (2021) 035004}
  [\href{https://arxiv.org/abs/2008.03261}{{\ttfamily 2008.03261}}].

\bibitem{Alok:2020jod}
A.~K. Alok, A.~Dighe, S.~Gangal and J.~Kumar, \emph{{The role of non-universal
  $Z$ couplings in explaining the $V_{us}$ anomaly}},
  \href{https://arxiv.org/abs/2010.12009}{{\ttfamily 2010.12009}}.

\bibitem{Crivellin:2020oup}
A.~Crivellin, C.~A. Manzari, M.~Alguero and J.~Matias, \emph{{Combined
  Explanation of the $Z\to b\bar b$ Forward-Backward Asymmetry, the Cabibbo
  Angle Anomaly, $\tau\to\mu\nu\nu$ and $b\to s\ell^+\ell^-$ Data}},
  \href{https://arxiv.org/abs/2010.14504}{{\ttfamily 2010.14504}}.

\bibitem{Crivellin:2020klg}
A.~Crivellin, F.~Kirk, C.~A. Manzari and L.~Panizzi, \emph{{Searching for
  Lepton Flavour (Universality) Violation and Collider Signals from a
  Singly-Charged Scalar Singlet}},
  \href{https://arxiv.org/abs/2012.09845}{{\ttfamily 2012.09845}}.

\bibitem{Crivellin:2021egp}
A.~Crivellin, D.~M\"uller and L.~Schnell, \emph{{Combined Constraints on First
  Generation Leptoquarks}},  \href{https://arxiv.org/abs/2101.07811}{{\ttfamily
  2101.07811}}.

\bibitem{Crivellin:2021njn}
A.~Crivellin, M.~Hoferichter and C.~A. Manzari, \emph{{The Fermi constant from
  muon decay versus electroweak fits and CKM unitarity}},
  \href{https://arxiv.org/abs/2102.02825}{{\ttfamily 2102.02825}}.

\bibitem{Crivellin:2021rbf}
A.~Crivellin, C.~A. Manzari and M.~Montull, \emph{{Correlating Non-Resonant
  Di-Electron Searches at the LHC to the Cabibbo-Angle Anomaly and Lepton
  Flavour Universality Violation}},
  \href{https://arxiv.org/abs/2103.12003}{{\ttfamily 2103.12003}}.

\bibitem{Buras:2013qja}
A.~J. Buras and J.~Girrbach, \emph{{Left-handed $Z'$ and $Z$ FCNC quark
  couplings facing new $b \to s \mu^+ \mu^-$ data}},
  \href{https://doi.org/10.1007/JHEP12(2013)009}{\emph{JHEP} {\bfseries 12}
  (2013) 009} [\href{https://arxiv.org/abs/1309.2466}{{\ttfamily 1309.2466}}].

\bibitem{Gauld:2013qba}
R.~Gauld, F.~Goertz and U.~Haisch, \emph{{On minimal $Z'$ explanations of the
  $B\to K^*\mu^+\mu^-$ anomaly}},
  \href{https://doi.org/10.1103/PhysRevD.89.015005}{\emph{Phys. Rev. D}
  {\bfseries 89} (2014) 015005}
  [\href{https://arxiv.org/abs/1308.1959}{{\ttfamily 1308.1959}}].

\bibitem{Gauld:2013qja}
R.~Gauld, F.~Goertz and U.~Haisch, \emph{{An explicit $Z^\prime$-boson
  explanation of the $B \to K^* \mu^+ \mu^-$ anomaly}},
  \href{https://doi.org/10.1007/JHEP01(2014)069}{\emph{JHEP} {\bfseries 01}
  (2014) 069} [\href{https://arxiv.org/abs/1310.1082}{{\ttfamily 1310.1082}}].

\bibitem{Crivellin:2015lwa}
A.~Crivellin, G.~D'Ambrosio and J.~Heeck, \emph{{Addressing the LHC flavor
  anomalies with horizontal gauge symmetries}},
  \href{https://doi.org/10.1103/PhysRevD.91.075006}{\emph{Phys. Rev. D}
  {\bfseries 91} (2015) 075006}
  [\href{https://arxiv.org/abs/1503.03477}{{\ttfamily 1503.03477}}].

\bibitem{Niehoff:2015bfa}
C.~Niehoff, P.~Stangl and D.~M. Straub, \emph{{Violation of lepton flavour
  universality in composite Higgs models}},
  \href{https://doi.org/10.1016/j.physletb.2015.05.063}{\emph{Phys. Lett. B}
  {\bfseries 747} (2015) 182}
  [\href{https://arxiv.org/abs/1503.03865}{{\ttfamily 1503.03865}}].

\bibitem{Carmona:2015ena}
A.~Carmona and F.~Goertz, \emph{{Lepton Flavor and Nonuniversality from Minimal
  Composite Higgs Setups}},
  \href{https://doi.org/10.1103/PhysRevLett.116.251801}{\emph{Phys. Rev. Lett.}
  {\bfseries 116} (2016) 251801}
  [\href{https://arxiv.org/abs/1510.07658}{{\ttfamily 1510.07658}}].

\bibitem{Falkowski:2015zwa}
A.~Falkowski, M.~Nardecchia and R.~Ziegler, \emph{{Lepton Flavor
  Non-Universality in B-meson Decays from a U(2) Flavor Model}},
  \href{https://doi.org/10.1007/JHEP11(2015)173}{\emph{JHEP} {\bfseries 11}
  (2015) 173} [\href{https://arxiv.org/abs/1509.01249}{{\ttfamily
  1509.01249}}].

\bibitem{Celis:2015eqs}
A.~Celis, W.-Z. Feng and D.~L\"ust, \emph{{Stringy explanation of b
  \textrightarrow{} s\ensuremath{\ell}$^{+}$ \ensuremath{\ell}$^{-}$
  anomalies}}, \href{https://doi.org/10.1007/JHEP02(2016)007}{\emph{JHEP}
  {\bfseries 02} (2016) 007}
  [\href{https://arxiv.org/abs/1512.02218}{{\ttfamily 1512.02218}}].

\bibitem{Celis:2015ara}
A.~Celis, J.~Fuentes-Martin, M.~Jung and H.~Serodio, \emph{{Family nonuniversal
  $Z^\prime$ models with protected flavor-changing interactions}},
  \href{https://doi.org/10.1103/PhysRevD.92.015007}{\emph{Phys. Rev. D}
  {\bfseries 92} (2015) 015007}
  [\href{https://arxiv.org/abs/1505.03079}{{\ttfamily 1505.03079}}].

\bibitem{Crivellin:2016ejn}
A.~Crivellin, J.~Fuentes-Martin, A.~Greljo and G.~Isidori, \emph{{Lepton Flavor
  Non-Universality in B decays from Dynamical Yukawas}},
  \href{https://doi.org/10.1016/j.physletb.2016.12.057}{\emph{Phys. Lett. B}
  {\bfseries 766} (2017) 77}
  [\href{https://arxiv.org/abs/1611.02703}{{\ttfamily 1611.02703}}].

\bibitem{GarciaGarcia:2016nvr}
I.~Garcia~Garcia, \emph{{LHCb anomalies from a natural perspective}},
  \href{https://doi.org/10.1007/JHEP03(2017)040}{\emph{JHEP} {\bfseries 03}
  (2017) 040} [\href{https://arxiv.org/abs/1611.03507}{{\ttfamily
  1611.03507}}].

\bibitem{Faisel:2017glo}
G.~Faisel and J.~Tandean, \emph{{Connecting $ b\to s\ell \overline{\ell} $
  anomalies to enhanced rare nonleptonic $ {\overline{B}}_s^0 $ decays in
  $Z^\prime$ model}},
  \href{https://doi.org/10.1007/JHEP02(2018)074}{\emph{JHEP} {\bfseries 02}
  (2018) 074} [\href{https://arxiv.org/abs/1710.11102}{{\ttfamily
  1710.11102}}].

\bibitem{King:2017anf}
S.~F. King, \emph{{Flavourful $Z^\prime$ models for $ {R}_{K^{\left(\ast
  \right)}} $}}, \href{https://doi.org/10.1007/JHEP08(2017)019}{\emph{JHEP}
  {\bfseries 08} (2017) 019}
  [\href{https://arxiv.org/abs/1706.06100}{{\ttfamily 1706.06100}}].

\bibitem{Chiang:2017hlj}
C.-W. Chiang, X.-G. He, J.~Tandean and X.-B. Yuan, \emph{{$R_{K^{(*)}}$ and
  related $b\to s\ell\bar\ell$ anomalies in minimal flavor violation framework
  with $Z'$ boson}},
  \href{https://doi.org/10.1103/PhysRevD.96.115022}{\emph{Phys. Rev. D}
  {\bfseries 96} (2017) 115022}
  [\href{https://arxiv.org/abs/1706.02696}{{\ttfamily 1706.02696}}].

\bibitem{DiChiara:2017cjq}
S.~Di~Chiara, A.~Fowlie, S.~Fraser, C.~Marzo, L.~Marzola, M.~Raidal et~al.,
  \emph{{Minimal flavor-changing $Z'$ models and muon $g-2$ after the $R_{K^*}$
  measurement}},
  \href{https://doi.org/10.1016/j.nuclphysb.2017.08.003}{\emph{Nucl. Phys. B}
  {\bfseries 923} (2017) 245}
  [\href{https://arxiv.org/abs/1704.06200}{{\ttfamily 1704.06200}}].

\bibitem{Ko:2017lzd}
P.~Ko, Y.~Omura, Y.~Shigekami and C.~Yu, \emph{{LHCb anomaly and B physics in
  flavored $Z^\prime$ models with flavored Higgs doublets}},
  \href{https://doi.org/10.1103/PhysRevD.95.115040}{\emph{Phys. Rev. D}
  {\bfseries 95} (2017) 115040}
  [\href{https://arxiv.org/abs/1702.08666}{{\ttfamily 1702.08666}}].

\bibitem{Sannino:2017utc}
F.~Sannino, P.~Stangl, D.~M. Straub and A.~E. Thomsen, \emph{{Flavor Physics
  and Flavor Anomalies in Minimal Fundamental Partial Compositeness}},
  \href{https://doi.org/10.1103/PhysRevD.97.115046}{\emph{Phys. Rev. D}
  {\bfseries 97} (2018) 115046}
  [\href{https://arxiv.org/abs/1712.07646}{{\ttfamily 1712.07646}}].

\bibitem{Falkowski:2018dsl}
A.~Falkowski, S.~F. King, E.~Perdomo and M.~Pierre, \emph{{Flavourful $Z'$
  portal for vector-like neutrino Dark Matter and $R_{K^{(*)}}$}},
  \href{https://doi.org/10.1007/JHEP08(2018)061}{\emph{JHEP} {\bfseries 08}
  (2018) 061} [\href{https://arxiv.org/abs/1803.04430}{{\ttfamily
  1803.04430}}].

\bibitem{Benavides:2018rgh}
R.~H. Benavides, L.~Mu\~noz, W.~A. Ponce, O.~Rodr\'\i{}guez and E.~Rojas,
  \emph{{Minimal $Z^\prime$ models for flavor anomalies}},
  \href{https://doi.org/10.1088/1361-6471/ab8d8d}{\emph{J. Phys. G} {\bfseries
  47} (2020) 075003} [\href{https://arxiv.org/abs/1812.05077}{{\ttfamily
  1812.05077}}].

\bibitem{Maji:2018gvz}
P.~Maji, P.~Nayek and S.~Sahoo, \emph{{Implication of family non-universal
  $Z^\prime$ model to rare exclusive $b \to s(l\bar l,\nu\bar\nu)$
  transitions}}, \href{https://doi.org/10.1093/ptep/ptz010}{\emph{PTEP}
  {\bfseries 2019} (2019) 033B06}
  [\href{https://arxiv.org/abs/1811.03869}{{\ttfamily 1811.03869}}].

\bibitem{Singirala:2018mio}
S.~Singirala, S.~Sahoo and R.~Mohanta, \emph{{Exploring dark matter, neutrino
  mass and $R_{K^{(*)},\phi}$ anomalies in $L_{\mu}-L_{\tau}$ model}},
  \href{https://doi.org/10.1103/PhysRevD.99.035042}{\emph{Phys. Rev. D}
  {\bfseries 99} (2019) 035042}
  [\href{https://arxiv.org/abs/1809.03213}{{\ttfamily 1809.03213}}].

\bibitem{Guadagnoli:2018ojc}
D.~Guadagnoli, M.~Reboud and O.~Sumensari, \emph{{A gauged horizontal $SU(2)$
  symmetry and $R_{K^{(\ast)}}$}},
  \href{https://doi.org/10.1007/JHEP11(2018)163}{\emph{JHEP} {\bfseries 11}
  (2018) 163} [\href{https://arxiv.org/abs/1807.03285}{{\ttfamily
  1807.03285}}].

\bibitem{Allanach:2018lvl}
B.~C. Allanach and J.~Davighi, \emph{{Third family hypercharge model for $
  {R}_{K^{\left(\ast \right)}} $ and aspects of the fermion mass problem}},
  \href{https://doi.org/10.1007/JHEP12(2018)075}{\emph{JHEP} {\bfseries 12}
  (2018) 075} [\href{https://arxiv.org/abs/1809.01158}{{\ttfamily
  1809.01158}}].

\bibitem{Duan:2018akc}
G.~H. Duan, X.~Fan, M.~Frank, C.~Han and J.~M. Yang, \emph{{A minimal
  $U(1)^\prime$ extension of MSSM in light of the B decay anomaly}},
  \href{https://doi.org/10.1016/j.physletb.2018.12.005}{\emph{Phys. Lett. B}
  {\bfseries 789} (2019) 54}
  [\href{https://arxiv.org/abs/1808.04116}{{\ttfamily 1808.04116}}].

\bibitem{King:2018fcg}
S.~F. King, \emph{{$ {R}_{K^{\left(*\right)}} $ and the origin of Yukawa
  couplings}}, \href{https://doi.org/10.1007/JHEP09(2018)069}{\emph{JHEP}
  {\bfseries 09} (2018) 069}
  [\href{https://arxiv.org/abs/1806.06780}{{\ttfamily 1806.06780}}].

\bibitem{Kohda:2018xbc}
M.~Kohda, T.~Modak and A.~Soffer, \emph{{Identifying a $Z'$ behind $b \to s
  \ell \ell$ anomalies at the LHC}},
  \href{https://doi.org/10.1103/PhysRevD.97.115019}{\emph{Phys. Rev. D}
  {\bfseries 97} (2018) 115019}
  [\href{https://arxiv.org/abs/1803.07492}{{\ttfamily 1803.07492}}].

\bibitem{Dwivedi:2019uqd}
S.~Dwivedi, D.~Kumar~Ghosh, A.~Falkowski and N.~Ghosh, \emph{{Associated
  $Z^\prime$ production in the flavorful $U(1)$ scenario for $R_{K^{(*)}}$}},
  \href{https://doi.org/10.1140/epjc/s10052-020-7810-4}{\emph{Eur. Phys. J. C}
  {\bfseries 80} (2020) 263}
  [\href{https://arxiv.org/abs/1908.03031}{{\ttfamily 1908.03031}}].

\bibitem{Foldenauer:2019vgn}
P.~Foldenauer, \emph{{Phenomenology of Extra Abelian Gauge Symmetries}}, Ph.D.
  thesis, U. Heidelberg (main), 7, 2019.
\newblock 10.11588/heidok.00026777.

\bibitem{Ko:2019tts}
P.~Ko, T.~Nomura and C.~Yu, \emph{{$b\rightarrow s \mu^+ \mu^-$ anomalies and
  related phenomenology in $U(1)_{B_3 - x_\mu L_\mu - x_\tau L_\tau}$ flavor
  gauge models}}, \href{https://doi.org/10.1007/JHEP04(2019)102}{\emph{JHEP}
  {\bfseries 04} (2019) 102}
  [\href{https://arxiv.org/abs/1902.06107}{{\ttfamily 1902.06107}}].

\bibitem{Allanach:2019iiy}
B.~C. Allanach and J.~Davighi, \emph{{Naturalising the third family hypercharge
  model for neutral current $B$-anomalies}},
  \href{https://doi.org/10.1140/epjc/s10052-019-7414-z}{\emph{Eur. Phys. J. C}
  {\bfseries 79} (2019) 908}
  [\href{https://arxiv.org/abs/1905.10327}{{\ttfamily 1905.10327}}].

\bibitem{Altmannshofer:2019xda}
W.~Altmannshofer, J.~Davighi and M.~Nardecchia, \emph{{Gauging the accidental
  symmetries of the standard model, and implications for the flavor
  anomalies}}, \href{https://doi.org/10.1103/PhysRevD.101.015004}{\emph{Phys.
  Rev. D} {\bfseries 101} (2020) 015004}
  [\href{https://arxiv.org/abs/1909.02021}{{\ttfamily 1909.02021}}].

\bibitem{Calibbi:2019lvs}
L.~Calibbi, A.~Crivellin, F.~Kirk, C.~A. Manzari and L.~Vernazza,
  \emph{{$Z^\prime$ models with less-minimal flavour violation}},
  \href{https://doi.org/10.1103/PhysRevD.101.095003}{\emph{Phys. Rev. D}
  {\bfseries 101} (2020) 095003}
  [\href{https://arxiv.org/abs/1910.00014}{{\ttfamily 1910.00014}}].

\bibitem{Aebischer:2019blw}
J.~Aebischer, A.~J. Buras, M.~Cerd\`a-Sevilla and F.~De~Fazio,
  \emph{{Quark-lepton connections in $Z^\prime$ mediated FCNC processes: gauge
  anomaly cancellations at work}},
  \href{https://doi.org/10.1007/JHEP02(2020)183}{\emph{JHEP} {\bfseries 02}
  (2020) 183} [\href{https://arxiv.org/abs/1912.09308}{{\ttfamily
  1912.09308}}].

\bibitem{Allanach:2020kss}
B.~C. Allanach, \emph{{$U(1)_{B_3-L_2}$ explanation of the neutral current
  $B$\ensuremath{-}anomalies}},
  \href{https://doi.org/10.1140/epjc/s10052-021-08855-w}{\emph{Eur. Phys. J. C}
  {\bfseries 81} (2021) 56} [\href{https://arxiv.org/abs/2009.02197}{{\ttfamily
  2009.02197}}].

\bibitem{Greljo:2021xmg}
A.~Greljo, P.~Stangl and A.~E. Thomsen, \emph{{A Model of Muon Anomalies}},
  \href{https://arxiv.org/abs/2103.13991}{{\ttfamily 2103.13991}}.

\bibitem{Aaij:2017vbb}
{\scshape LHCb} collaboration, \emph{{Test of lepton universality with $B^{0}
  \rightarrow K^{*0}\ell^{+}\ell^{-}$ decays}},
  \href{https://doi.org/10.1007/JHEP08(2017)055}{\emph{JHEP} {\bfseries 08}
  (2017) 055} [\href{https://arxiv.org/abs/1705.05802}{{\ttfamily
  1705.05802}}].

\bibitem{Aaij:2019wad}
{\scshape LHCb} collaboration, \emph{{Search for lepton-universality violation
  in $B^+\to K^+\ell^+\ell^-$ decays}},
  \href{https://doi.org/10.1103/PhysRevLett.122.191801}{\emph{Phys. Rev. Lett.}
  {\bfseries 122} (2019) 191801}
  [\href{https://arxiv.org/abs/1903.09252}{{\ttfamily 1903.09252}}].

\bibitem{Capdevila:2017bsm}
B.~Capdevila, A.~Crivellin, S.~Descotes-Genon, J.~Matias and J.~Virto,
  \emph{{Patterns of New Physics in $b\to s\ell^+\ell^-$ transitions in the
  light of recent data}},
  \href{https://doi.org/10.1007/JHEP01(2018)093}{\emph{JHEP} {\bfseries 01}
  (2018) 093} [\href{https://arxiv.org/abs/1704.05340}{{\ttfamily
  1704.05340}}].

\bibitem{Altmannshofer:2017yso}
W.~Altmannshofer, P.~Stangl and D.~M. Straub, \emph{{Interpreting Hints for
  Lepton Flavor Universality Violation}},
  \href{https://doi.org/10.1103/PhysRevD.96.055008}{\emph{Phys. Rev. D}
  {\bfseries 96} (2017) 055008}
  [\href{https://arxiv.org/abs/1704.05435}{{\ttfamily 1704.05435}}].

\bibitem{DAmico:2017mtc}
G.~D'Amico, M.~Nardecchia, P.~Panci, F.~Sannino, A.~Strumia, R.~Torre et~al.,
  \emph{{Flavour anomalies after the $R_{K^*}$ measurement}},
  \href{https://doi.org/10.1007/JHEP09(2017)010}{\emph{JHEP} {\bfseries 09}
  (2017) 010} [\href{https://arxiv.org/abs/1704.05438}{{\ttfamily
  1704.05438}}].

\bibitem{Ciuchini:2017mik}
M.~Ciuchini, A.~M. Coutinho, M.~Fedele, E.~Franco, A.~Paul, L.~Silvestrini
  et~al., \emph{{On Flavourful Easter eggs for New Physics hunger and Lepton
  Flavour Universality violation}},
  \href{https://doi.org/10.1140/epjc/s10052-017-5270-2}{\emph{Eur. Phys. J. C}
  {\bfseries 77} (2017) 688}
  [\href{https://arxiv.org/abs/1704.05447}{{\ttfamily 1704.05447}}].

\bibitem{Hiller:2017bzc}
G.~Hiller and I.~Nisandzic, \emph{{$R_K$ and $R_{K^{\ast}}$ beyond the standard
  model}}, \href{https://doi.org/10.1103/PhysRevD.96.035003}{\emph{Phys. Rev.
  D} {\bfseries 96} (2017) 035003}
  [\href{https://arxiv.org/abs/1704.05444}{{\ttfamily 1704.05444}}].

\bibitem{Geng:2017svp}
L.-S. Geng, B.~Grinstein, S.~J\"ager, J.~Martin~Camalich, X.-L. Ren and R.-X.
  Shi, \emph{{Towards the discovery of new physics with lepton-universality
  ratios of $b\to s\ell\ell$ decays}},
  \href{https://doi.org/10.1103/PhysRevD.96.093006}{\emph{Phys. Rev. D}
  {\bfseries 96} (2017) 093006}
  [\href{https://arxiv.org/abs/1704.05446}{{\ttfamily 1704.05446}}].

\bibitem{Hurth:2017hxg}
T.~Hurth, F.~Mahmoudi, D.~Martinez~Santos and S.~Neshatpour, \emph{{Lepton
  nonuniversality in exclusive $b{\rightarrow}s{\ell}{\ell}$ decays}},
  \href{https://doi.org/10.1103/PhysRevD.96.095034}{\emph{Phys. Rev. D}
  {\bfseries 96} (2017) 095034}
  [\href{https://arxiv.org/abs/1705.06274}{{\ttfamily 1705.06274}}].

\bibitem{Alguero:2019ptt}
M.~Alguer\'o, B.~Capdevila, A.~Crivellin, S.~Descotes-Genon, P.~Masjuan,
  J.~Matias et~al., \emph{{Emerging patterns of New Physics with and without
  Lepton Flavour Universal contributions}},
  \href{https://doi.org/10.1140/epjc/s10052-019-7216-3}{\emph{Eur. Phys. J. C}
  {\bfseries 79} (2019) 714}
  [\href{https://arxiv.org/abs/1903.09578}{{\ttfamily 1903.09578}}].

\bibitem{Aebischer:2019mlg}
J.~Aebischer, W.~Altmannshofer, D.~Guadagnoli, M.~Reboud, P.~Stangl and D.~M.
  Straub, \emph{{$B$-decay discrepancies after Moriond 2019}},
  \href{https://doi.org/10.1140/epjc/s10052-020-7817-x}{\emph{Eur. Phys. J. C}
  {\bfseries 80} (2020) 252}
  [\href{https://arxiv.org/abs/1903.10434}{{\ttfamily 1903.10434}}].

\bibitem{Ciuchini:2019usw}
M.~Ciuchini, A.~M. Coutinho, M.~Fedele, E.~Franco, A.~Paul, L.~Silvestrini
  et~al., \emph{{New Physics in $b \to s \ell^+ \ell^-$ confronts new data on
  Lepton Universality}},
  \href{https://doi.org/10.1140/epjc/s10052-019-7210-9}{\emph{Eur. Phys. J. C}
  {\bfseries 79} (2019) 719}
  [\href{https://arxiv.org/abs/1903.09632}{{\ttfamily 1903.09632}}].

\bibitem{Arbey:2019duh}
A.~Arbey, T.~Hurth, F.~Mahmoudi, D.~M. Santos and S.~Neshatpour, \emph{{Update
  on the b\textrightarrow{}s anomalies}},
  \href{https://doi.org/10.1103/PhysRevD.100.015045}{\emph{Phys. Rev. D}
  {\bfseries 100} (2019) 015045}
  [\href{https://arxiv.org/abs/1904.08399}{{\ttfamily 1904.08399}}].

\bibitem{Matias:2012xw}
J.~Matias, F.~Mescia, M.~Ramon and J.~Virto, \emph{{Complete Anatomy of
  $\bar{B}_d \to \bar{K}^{* 0} (\to K \pi)l^+l^-$ and its angular
  distribution}}, \href{https://doi.org/10.1007/JHEP04(2012)104}{\emph{JHEP}
  {\bfseries 04} (2012) 104} [\href{https://arxiv.org/abs/1202.4266}{{\ttfamily
  1202.4266}}].

\bibitem{Descotes-Genon:2013vna}
S.~Descotes-Genon, T.~Hurth, J.~Matias and J.~Virto, \emph{{Optimizing the
  basis of $B\to K^*ll$ observables in the full kinematic range}},
  \href{https://doi.org/10.1007/JHEP05(2013)137}{\emph{JHEP} {\bfseries 05}
  (2013) 137} [\href{https://arxiv.org/abs/1303.5794}{{\ttfamily 1303.5794}}].

\bibitem{Aaij:2015oid}
{\scshape LHCb} collaboration, \emph{{Angular analysis of the $B^{0} \to K^{*0}
  \mu^{+} \mu^{-}$ decay using 3 fb$^{-1}$ of integrated luminosity}},
  \href{https://doi.org/10.1007/JHEP02(2016)104}{\emph{JHEP} {\bfseries 02}
  (2016) 104} [\href{https://arxiv.org/abs/1512.04442}{{\ttfamily
  1512.04442}}].

\bibitem{Aaij:2020nrf}
{\scshape LHCb} collaboration, \emph{{Measurement of $CP$-Averaged Observables
  in the $B^{0}\rightarrow K^{*0}\mu^{+}\mu^{-}$ Decay}},
  \href{https://doi.org/10.1103/PhysRevLett.125.011802}{\emph{Phys. Rev. Lett.}
  {\bfseries 125} (2020) 011802}
  [\href{https://arxiv.org/abs/2003.04831}{{\ttfamily 2003.04831}}].

\bibitem{Ciuchini:2020gvn}
M.~Ciuchini, M.~Fedele, E.~Franco, A.~Paul, L.~Silvestrini and M.~Valli,
  \emph{{Lessons from the $B^{0,+}\to K^{*0,+}\mu^+\mu^-$ angular analyses}},
  \href{https://doi.org/10.1103/PhysRevD.103.015030}{\emph{Phys. Rev. D}
  {\bfseries 103} (2021) 015030}
  [\href{https://arxiv.org/abs/2011.01212}{{\ttfamily 2011.01212}}].

\bibitem{Altmannshofer:2021qrr}
W.~Altmannshofer and P.~Stangl, \emph{{New Physics in Rare B Decays after
  Moriond 2021}},  \href{https://arxiv.org/abs/2103.13370}{{\ttfamily
  2103.13370}}.

\bibitem{Bobeth:2016llm}
C.~Bobeth, A.~J. Buras, A.~Celis and M.~Jung, \emph{{Patterns of Flavour
  Violation in Models with Vector-Like Quarks}},
  \href{https://doi.org/10.1007/JHEP04(2017)079}{\emph{JHEP} {\bfseries 04}
  (2017) 079} [\href{https://arxiv.org/abs/1609.04783}{{\ttfamily
  1609.04783}}].

\bibitem{deBlas:2019okz}
J.~De~Blas et~al., \emph{{$\texttt{HEPfit}$: a code for the combination of
  indirect and direct constraints on high energy physics models}},
  \href{https://doi.org/10.1140/epjc/s10052-020-7904-z}{\emph{Eur. Phys. J. C}
  {\bfseries 80} (2020) 456}
  [\href{https://arxiv.org/abs/1910.14012}{{\ttfamily 1910.14012}}].

\bibitem{delAguila:2010mx}
F.~del Aguila, J.~de~Blas and M.~Perez-Victoria, \emph{{Electroweak Limits on
  General New Vector Bosons}},
  \href{https://doi.org/10.1007/JHEP09(2010)033}{\emph{JHEP} {\bfseries 09}
  (2010) 033} [\href{https://arxiv.org/abs/1005.3998}{{\ttfamily 1005.3998}}].

\bibitem{deBlas:2012qp}
J.~de~Blas, J.~M. Lizana and M.~Perez-Victoria, \emph{{Combining searches of
  $Z^\prime$ and $W^\prime$ bosons}},
  \href{https://doi.org/10.1007/JHEP01(2013)166}{\emph{JHEP} {\bfseries 01}
  (2013) 166} [\href{https://arxiv.org/abs/1211.2229}{{\ttfamily 1211.2229}}].

\bibitem{Alonso:2018bcg}
R.~Alonso, A.~Carmona, B.~M. Dillon, J.~F. Kamenik, J.~Martin~Camalich and
  J.~Zupan, \emph{{A clockwork solution to the flavor puzzle}},
  \href{https://doi.org/10.1007/JHEP10(2018)099}{\emph{JHEP} {\bfseries 10}
  (2018) 099} [\href{https://arxiv.org/abs/1807.09792}{{\ttfamily
  1807.09792}}].

\bibitem{Smolkovic:2019jow}
A.~Smolkovi\v{c}, M.~Tammaro and J.~Zupan, \emph{{Anomaly free Froggatt-Nielsen
  models of flavor}},
  \href{https://doi.org/10.1007/JHEP10(2019)188}{\emph{JHEP} {\bfseries 10}
  (2019) 188} [\href{https://arxiv.org/abs/1907.10063}{{\ttfamily
  1907.10063}}].

\bibitem{Ellis:2017nrp}
J.~Ellis, M.~Fairbairn and P.~Tunney, \emph{{Anomaly-Free Models for Flavour
  Anomalies}}, \href{https://doi.org/10.1140/epjc/s10052-018-5725-0}{\emph{Eur.
  Phys. J. C} {\bfseries 78} (2018) 238}
  [\href{https://arxiv.org/abs/1705.03447}{{\ttfamily 1705.03447}}].

\bibitem{Marciano:2005ec}
W.~J. Marciano and A.~Sirlin, \emph{{Improved calculation of electroweak
  radiative corrections and the value of V(ud)}},
  \href{https://doi.org/10.1103/PhysRevLett.96.032002}{\emph{Phys. Rev. Lett.}
  {\bfseries 96} (2006) 032002}
  [\href{https://arxiv.org/abs/hep-ph/0510099}{{\ttfamily hep-ph/0510099}}].

\bibitem{Seng:2018yzq}
C.-Y. Seng, M.~Gorchtein, H.~H. Patel and M.~J. Ramsey-Musolf, \emph{{Reduced
  Hadronic Uncertainty in the Determination of $V_{ud}$}},
  \href{https://doi.org/10.1103/PhysRevLett.121.241804}{\emph{Phys. Rev. Lett.}
  {\bfseries 121} (2018) 241804}
  [\href{https://arxiv.org/abs/1807.10197}{{\ttfamily 1807.10197}}].

\bibitem{Seng:2018qru}
C.~Y. Seng, M.~Gorchtein and M.~J. Ramsey-Musolf, \emph{{Dispersive evaluation
  of the inner radiative correction in neutron and nuclear $\beta$ decay}},
  \href{https://doi.org/10.1103/PhysRevD.100.013001}{\emph{Phys. Rev. D}
  {\bfseries 100} (2019) 013001}
  [\href{https://arxiv.org/abs/1812.03352}{{\ttfamily 1812.03352}}].

\bibitem{Gorchtein:2018fxl}
M.~Gorchtein, \emph{{\ensuremath{\gamma}W Box Inside Out: Nuclear
  Polarizabilities Distort the Beta Decay Spectrum}},
  \href{https://doi.org/10.1103/PhysRevLett.123.042503}{\emph{Phys. Rev. Lett.}
  {\bfseries 123} (2019) 042503}
  [\href{https://arxiv.org/abs/1812.04229}{{\ttfamily 1812.04229}}].

\bibitem{Czarnecki:2019mwq}
A.~Czarnecki, W.~J. Marciano and A.~Sirlin, \emph{{Radiative Corrections to
  Neutron and Nuclear Beta Decays Revisited}},
  \href{https://doi.org/10.1103/PhysRevD.100.073008}{\emph{Phys. Rev. D}
  {\bfseries 100} (2019) 073008}
  [\href{https://arxiv.org/abs/1907.06737}{{\ttfamily 1907.06737}}].

\bibitem{Hayen:2020cxh}
L.~Hayen, \emph{{Standard Model $\mathcal{O}(\alpha)$ renormalization of $g_A$
  and its impact on new physics searches}},
  \href{https://arxiv.org/abs/2010.07262}{{\ttfamily 2010.07262}}.

\bibitem{Hardy:2020qwl}
J.~C. Hardy and I.~S. Towner, \emph{{Superallowed $0^+ \to 0^+$ nuclear $\beta$
  decays: 2020 critical survey, with implications for V$_{ud}$ and CKM
  unitarity}}, \href{https://doi.org/10.1103/PhysRevC.102.045501}{\emph{Phys.
  Rev. C} {\bfseries 102} (2020) 045501}.

\bibitem{Moulson:2017ive}
M.~Moulson, \emph{{Experimental determination of $V_{us}$ from kaon decays}},
  \href{https://doi.org/10.22323/1.291.0033}{\emph{PoS} {\bfseries CKM2016}
  (2017) 033} [\href{https://arxiv.org/abs/1704.04104}{{\ttfamily
  1704.04104}}].

\bibitem{Zyla:2020zbs}
{\scshape Particle Data Group} collaboration, \emph{{Review of Particle
  Physics}}, \href{https://doi.org/10.1093/ptep/ptaa104}{\emph{PTEP} {\bfseries
  2020} (2020) 083C01}.

\bibitem{Buras:2011zb}
A.~J. Buras, L.~Merlo and E.~Stamou, \emph{{The Impact of Flavour Changing
  Neutral Gauge Bosons on $\bar{B} \to X_s \gamma$}},
  \href{https://doi.org/10.1007/JHEP08(2011)124}{\emph{JHEP} {\bfseries 08}
  (2011) 124} [\href{https://arxiv.org/abs/1105.5146}{{\ttfamily 1105.5146}}].

\bibitem{Crivellin:2018qmi}
A.~Crivellin, M.~Hoferichter and P.~Schmidt-Wellenburg, \emph{{Combined
  explanations of $(g-2)_{\mu,e}$ and implications for a large muon EDM}},
  \href{https://doi.org/10.1103/PhysRevD.98.113002}{\emph{Phys. Rev. D}
  {\bfseries 98} (2018) 113002}
  [\href{https://arxiv.org/abs/1807.11484}{{\ttfamily 1807.11484}}].

\bibitem{Lindner:2016bgg}
M.~Lindner, M.~Platscher and F.~S. Queiroz, \emph{{A Call for New Physics : The
  Muon Anomalous Magnetic Moment and Lepton Flavor Violation}},
  \href{https://doi.org/10.1016/j.physrep.2017.12.001}{\emph{Phys. Rept.}
  {\bfseries 731} (2018) 1} [\href{https://arxiv.org/abs/1610.06587}{{\ttfamily
  1610.06587}}].

\bibitem{Bertl:2006up}
{\scshape SINDRUM II} collaboration, \emph{{A Search for muon to electron
  conversion in muonic gold}},
  \href{https://doi.org/10.1140/epjc/s2006-02582-x}{\emph{Eur. Phys. J. C}
  {\bfseries 47} (2006) 337}.

\bibitem{Aubert:2009ag}
{\scshape BaBar} collaboration, \emph{{Searches for Lepton Flavor Violation in
  the Decays $\tau^+\to e^+\gamma$ and $\tau^+\to \mu^+\gamma$ }},
  \href{https://doi.org/10.1103/PhysRevLett.104.021802}{\emph{Phys. Rev. Lett.}
  {\bfseries 104} (2010) 021802}
  [\href{https://arxiv.org/abs/0908.2381}{{\ttfamily 0908.2381}}].

\bibitem{TheMEG:2016wtm}
{\scshape MEG} collaboration, \emph{{Search for the lepton flavour violating
  decay $\mu ^+ \rightarrow \mathrm {e}^+ \gamma $ with the full dataset of the
  MEG experiment}},
  \href{https://doi.org/10.1140/epjc/s10052-016-4271-x}{\emph{Eur. Phys. J. C}
  {\bfseries 76} (2016) 434}
  [\href{https://arxiv.org/abs/1605.05081}{{\ttfamily 1605.05081}}].

\bibitem{Kou:2018nap}
{\scshape Belle-II} collaboration, \emph{{The Belle II Physics Book}},
  \href{https://doi.org/10.1093/ptep/ptz106}{\emph{PTEP} {\bfseries 2019}
  (2019) 123C01} [\href{https://arxiv.org/abs/1808.10567}{{\ttfamily
  1808.10567}}].

\bibitem{Baldini:2018nnn}
{\scshape MEG II} collaboration, \emph{{The design of the MEG II experiment}},
  \href{https://doi.org/10.1140/epjc/s10052-018-5845-6}{\emph{Eur. Phys. J. C}
  {\bfseries 78} (2018) 380}
  [\href{https://arxiv.org/abs/1801.04688}{{\ttfamily 1801.04688}}].

\bibitem{Hanneke:2008tm}
D.~Hanneke, S.~Fogwell and G.~Gabrielse, \emph{{New Measurement of the Electron
  Magnetic Moment and the Fine Structure Constant}},
  \href{https://doi.org/10.1103/PhysRevLett.100.120801}{\emph{Phys. Rev. Lett.}
  {\bfseries 100} (2008) 120801}
  [\href{https://arxiv.org/abs/0801.1134}{{\ttfamily 0801.1134}}].

\bibitem{Aoyama:2017uqe}
T.~Aoyama, T.~Kinoshita and M.~Nio, \emph{{Revised and Improved Value of the
  QED Tenth-Order Electron Anomalous Magnetic Moment}},
  \href{https://doi.org/10.1103/PhysRevD.97.036001}{\emph{Phys. Rev. D}
  {\bfseries 97} (2018) 036001}
  [\href{https://arxiv.org/abs/1712.06060}{{\ttfamily 1712.06060}}].

\bibitem{Laporta:2017okg}
S.~Laporta, \emph{{High-precision calculation of the 4-loop contribution to the
  electron g-2 in QED}},
  \href{https://doi.org/10.1016/j.physletb.2017.06.056}{\emph{Phys. Lett. B}
  {\bfseries 772} (2017) 232}
  [\href{https://arxiv.org/abs/1704.06996}{{\ttfamily 1704.06996}}].

\bibitem{Parker:2018vye}
R.~H. Parker, C.~Yu, W.~Zhong, B.~Estey and H.~M\"uller, \emph{{Measurement of
  the fine-structure constant as a test of the Standard Model}},
  \href{https://doi.org/10.1126/science.aap7706}{\emph{Science} {\bfseries 360}
  (2018) 191} [\href{https://arxiv.org/abs/1812.04130}{{\ttfamily
  1812.04130}}].

\bibitem{Morel:2020dww}
L.~Morel, Z.~Yao, P.~Clad\'e and S.~Guellati-Kh\'elifa, \emph{{Determination of
  the fine-structure constant with an accuracy of 81 parts per trillion}},
  \href{https://doi.org/10.1038/s41586-020-2964-7}{\emph{Nature} {\bfseries
  588} (2020) 61}.

\bibitem{Borsanyi:2020mff}
S.~Borsanyi et~al., \emph{{Leading hadronic contribution to the muon 2 magnetic
  moment from lattice QCD}},
  \href{https://arxiv.org/abs/2002.12347}{{\ttfamily 2002.12347}}.

\bibitem{Hoferichter:2019gzf}
M.~Hoferichter, B.-L. Hoid and B.~Kubis, \emph{{Three-pion contribution to
  hadronic vacuum polarization}},
  \href{https://doi.org/10.1007/JHEP08(2019)137}{\emph{JHEP} {\bfseries 08}
  (2019) 137} [\href{https://arxiv.org/abs/1907.01556}{{\ttfamily
  1907.01556}}].

\bibitem{Davier:2019can}
M.~Davier, A.~Hoecker, B.~Malaescu and Z.~Zhang, \emph{{A new evaluation of the
  hadronic vacuum polarisation contributions to the muon anomalous magnetic
  moment and to $\alpha(M_Z^2)$}},
  \href{https://doi.org/10.1140/epjc/s10052-020-7792-2}{\emph{Eur. Phys. J. C}
  {\bfseries 80} (2020) 241}
  [\href{https://arxiv.org/abs/1908.00921}{{\ttfamily 1908.00921}}].

\bibitem{Colangelo:2018mtw}
G.~Colangelo, M.~Hoferichter and P.~Stoffer, \emph{{Two-pion contribution to
  hadronic vacuum polarization}},
  \href{https://doi.org/10.1007/JHEP02(2019)006}{\emph{JHEP} {\bfseries 02}
  (2019) 006} [\href{https://arxiv.org/abs/1810.00007}{{\ttfamily
  1810.00007}}].

\bibitem{Ananthanarayan:2018nyx}
B.~Ananthanarayan, I.~Caprini and D.~Das, \emph{{Pion electromagnetic form
  factor at high precision with implications to $a_\mu^{\pi\pi}$ and the onset
  of perturbative QCD}},
  \href{https://doi.org/10.1103/PhysRevD.98.114015}{\emph{Phys. Rev. D}
  {\bfseries 98} (2018) 114015}
  [\href{https://arxiv.org/abs/1810.09265}{{\ttfamily 1810.09265}}].

\bibitem{Davier:2017zfy}
M.~Davier, A.~Hoecker, B.~Malaescu and Z.~Zhang, \emph{{Reevaluation of the
  hadronic vacuum polarisation contributions to the Standard Model predictions
  of the muon $g-2$ and ${\alpha (m_Z^2)}$ using newest hadronic cross-section
  data}}, \href{https://doi.org/10.1140/epjc/s10052-017-5161-6}{\emph{Eur.
  Phys. J. C} {\bfseries 77} (2017) 827}
  [\href{https://arxiv.org/abs/1706.09436}{{\ttfamily 1706.09436}}].

\bibitem{Keshavarzi:2018mgv}
A.~Keshavarzi, D.~Nomura and T.~Teubner, \emph{{Muon $g-2$ and $\alpha(M_Z^2)$:
  a new data-based analysis}},
  \href{https://doi.org/10.1103/PhysRevD.97.114025}{\emph{Phys. Rev. D}
  {\bfseries 97} (2018) 114025}
  [\href{https://arxiv.org/abs/1802.02995}{{\ttfamily 1802.02995}}].

\bibitem{Keshavarzi:2019abf}
A.~Keshavarzi, D.~Nomura and T.~Teubner, \emph{{$g-2$ of charged leptons,
  $\alpha (M^2_Z)$ , and the hyperfine splitting of muonium}},
  \href{https://doi.org/10.1103/PhysRevD.101.014029}{\emph{Phys. Rev. D}
  {\bfseries 101} (2020) 014029}
  [\href{https://arxiv.org/abs/1911.00367}{{\ttfamily 1911.00367}}].

\bibitem{Crivellin:2020zul}
A.~Crivellin, M.~Hoferichter, C.~A. Manzari and M.~Montull, \emph{{Hadronic
  Vacuum Polarization: $(g-2)_\mu$ versus Global Electroweak Fits}},
  \href{https://doi.org/10.1103/PhysRevLett.125.091801}{\emph{Phys. Rev. Lett.}
  {\bfseries 125} (2020) 091801}
  [\href{https://arxiv.org/abs/2003.04886}{{\ttfamily 2003.04886}}].

\bibitem{Keshavarzi:2020bfy}
A.~Keshavarzi, W.~J. Marciano, M.~Passera and A.~Sirlin, \emph{{Muon $g-2$ and
  $\Delta \alpha$ connection}},
  \href{https://doi.org/10.1103/PhysRevD.102.033002}{\emph{Phys. Rev. D}
  {\bfseries 102} (2020) 033002}
  [\href{https://arxiv.org/abs/2006.12666}{{\ttfamily 2006.12666}}].

\bibitem{Bennett:2008dy}
{\scshape Muon (g-2)} collaboration, \emph{{An Improved Limit on the Muon
  Electric Dipole Moment}},
  \href{https://doi.org/10.1103/PhysRevD.80.052008}{\emph{Phys. Rev. D}
  {\bfseries 80} (2009) 052008}
  [\href{https://arxiv.org/abs/0811.1207}{{\ttfamily 0811.1207}}].

\bibitem{Baron:2013eja}
{\scshape ACME} collaboration, \emph{{Order of Magnitude Smaller Limit on the
  Electric Dipole Moment of the Electron}},
  \href{https://doi.org/10.1126/science.1248213}{\emph{Science} {\bfseries 343}
  (2014) 269} [\href{https://arxiv.org/abs/1310.7534}{{\ttfamily 1310.7534}}].

\bibitem{Andreev:2018ayy}
{\scshape ACME} collaboration, \emph{{Improved limit on the electric dipole
  moment of the electron}},
  \href{https://doi.org/10.1038/s41586-018-0599-8}{\emph{Nature} {\bfseries
  562} (2018) 355}.

\bibitem{Inami:2002ah}
{\scshape Belle} collaboration, \emph{{Search for the electric dipole moment of
  the tau lepton}},
  \href{https://doi.org/10.1016/S0370-2693(02)02984-2}{\emph{Phys. Lett. B}
  {\bfseries 551} (2003) 16}
  [\href{https://arxiv.org/abs/hep-ex/0210066}{{\ttfamily hep-ex/0210066}}].

\bibitem{Abdallah:2003xd}
{\scshape DELPHI} collaboration, \emph{{Study of tau-pair production in
  photon-photon collisions at LEP and limits on the anomalous electromagnetic
  moments of the tau lepton}},
  \href{https://doi.org/10.1140/epjc/s2004-01852-y}{\emph{Eur. Phys. J. C}
  {\bfseries 35} (2004) 159}
  [\href{https://arxiv.org/abs/hep-ex/0406010}{{\ttfamily hep-ex/0406010}}].

\bibitem{Eidelman:2007sb}
S.~Eidelman and M.~Passera, \emph{{Theory of the tau lepton anomalous magnetic
  moment}}, \href{https://doi.org/10.1142/S0217732307022694}{\emph{Mod. Phys.
  Lett. A} {\bfseries 22} (2007) 159}
  [\href{https://arxiv.org/abs/hep-ph/0701260}{{\ttfamily hep-ph/0701260}}].

\bibitem{Adelmann:2021udj}
A.~Adelmann et~al., \emph{{Search for a muon EDM using the frozen-spin
  technique}},  \href{https://arxiv.org/abs/2102.08838}{{\ttfamily
  2102.08838}}.

\bibitem{Bellgardt:1987du}
{\scshape SINDRUM} collaboration, \emph{{Search for the Decay $\mu^+\to e^+ e^+
  e^-$}}, \href{https://doi.org/10.1016/0550-3213(88)90462-2}{\emph{Nucl. Phys.
  B} {\bfseries 299} (1988) 1}.

\bibitem{Hayasaka:2010np}
K.~Hayasaka et~al., \emph{{Search for Lepton Flavor Violating Tau Decays into
  Three Leptons with 719 Million Produced Tau+Tau- Pairs}},
  \href{https://doi.org/10.1016/j.physletb.2010.03.037}{\emph{Phys. Lett. B}
  {\bfseries 687} (2010) 139}
  [\href{https://arxiv.org/abs/1001.3221}{{\ttfamily 1001.3221}}].

\bibitem{Lees:2010ez}
{\scshape BaBar} collaboration, \emph{{Limits on tau Lepton-Flavor Violating
  Decays in three charged leptons}},
  \href{https://doi.org/10.1103/PhysRevD.81.111101}{\emph{Phys. Rev. D}
  {\bfseries 81} (2010) 111101}
  [\href{https://arxiv.org/abs/1002.4550}{{\ttfamily 1002.4550}}].

\bibitem{Aaij:2014azz}
{\scshape LHCb} collaboration, \emph{{Search for the lepton flavour violating
  decay \ensuremath{\tau}$^{-}$ \textrightarrow{} \ensuremath{\mu}$^{-}$
  \ensuremath{\mu}$^{+}$ \ensuremath{\mu}$^{-}$}},
  \href{https://doi.org/10.1007/JHEP02(2015)121}{\emph{JHEP} {\bfseries 02}
  (2015) 121} [\href{https://arxiv.org/abs/1409.8548}{{\ttfamily 1409.8548}}].

\bibitem{Cerri:2018ypt}
A.~Cerri et~al., \emph{{Report from Working Group 4}: {Opportunities in Flavour
  Physics at the HL-LHC and HE-LHC}},
  \href{https://doi.org/10.23731/CYRM-2019-007.867}{\emph{CERN Yellow Rep.
  Monogr.} {\bfseries 7} (2019) 867}
  [\href{https://arxiv.org/abs/1812.07638}{{\ttfamily 1812.07638}}].

\bibitem{Blondel:2013ia}
A.~Blondel et~al., \emph{{Research Proposal for an Experiment to Search for the
  Decay $\mu \to eee$}},  \href{https://arxiv.org/abs/1301.6113}{{\ttfamily
  1301.6113}}.

\bibitem{Perrevoort:2016nuv}
{\scshape Mu3e} collaboration, \emph{{Status of the Mu3e Experiment at PSI}},
  \href{https://doi.org/10.1051/epjconf/201611801028}{\emph{EPJ Web Conf.}
  {\bfseries 118} (2016) 01028}
  [\href{https://arxiv.org/abs/1605.02906}{{\ttfamily 1605.02906}}].

\bibitem{Kitano:2002mt}
R.~Kitano, M.~Koike and Y.~Okada, \emph{{Detailed calculation of lepton flavor
  violating muon electron conversion rate for various nuclei}},
  \href{https://doi.org/10.1103/PhysRevD.76.059902}{\emph{Phys. Rev. D}
  {\bfseries 66} (2002) 096002}
  [\href{https://arxiv.org/abs/hep-ph/0203110}{{\ttfamily hep-ph/0203110}}].

\bibitem{Suzuki:1987jf}
T.~Suzuki, D.~F. Measday and J.~P. Roalsvig, \emph{{Total Nuclear Capture Rates
  for Negative Muons}},
  \href{https://doi.org/10.1103/PhysRevC.35.2212}{\emph{Phys. Rev. C}
  {\bfseries 35} (1987) 2212}.

\bibitem{Baldini:2018uhj}
A.~Baldini et~al., \emph{{A submission to the 2020 update of the European
  Strategy for Particle Physics on behalf of the COMET, MEG, Mu2e and Mu3e
  collaborations}},  \href{https://arxiv.org/abs/1812.06540}{{\ttfamily
  1812.06540}}.

\bibitem{ALEPH:2005ab}
{\scshape ALEPH, DELPHI, L3, OPAL, SLD, LEP Electroweak Working Group, SLD
  Electroweak Group, SLD Heavy Flavour Group} collaboration, \emph{{Precision
  electroweak measurements on the $Z$ resonance}},
  \href{https://doi.org/10.1016/j.physrep.2005.12.006}{\emph{Phys. Rept.}
  {\bfseries 427} (2006) 257}
  [\href{https://arxiv.org/abs/hep-ex/0509008}{{\ttfamily hep-ex/0509008}}].

\bibitem{Tanabashi:2018oca}
{\scshape Particle Data Group} collaboration, \emph{{Review of Particle
  Physics}}, \href{https://doi.org/10.1103/PhysRevD.98.030001}{\emph{Phys. Rev.
  D} {\bfseries 98} (2018) 030001}.

\bibitem{Aaltonen:2016nuy}
{\scshape CDF} collaboration, \emph{{Measurement of $\sin^2\theta^{\rm
  lept}_{\rm eff}$ using $e^+e^-$ pairs from $\gamma^*/Z$ bosons produced in
  $p\bar{p}$ collisions at a center-of-momentum energy of 1.96 TeV}},
  \href{https://doi.org/10.1103/PhysRevD.93.112016}{\emph{Phys. Rev. D}
  {\bfseries 93} (2016) 112016}
  [\href{https://arxiv.org/abs/1605.02719}{{\ttfamily 1605.02719}}].

\bibitem{Chatrchyan:2011ya}
{\scshape CMS} collaboration, \emph{{Measurement of the weak mixing angle with
  the Drell-Yan process in proton-proton collisions at the LHC}},
  \href{https://doi.org/10.1103/PhysRevD.84.112002}{\emph{Phys. Rev. D}
  {\bfseries 84} (2011) 112002}
  [\href{https://arxiv.org/abs/1110.2682}{{\ttfamily 1110.2682}}].

\bibitem{Aaij:2015lka}
{\scshape LHCb} collaboration, \emph{{Measurement of the forward-backward
  asymmetry in $Z/\gamma^{\ast} \rightarrow \mu^{+}\mu^{-}$ decays and
  determination of the effective weak mixing angle}},
  \href{https://doi.org/10.1007/JHEP11(2015)190}{\emph{JHEP} {\bfseries 11}
  (2015) 190} [\href{https://arxiv.org/abs/1509.07645}{{\ttfamily
  1509.07645}}].

\bibitem{Haisch:2011up}
U.~Haisch and S.~Westhoff, \emph{{Massive Color-Octet Bosons: Bounds on Effects
  in Top-Quark Pair Production}},
  \href{https://doi.org/10.1007/JHEP08(2011)088}{\emph{JHEP} {\bfseries 08}
  (2011) 088} [\href{https://arxiv.org/abs/1106.0529}{{\ttfamily 1106.0529}}].

\bibitem{Aaboud:2018wps}
{\scshape ATLAS} collaboration, \emph{{Measurement of the Higgs boson mass in
  the $H\rightarrow ZZ^* \rightarrow 4\ell$ and $H \rightarrow \gamma\gamma$
  channels with $\sqrt{s}=13$ TeV $pp$ collisions using the ATLAS detector}},
  \href{https://doi.org/10.1016/j.physletb.2018.07.050}{\emph{Phys. Lett. B}
  {\bfseries 784} (2018) 345}
  [\href{https://arxiv.org/abs/1806.00242}{{\ttfamily 1806.00242}}].

\bibitem{CMS:2019drq}
{\scshape CMS} collaboration, \emph{{A measurement of the Higgs boson mass in
  the diphoton decay channel}}, .

\bibitem{TevatronElectroweakWorkingGroup:2016lid}
{\scshape CDF, D0} collaboration, \emph{{Combination of CDF and D0 results on
  the mass of the top quark using up $9.7\:{\rm fb}^{-1}$ at the Tevatron}},
  \href{https://arxiv.org/abs/1608.01881}{{\ttfamily 1608.01881}}.

\bibitem{Aaboud:2018zbu}
{\scshape ATLAS} collaboration, \emph{{Measurement of the top quark mass in the
  $t\bar{t}\rightarrow $ lepton+jets channel from $\sqrt{s}=8$ TeV ATLAS data
  and combination with previous results}},
  \href{https://doi.org/10.1140/epjc/s10052-019-6757-9}{\emph{Eur. Phys. J. C}
  {\bfseries 79} (2019) 290}
  [\href{https://arxiv.org/abs/1810.01772}{{\ttfamily 1810.01772}}].

\bibitem{Sirunyan:2018mlv}
{\scshape CMS} collaboration, \emph{{Measurement of the top quark mass in the
  all-jets final state at $\sqrt{s} =$ 13 TeV and combination with the
  lepton+jets channel}},
  \href{https://doi.org/10.1140/epjc/s10052-019-6788-2}{\emph{Eur. Phys. J. C}
  {\bfseries 79} (2019) 313}
  [\href{https://arxiv.org/abs/1812.10534}{{\ttfamily 1812.10534}}].

\bibitem{Aad:2020gkd}
{\scshape ATLAS} collaboration, \emph{{Charged-lepton-flavour violation at the
  LHC: a search for $Z\to e\tau/\mu\tau$ decays with the ATLAS detector}},
  \href{https://arxiv.org/abs/2010.02566}{{\ttfamily 2010.02566}}.

\bibitem{Akers:1995gz}
{\scshape OPAL} collaboration, \emph{{A Search for lepton flavor violating Z0
  decays}}, \href{https://doi.org/10.1007/BF01553981}{\emph{Z. Phys. C}
  {\bfseries 67} (1995) 555}.

\bibitem{Abreu:1996mj}
{\scshape DELPHI} collaboration, \emph{{Search for lepton flavor number
  violating Z0 decays}}, \href{https://doi.org/10.1007/s002880050313}{\emph{Z.
  Phys. C} {\bfseries 73} (1997) 243}.

\bibitem{Aad:2014bca}
{\scshape ATLAS} collaboration, \emph{{Search for the lepton flavor violating
  decay Z\textrightarrow{}e\ensuremath{\mu} in pp collisions at $\sqrt{s}$ TeV
  with the ATLAS detector}},
  \href{https://doi.org/10.1103/PhysRevD.90.072010}{\emph{Phys. Rev. D}
  {\bfseries 90} (2014) 072010}
  [\href{https://arxiv.org/abs/1408.5774}{{\ttfamily 1408.5774}}].

\bibitem{Geiregat:1990gz}
{\scshape CHARM-II} collaboration, \emph{{First observation of neutrino trident
  production}}, \href{https://doi.org/10.1016/0370-2693(90)90146-W}{\emph{Phys.
  Lett. B} {\bfseries 245} (1990) 271}.

\bibitem{Mishra:1991bv}
{\scshape CCFR} collaboration, \emph{{Neutrino tridents and W Z interference}},
  \href{https://doi.org/10.1103/PhysRevLett.66.3117}{\emph{Phys. Rev. Lett.}
  {\bfseries 66} (1991) 3117}.

\bibitem{Adams:1998yf}
{\scshape NuTeV} collaboration, \emph{{Neutrino trident production from
  NuTeV}},  in \emph{{29th International Conference on High-Energy Physics}},
  pp.~631--634, 7, 1998, \href{https://arxiv.org/abs/hep-ex/9811012}{{\ttfamily
  hep-ex/9811012}}.

\bibitem{Caldwell:2008fw}
A.~Caldwell, D.~Kollar and K.~Kroninger, \emph{{BAT: The Bayesian Analysis
  Toolkit}}, \href{https://doi.org/10.1016/j.cpc.2009.06.026}{\emph{Comput.
  Phys. Commun.} {\bfseries 180} (2009) 2197}
  [\href{https://arxiv.org/abs/0808.2552}{{\ttfamily 0808.2552}}].

\bibitem{Abada:2019zxq}
{\scshape FCC} collaboration, \emph{{FCC-ee: The Lepton Collider}: {Future
  Circular Collider Conceptual Design Report Volume 2}},
  \href{https://doi.org/10.1140/epjst/e2019-900045-4}{\emph{Eur. Phys. J. ST}
  {\bfseries 228} (2019) 261}.

\bibitem{CEPCStudyGroup:2018ghi}
{\scshape CEPC Study Group} collaboration, \emph{{CEPC Conceptual Design
  Report: Volume 2 - Physics \& Detector}},
  \href{https://arxiv.org/abs/1811.10545}{{\ttfamily 1811.10545}}.

\bibitem{Crivellin:2017rmk}
A.~Crivellin, S.~Davidson, G.~M. Pruna and A.~Signer,
  \emph{{Renormalisation-group improved analysis of $\mu\to e$ processes in a
  systematic effective-field-theory approach}},
  \href{https://doi.org/10.1007/JHEP05(2017)117}{\emph{JHEP} {\bfseries 05}
  (2017) 117} [\href{https://arxiv.org/abs/1702.03020}{{\ttfamily
  1702.03020}}].

\bibitem{Pruna:2014asa}
G.~M. Pruna and A.~Signer, \emph{{The $\mu\to e\gamma$ decay in a systematic
  effective field theory approach with dimension 6 operators}},
  \href{https://doi.org/10.1007/JHEP10(2014)014}{\emph{JHEP} {\bfseries 10}
  (2014) 014} [\href{https://arxiv.org/abs/1408.3565}{{\ttfamily 1408.3565}}].

\bibitem{Aebischer:2017gaw}
J.~Aebischer, M.~Fael, C.~Greub and J.~Virto, \emph{{B physics Beyond the
  Standard Model at One Loop: Complete Renormalization Group Evolution below
  the Electroweak Scale}},
  \href{https://doi.org/10.1007/JHEP09(2017)158}{\emph{JHEP} {\bfseries 09}
  (2017) 158} [\href{https://arxiv.org/abs/1704.06639}{{\ttfamily
  1704.06639}}].

\end{thebibliography}\endgroup

\end{document}